\UseRawInputEncoding
\documentclass[twocolumn,
nofootinbib,superscriptaddress,aps,10pt,longbibliography]{revtex4-1}

\usepackage{bbold}

\usepackage[usenames,dvipsnames]{xcolor}
\usepackage{amsmath}
\usepackage{amsthm, amssymb}
\usepackage{enumitem}
\usepackage{slashed}
\usepackage{graphicx}
\usepackage{tikz}
\usetikzlibrary{calc,fadings,decorations.pathreplacing,shapes,shapes.multipart,arrows,shapes.misc,intersections,positioning,patterns}
\usepackage{bm}
\usepackage{dsfont}
\usepackage{changepage}
\usepackage{array}

\definecolor{bluepurple2}{rgb}{0.06,0,0.6}
\usepackage[colorlinks=true,citecolor=blue,linkcolor=bluepurple2]{hyperref}

\usepackage{bm}
\renewcommand{\vec}[1]{\boldsymbol{\mathbf{#1}}}

\newcommand{\bit}{\begin{itemize}}
\newcommand{\eit}{\end{itemize}}

\newcommand{\f}{\frac}
\renewcommand{\>}{\right\rangle}
\newcommand{\<}{\left\langle}
\newcommand{\ba}{\begin{align}}
\newcommand{\ea}{\end{align}}
\newcommand{\be}{\begin{equation}}
\newcommand{\ee}{\end{equation}}
\newcommand{\bi}{\begin{itemize}}
\newcommand{\ei}{\end{itemize}}
\newcommand{\lf}{\left(}
\newcommand{\ri}{\right)}
\newcommand{\dd}{\mathrm{d}}

\DeclareRobustCommand{\dotK}{\raisebox{-0.1em}{\includegraphics[height=0.6em]{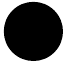}}}
\DeclareRobustCommand{\doublelineK}{\raisebox{0.03em}{\includegraphics[height=0.35em]{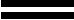}}}

\begin{document}
\date{\today}

\newcommand{\bra}[1]{\< #1 \right|}
\newcommand{\ket}[1]{\left| #1 \>}

\title{Self-dual criticality in three-dimensional $\mathbb{Z}_2$ gauge theory with matter}

\author{Andr\'es M. Somoza}
\affiliation{Departamento de F\'isica -- CIOyN, Universidad de Murcia, Murcia 30.071, Spain}
\author{Pablo Serna}
\affiliation{Departamento de F\'isica -- CIOyN, Universidad de Murcia, Murcia 30.071, Spain}
\affiliation{Laboratoire de Physique de l'\'Ecole Normale Sup\'erieure, ENS, Universit\'e PSL, CNRS, Sorbonne Universit\'e,
Universit\'e Paris-Diderot, Sorbonne Paris Cit\'e, Paris, France.}
\author{Adam Nahum}
\affiliation{
Theoretical Physics, University of Oxford, Parks Road, Oxford OX1 3PU, United Kingdom}

\date{\today}

\begin{abstract}
The simplest topologically ordered phase in 2+1D is the deconfined phase of $\mathbb{Z}_2$ gauge theory,  realized for example in the toric code. 
This phase permits a duality that exchanges electric and magnetic excitations (``$e$'' and ``$m$'' particles). 
Condensing either particle while the other remains gapped yields a phase transition with 3D Ising exponents.
More mysterious, however, is the transition out of the deconfined phase when self-duality symmetry is preserved.
If this transition is continuous, which has so far been unclear, then it may be the simplest critical point for which we still lack any useful continuum Lagrangian description.
This transition also has a soft matter interpretation, as a multicritical point for classical membranes in 3D.

We study the self-dual transition with Monte Carlo simulations of the $\mathbb{Z}_2$ gauge-Higgs model on cubic lattices of linear size $L\leq 96$.
Our results indicate a continuous transition: for example, cumulants show a striking parameter-free scaling collapse.
We estimate scaling dimensions by using duality symmetry to distinguish the leading duality-odd/duality-even scaling operators $A$ and $S$. 
All local operators have large scaling dimensions, making standard techniques for locating the critical point ineffective. 
We develop an alternative using ``renormalization group trajectories'' of cumulants.  We check that two- and three-point functions, and temporal correlators in the Monte-Carlo dynamics, are scale-invariant, with scaling dimensions $x_A$ and $x_S$ and dynamical exponent $z$. 

We also give a picture for emergence of 1-form symmetries, in some parts of the phase diagram, in terms of ``patching'' of membranes/worldsurfaces. 
We relate this to the percolation of anyon worldlines in spacetime. 
Analyzing percolation yields a fourth exponent for the self-dual transition. We propose variations of the model for further investigation.
 \end{abstract}
\maketitle

\section{Introduction}
\label{sec:intro}

Continuum field theory provides a language for a huge range of classical and quantum phase transitions \cite{fisher1974renormalization,sachdev2007quantum}. 
This includes many cases for which a simple Landau-Ginsburg formulation is insufficient
 \cite{wegner1971duality,kosterlitz1973ordering, de1972exponents, huse1991sponge, lammert1993topology,francesco2012conformal, fradkin2013field, senthil2004deconfined}.
For example, a wide range of topological phase transitions, 
lacking any local order parameter \cite{wegner1971duality}, may  be brought under some measure of analytical control
using the language of continuum gauge theory, together with various kinds of perturbative expansion ($\epsilon$ expansions, large $N$ expansions, etcetera).
However, despite the wild success of the field theory approach to critical phenomena, 
there exist phase transitions in simple and natural models
that still remain out of reach of field theory tools.
This paper characterizes what we suggest is the paradigmatic example of these mysterious transitions.
This is the ``self-dual'' phase transition between confined and deconfined phases of $\mathbb{Z}_2$ gauge theory in three dimensions \cite{fradkin1979phase,jongeward1980monte,TupitsynTopological,vidal2009low}.

The $\mathbb{Z}_2$ gauge--Higgs model \cite{wegner1971duality, kogut1979introduction, fradkin1979phase} has a stable deconfined phase, as well as a trivial phase, in three dimensions.
In the context of quantum systems in 2+1D
(the model also has a 3D classical interpretation that we discuss below)
the deconfined phase is the simplest $\mathbb{Z}_2$ spin liquid \cite{read1989statistics, kivelson1989statistics,
read1991large, wen1991mean, senthil2000z, moessner2001short}: 
the phase of matter  realized, for example, by the toric code \cite{kitaev2003fault}. 
The anyon excitations of this phase include quasiparticles denoted  
 $e$ and $m$,  with nontrivial mutual statistics, which
 correspond to charge and flux in the gauge theory.

\begin{figure}[b]
\includegraphics[width=0.7\linewidth]{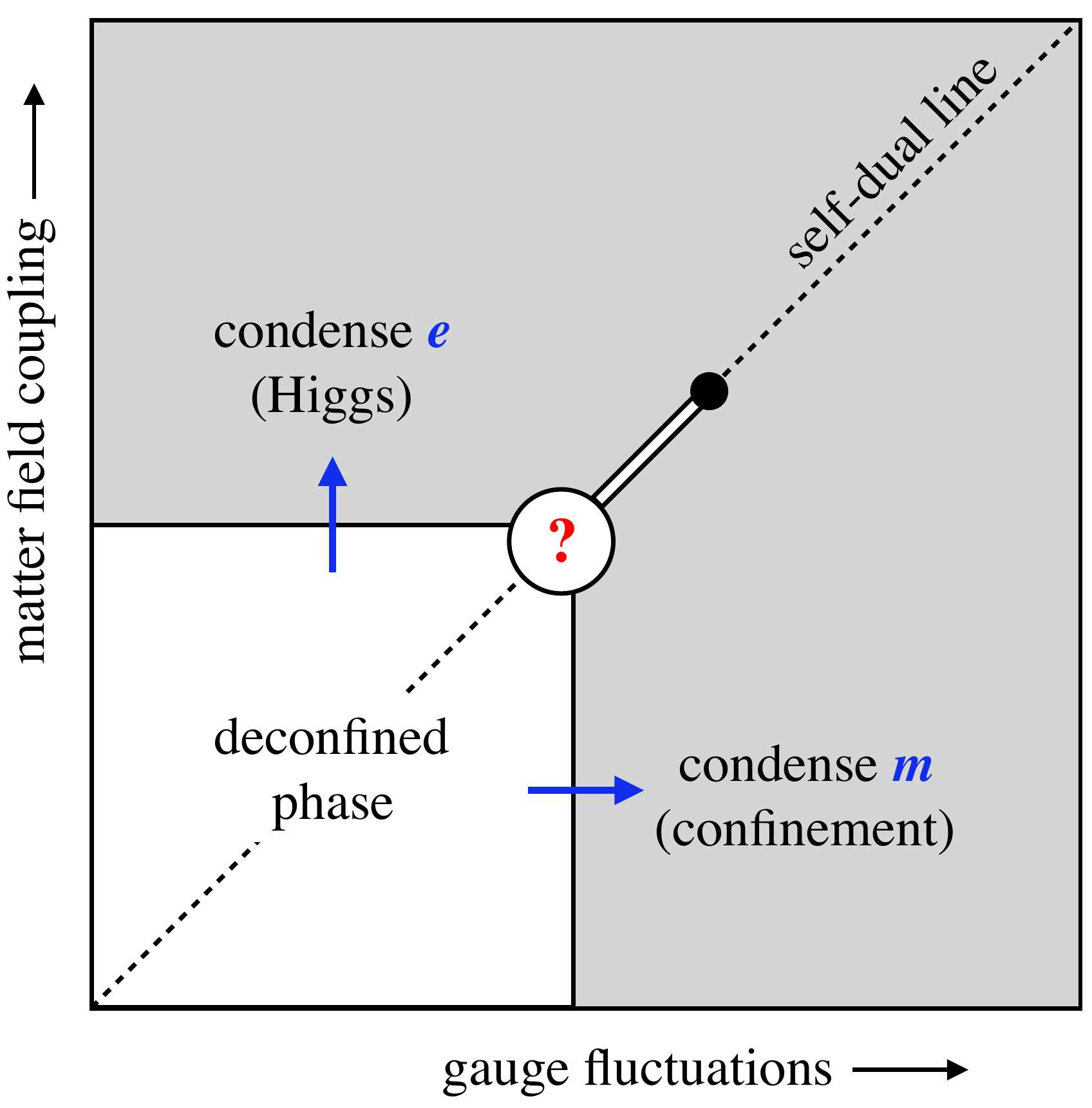}
\caption{Topology of the phase diagram of $\mathbb{Z}_2$ gauge theory with matter.
The shaded region is the trivial phase. 
The double line ($\doublelineK$) represents a first-order line, ending at a standard critical endpoint ($\dotK$).
The Higgs and confinement transitions have Ising exponents. 
The question mark shows the region studied in this paper. We give evidence for a scale-invariant, self-dual critical point.}
\label{fig:schematic_phase_diag}
\end{figure}

The simplest lattice field theory formulation of the $\mathbb{Z}_2$ gauge-Higgs model has a two-dimensional parameter space 
\cite{wegner1971duality, balian1974gauge, kogut1979introduction, fradkin1979phase,jongeward1980monte,TupitsynTopological}.
In the quantum language, these two couplings allow us to separately control the masses of both $e$ and $m$ excitations. 
In the language of the lattice gauge theory, one of the couplings controls the ``stiffness'' associated with fluctuations of the matter field, and the other the stiffness of gauge field:  see the schematic Fig.~\ref{fig:schematic_phase_diag}. 

While there are only two stable phases in Fig.~\ref{fig:schematic_phase_diag}, there are various possibilities for the transition between them  \cite{fradkin1979phase,jongeward1980monte,huse1991sponge,vidal2009low,TupitsynTopological}.
The Higgs and confinement transitions correspond to condensation of the $e$ particle, and of the $m$ particle, respectively.
These two lines of transitions are in fact completely equivalent, as they are mapped into each other by the crucial duality transformation, which exchanges the two kinds of particle. 
They are subtle phase transitions with no local order parameter \cite{wegner1971duality}. 
Nevertheless, they \textit{are} amenable to field theory tools.
For example, gauge fluctuations are in fact irrelevant at the Higgs transition \cite{fradkin1979phase}.
Its universal scaling is therefore same as in the limit of infinite gauge stiffness, 
where the partition function is simply related to that of the standard Ising model (with a sum over boundary conditions): in a sense, we can define a ``fictitious'' Ising order parameter.
In the language of anyons, the reason that this  transition (where $e$ condenses) is relatively conventional is because $m$ remains massive: this ensures that nontrivial braiding processes are not important at low energies.

By contrast, the nature of the transition out of the deconfined phase \textit{on the self-dual line}, where there is a symmetry between $e$ and $m$,
has not been understood.
 Previous Monte-Carlo \cite{TupitsynTopological} and series expansion \cite{vidal2009low} studies gave some evidence for a multicritical point here, but the order of the transition, and the structure of the phase diagram close to this ``corner'' of the deconfined phase, have not been definitively resolved \cite{TupitsynTopological,wu2012phase}.
The argument above, which relates the Higgs transition to a simple Landau theory, no longer applies, so that we really have to confront the issue of coarse-graining a discrete gauge field, 
whose low-lying excitations have nontrivial mutual statistics.

The basic challenge can also be understood in geometrical terms.
The gauge-Higgs model describes various phases of fluctuating membranes in three spatial dimensions \cite{huse1991sponge}.
This is a fascinating system in its own right, relevant to experiments on amphiphilic membranes, where  the  deconfined phase is known as the ``symmetric sponge'' phase \cite{huse1988phase,cates1988random,david1989n,roux1990light,roux1992sponge,roux1995sponge,pelitiamphiphilic}.
We argue below that, if the ``holes'' in these membranes are ``small'', and disappear under coarse-graining, the membranes are effectively \textit{closed} surfaces. 
Mapping them to Ising domain walls is then one way to think about the  fictitious Ising order  parameter described above. 
But as the self-dual point is approached, 
the holes become large (we demonstrate this explicitly),
so that this way of thinking breaks down.

In this paper we determine many of the properties of the self-dual transition using extensive Monte Carlo simulations and arguments based on the renormalization group and symmetry. 
Our numerical results include the first demonstration of scale-invariance at this transition, via scaling collapse of numerous observables.
Our results for exponents also raise intriguing theoretical questions about how to understand this transition.

First, we give strong evidence that the transition is governed by a scale-invariant fixed point,
for example via a striking  scaling collapse that does not require any fitting parameters. 
We classify the leading local operators $S(r)$ and $A(r)$ as even and odd under duality symmetry respectively, 
and estimate their scaling dimensions $x_S$ and $x_A$ using scaling collapses and two-point functions. 
We check that three point functions are compatible with conformal invariance.

We also address some fundamental aspects of the anyon condensation transitions away from the self-dual line. As  noted above,  a key feature of the Higgs and confinement transitions is the possibility of using a Landau theory for a ``fictitious'' Ising order parameter.  (These are sometimes referred to as ``Ising$^*$'' transitions \cite{nandkishore2012orthogonal}.)
The emergence of this order parameter may be related to the  question of where in the phase diagram certain  ``one-form'' symmetries \cite{kapustin2017higher,gaiotto2015generalized, wen2019emergent} emerge under coarse-graining. 
We propose an explicit  construction of the fictitious Ising field (and of the string operators of the one-form symmetry). This construction is based on  ``repairing'' or ``patching'' the membranes that appear in a geometrical representation of the partition function.
We relate the feasibility of this patching operation to the question of whether $e$ and $m$ worldlines ``percolate''  in spacetime,
and obtain the phase diagram for this percolation \cite{huse1991sponge} numerically.
This shows that the fictitious Ising fields can be constructed on the Ising$^*$ transition lines, but not at the self-dual critical point.
However, we find that scale invariance at the self-dual transition can be diagnosed via the percolation of worldlines,
and compute their universal fractal dimension $d_f$.
(The result hints at a possible relation between exponents.)

We discuss the role of self-duality symmetry, arguing that it becomes an emergent internal symmetry in the IR. 
While our numerical analysis here is for the standard gauge-Higgs model, we also propose a modified lattice model, with a simpler action of duality, which it would be interesting to study further. 

The dynamics of the Monte-Carlo algorithm  (in Monte-Carlo time) correspond to a physically sensible  universality class for the stochastic dynamics of membranes in 3D.
We find that the dynamical exponent for this universality class is $z\simeq 2.48$ 
(not to be confused with the dynamical exponent $z_\text{QM}=1$ of the zero-temperature   quantum dynamics in the 2+1D interpretation) and show that two-time correlation functions are another way to obtain $x_{A,S}$.
The fact that the dynamical exponent is  large is one of the challenges in simulating this model:  unlike in many simple ordering transitions \cite{newman1999monte}, no efficient nonlocal Monte Carlo update, that reduces $z$ to a small value, is known for this problem.

Various features of the fixed point make standard approaches
 to  determining the precise location of the phase transition point, and the order of the transition,  ineffective.
These features include the structure of Binder cumulants close to the transition,
the lack of any local operator with a small scaling dimension,
and the fact that $x_S$ is very close to 1.5. (This is the threshold separating divergence and convergence of the heat capacity, and the proximity  to this threshold leads to large scaling corrections in this quantity.)
These features were the bane of our initial attempts at data analysis. 
We describe how they may be overcome, for example by focusing on appropriate dimensionless observables that allow a parameter-free scaling collapse.

Our numerical estimates for the exponents $x_S$ and $x_A$ turn out to be close to certain exponents in the XY universality class. 
This is remarkable in view of the mutual statistics of the condensing quasiparticles \cite{vidal2009low,TupitsynTopological, geraedts2012monte,burnell2018anyon}, which  would make any relationship with the XY fixed point very surprising (Sec.~\ref{sec:XY}).
The fixed point studied here is certainly distinct from the XY fixed point, as implied  for example by the very different universal properties of the adjacent phases.
On the other hand, it is not hard to find examples of pairs of 3D fixed points with exponents that are fairly close, but distinct.
This issue requires further investigation.
There are also many variations of the present model that remain to be studied (Secs.~\ref{sec:perturbations},~\ref{sec:dimcrossover},~\ref{sec:outlook}).

Textbook discussions of critical phenomena sometimes give the impression that studying universality in phase transitions
is synonymous with studying  Lagrangian quantum field theory. 
Therefore it is important to remember that there are critical points for which we so far lack any useful continuum Lagrangian (Sec.~\ref{sec:outlook}). 
Given that the self-dual transition is second-order, as previously suspected \cite{TupitsynTopological, vidal2009low} and as the numerical evidence presented here shows, then it is perhaps the simplest example of one of these untamed beasts.

However, a rich variety of other topological transitions, with distinct (nontrivially braiding) anyons simultaneously becoming massless, are possible,
with other discrete gauge theories providing further simple examples. Models with $\mathrm{U}(1)$ symmmetry  are also interesting \cite{geraedts2012monte,geraedts2012phases,lee2016monte,geraedts2012line,geraedts2012monte}, though they are more closely connected to continuum  $\mathrm{U}(1)$ gauge theory (perhaps with Chern-Simons terms). 
A systematic program to understand all of these transitions would be valuable. Past results on the formulation of field theories for deconfined phase transitions \cite{senthil2004deconfined,levin2004deconfined,motrunich2004emergent,senthil2005deconfined,tanaka2005many,senthil2006competing,sandvik2007evidence,bhattacharjee2011quantum,wang2017deconfined,gazit2018confinement,jiang2019ising}, where mutually nonlocal fields and Berry phases connecting different gapless degrees of freedom also play a key role, may provide some tools.

The $\mathbb{Z}_2$ deconfined phase is adjacent to another family of critical ``quantum loop models'' \cite{freedman2004class,freedman2005line,freedman2008lieb,troyer2008local,dai2020quantum} with no known Lagrangian description \cite{dai2020quantum}. 
Interestingly, while these critical points may again be viewed in terms of membranes in spacetime,  the obstacle to a continuum description is different  there: a topological constraint on the dynamics, rather than the existence of massless particles with nontrivial braiding.

These different kinds of examples suggest that statistical ensembles of membranes \cite{nelson2004statistical} in three and four dimensions (elementary ``string field theories'' \cite{polchinski1998string}) still hold many lessons for critical phenomena.

\tableofcontents

\section{The Ising gauge model }
\label{sec:model}

The gauge-Higgs model has many guises. 
We begin by reviewing several equivalent formulations of the partition function we study and the basic features of the phase diagram. 
Readers should skip topics with which they are familiar.

\subsection{As a lattice gauge theory}
\label{sec:modelasspins}

This is the standard formulation of $\mathbb{Z}_2$ gauge theory, with matter, on a cubic lattice \cite{wegner1971duality,kogut1979introduction}. ($\mathbb{Z}_2$ gauge theory is also referred to as ``Ising'' gauge theory.)
The degrees of freedom are classical Ising matter fields ${\tau_i=\pm 1}$ on the sites $i$ of the lattice and  gauge fields ${\sigma_{ij}= \pm 1}$ on the links $\<ij\>$.
The action includes a stiffness $K$ for the gauge fluctuations, and a coupling $J$ for the matter field. If $\square$ denotes a square plaquette, the partition function is:
\ba\label{eq:Z3Dgaugerep}
Z \propto 
\sum_{\{\sigma\}, \{\tau\}}
\exp \lf 
K \sum_{\square} \lf \prod_{\<ij\> \in \square} \sigma_{ij} \ri
+ 
J 
\sum_{\<ij\>} \tau_i \sigma_{ij} \, \tau_j \ri.
\end{align}
We work throughout on an ${L\times L\times L}$ torus.\footnote{We have used a $\propto$ sign rather than an equality because we will choose to absorb a trivial constant into the definition of $Z$.} 
The action is invariant under the $\mathbb{Z}_2$ gauge transformation ${\tau_i \rightarrow \tau_i \chi_i}$,
${\sigma_{ij} \rightarrow \chi_i \sigma_{ij} \chi_j}$ with ${\chi_i=\pm 1}$. 
If desired, we can choose the gauge ${\tau_i=1}$, leaving a lattice model for the $\sigma$ spins on the links only, with terms $J\sigma$ on the links:
this emphasizes that Eq.~\ref{eq:Z3Dgaugerep} is a lattice model with no  
internal global symmetries \cite{wegner1971duality}. However along a line in the phase diagram it has a self-duality symmetry, as discussed below. 
In parts of the phase diagram the model also has one-form symmetries, either microscopic or emergent, 
which we discuss in Sec.~\ref{sec:percolation}.

\begin{figure}[t]
\includegraphics[width=0.98\linewidth]{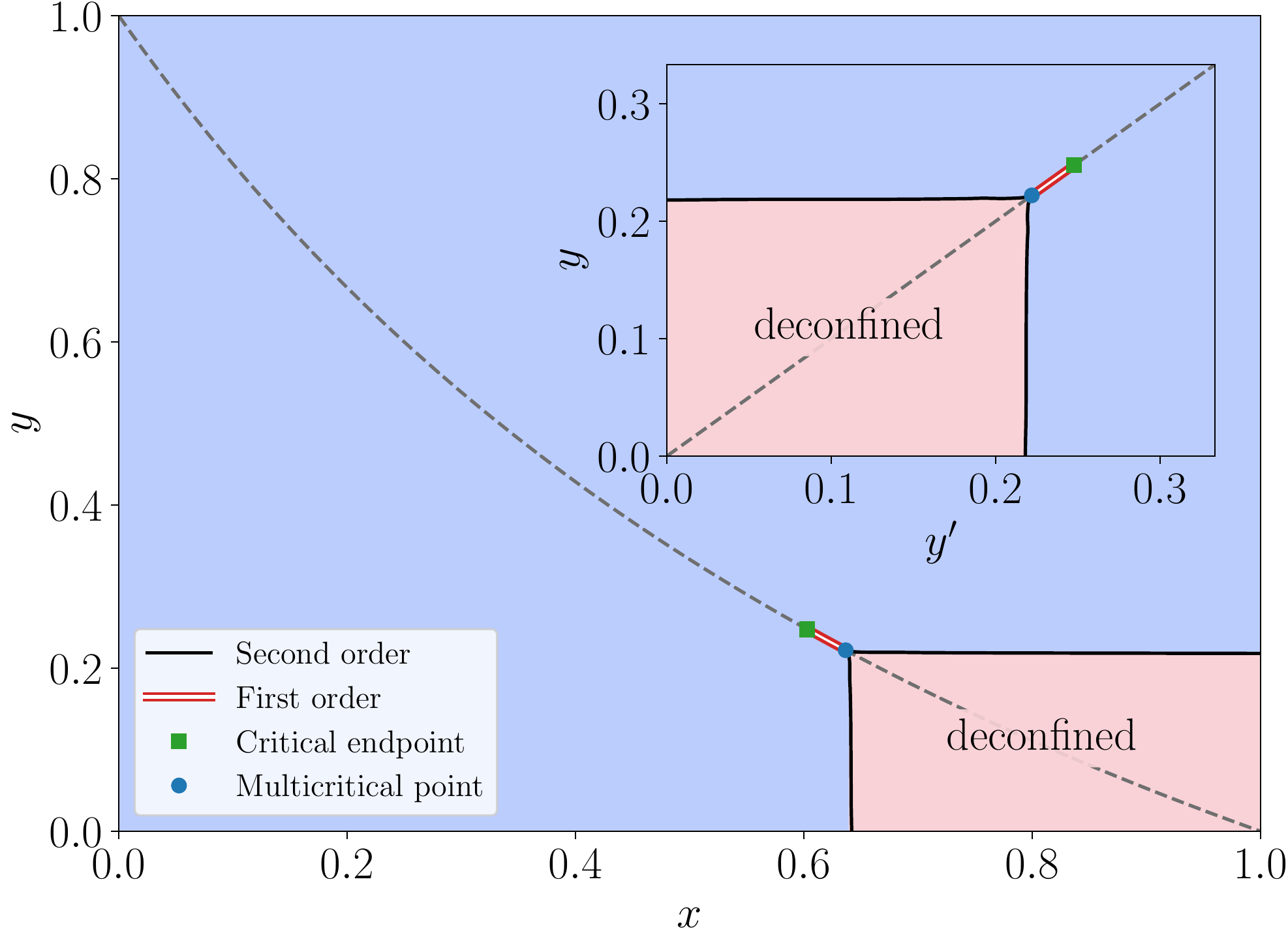}
\caption{Sketch of phase diagram in the plane $(x,y)$. 
The gauge stiffness $K=\tanh^{-1} x$ increases to the right, and the matter field coupling $J=\tanh^{-1} y$ increases upwards (self-dual line shown dashed).  Inset: same phase diagram in the $(y,y')$ coordinates, where duality acts as a reflection.
(The exponents below imply that the $e$ and $m$ condensation lines are asymptotically parallel as they approach the self-dual critical point, though the curvature is not visible at this scale.)
}
\label{fig:phase_diag}
\end{figure}

It will be convenient to define \cite{wegner1971duality,balian1974gauge}
\ba\label{eq:definexy}
x & = \tanh K, 
& 
y & = \tanh J.
\end{align}
The phase diagram in this parameterization is shown in the main panel of Fig.~\ref{fig:phase_diag}.
The dashed line is where the model is self-dual. 
The approximately rectangular region in the bottom right corner, 
at large gauge stiffness $K$ and small matter field coupling $J$, 
is the deconfined phase supporting deconfined anyons (Sec.~\ref{sec:quantummodel}).

\subsection{As a model of membranes}
\label{sec:membranepicture}

The model can be mapped to a statistical ensemble of ``membranes'' on the cubic lattice \cite{huse1991sponge,gregor2011diagnosing}.
In this picture, the parameters $x$ and $y$ control the microscopic \textit{surface} tension for the membrane, 
and the microscopic \textit{line} tension for membrane boundary respectively. 
The partition function is (see App.~\ref{app:membraneexpansion} for details):
\be\label{eq:partitionfunctionmembranes}
Z(x,y) = \sum_{\mathcal{M}} \, x^{|\mathcal{M}|} \, y^{|\partial \mathcal{M}|}.
\ee
Here a membrane configuration $\mathcal{M}$ is simply any set of  plaquettes of the cubic lattice: we call the plaquettes in $\mathcal{M}$ ``occupied''.
The energy of a configuration depends on the total number $|\mathcal{M}|$ of occupied plaquettes in the configuration,
and on the total length of membrane ``boundary'',  ${|\partial\mathcal{M}|}$.
This is the number of links where an \textit{odd} number of occupied plaquettes meet. 
We refer to these as occupied links.

Note that the deconfined phase occurs in the regime where membrane surface is cheap, but membrane boundary is expensive. 
The extreme limit of the deconfined phase is $x=1$, $y=0$, where we have an ensemble of \textit{closed membranes} with vanishing surface tension. 
We may exit the deconfined phase either by suppressing membrane area (decreasing $x$ sufficiently) or by 
tearing holes in the membranes (increasing $y$ sufficiently) \cite{huse1991sponge}. 

\begin{figure}[t]
\centering
\includegraphics[width=0.8\linewidth]{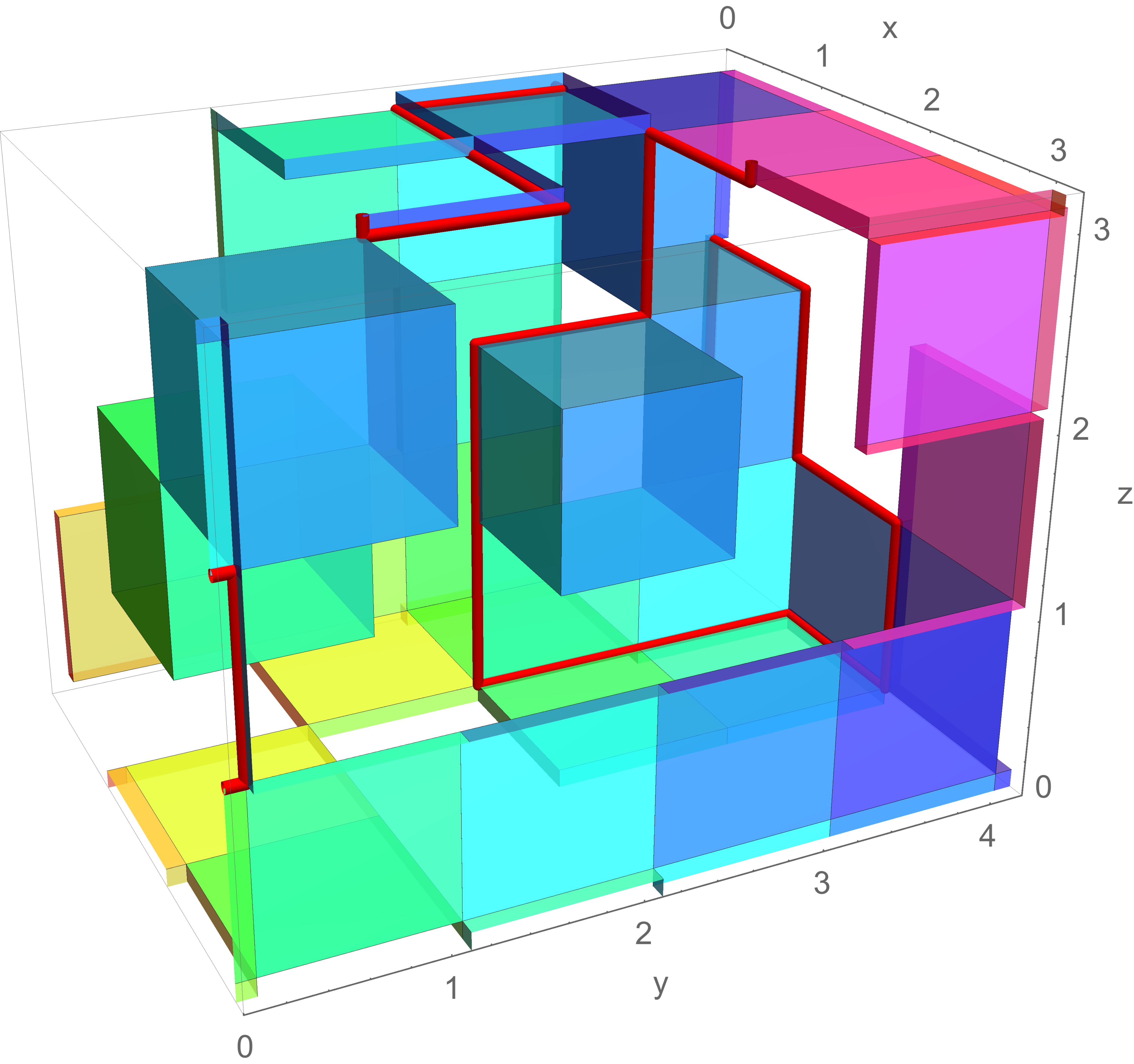}
\caption{Membranes with boundary: part of a configuration, close to the critical point on the self-dual line. Occupied plaquettes are shown coloured (colours have no meaning). Occupied links, where an odd number of plaquettes meet, are shown in red, thick.}
\label{fig:plaquetteconfig}
\end{figure}

Fig.~\ref{fig:plaquetteconfig} shows a part of a membrane configuration taken from a simulation close to the self-dual critical point that we study. Plaquettes in $\mathcal{M}$ have been coloured (arbitrarily) and the boundary links in ${\partial\mathcal{M}}$ have been marked in red. 
The membrane representation suggests investigating ``geometrical'' (percolation-like) observables close to the critical point, as well as thermodynamic ones \cite{huse1991sponge}. 
We discuss this in Sec.~\ref{sec:percolation}, showing that the loops in $\partial\mathcal{M}$ form a scale-invariant ensemble at the self-dual critical point.

The membrane picture is one way to see the duality property of the model \cite{wegner1971duality}. 
In Eq.~\ref{eq:partitionfunctionmembranes} the partition function is expressed as a sum over membrane configurations on the original cubic lattice. 
An alternative graphical representation yields an ensemble of precisely the same form, but for membranes on the dual cubic lattice (App.~\ref{app:membraneexpansion}), with dual values of the plaquette and link fugacities:
\ba\label{eq:dualvarsdefn}
x' & \equiv \f{1-y}{1+y}, 
& 
y' & \equiv  \f{1-x}{1+x}.
\end{align}
This pair of mappings shows that $Z(x,y)$ is equal, up to a trivial constant, to $Z(x', y')$.\footnote{Explicitly, $Z(x,y)=c^{3L^3}Z(x',y')$ with ${c=\f{(1+x) (1+y)}{2}}$.} 
Below we will see that duality can also be thought of as a conventional symmetry operation.
This symmetry is not manifest in the formulations above, but may be made apparent in an alternate representation of the partition sum in terms of worldlines of $e$ and $m$ particles (Sec.~\ref{sec:manifestlyselfdual}).

Note that, in view of Eq.~\ref{eq:dualvarsdefn}, we are free to choose $(y,y')$ as coordinates for the phase diagram, as in the inset to Fig.~\ref{fig:phase_diag}. The line $y=y'$ is then the self-dual line, where the Boltzmann weights are invariant under duality symmetry.

\subsection{Manifestly self-dual loop representation}
\label{sec:manifestlyselfdual}

The quantum interpretation reviewed in the next subsection motivates yet another representation of the path integral, in terms of two species of ``loops'', which represent worldlines of both $e$ and $m$ particles.
This representation  makes self-duality manifest.  The price we pay is minus signs in the Boltzmann weight, which encode the mutual semion statistics of the anyons $e$ and $m$.

\begin{figure}[t]
\includegraphics[width=0.4\linewidth]{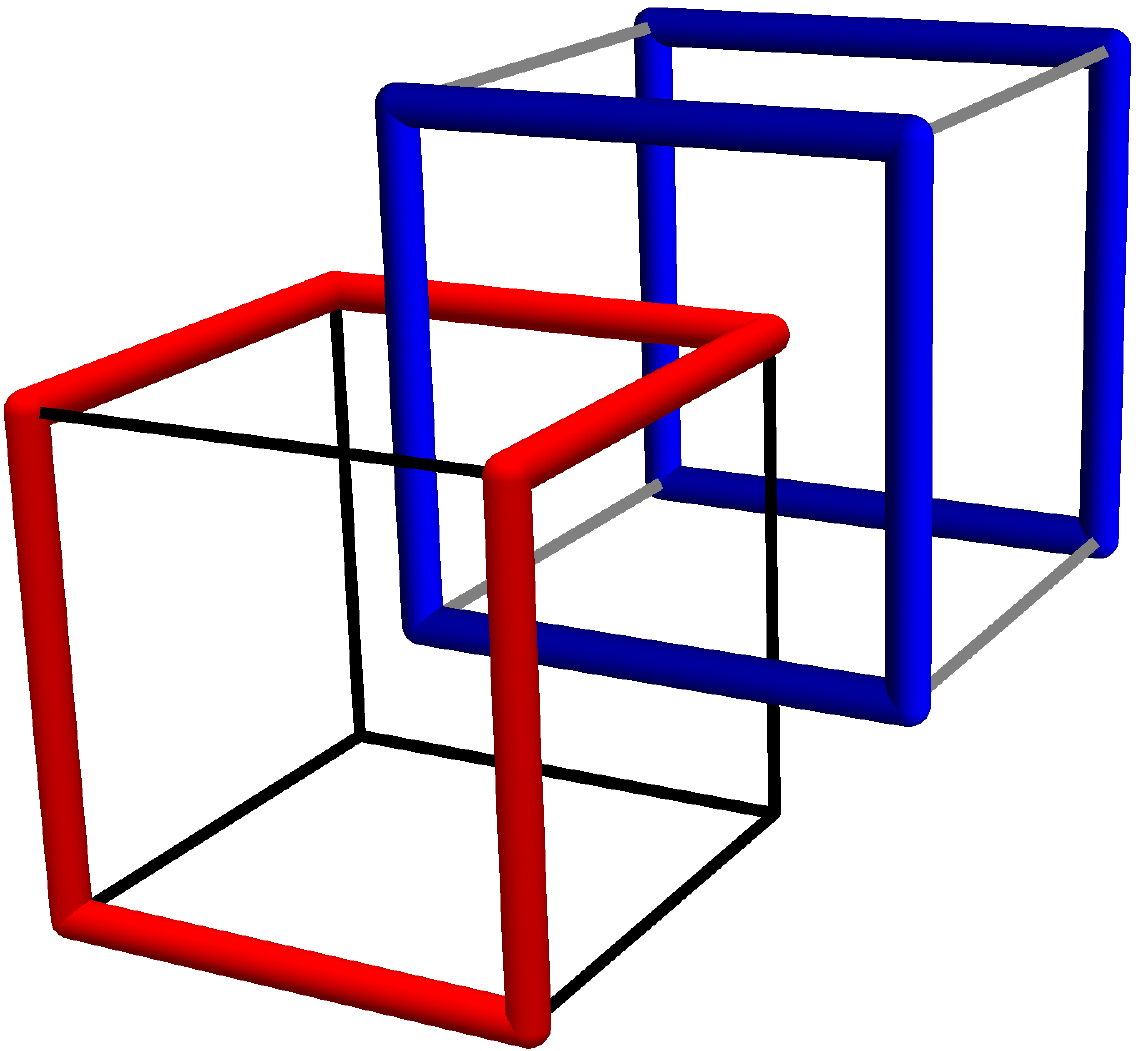}
\caption{Cubes of the original and dual lattices, with $e$ worldlines (red) and $m$ worldlines (blue) on the former and the latter respectively. (This configuration has linking number $X=1$.)}
\label{fig:linkedlattices}
\end{figure}

The partition function can be written as (see App.~\ref{app:manifestlyselfdual} for details):\footnote{The omitted proportionality constant is a trivial (nonuniversal) extensive contribution to the free energy. The factor of 4 is universal (and equal to the ground state degeneracy of the 2D quantum system in its deconfined phase).}
\be\label{eq:Zhybridrep}
Z \propto
4 \sum_{\mathcal{C}_e, \, \mathcal{C}_m}
(-1)^{X(\mathcal{C}_e, \mathcal{C}_m)} 
y^{|\mathcal{C}_e|}
{y'}^{|\mathcal{C}_m|}.
\ee
The electric and magnetic worldline configurations ${\mathcal{C}_e}$ and ${\mathcal{C}_m}$, which we refer to as loop configurations, 
are sets of ``occupied'' links on the original and dual lattices respectively. See Fig.~\ref{fig:linkedlattices}. 
Any even number of occupied links may be adjacent to each site, so the term ``loops'' is used loosely (see the footnote\footnote{${\mathcal{C}_e}$ obeys the same restrictions as   $\partial \mathcal{M}$: it must make sense to regard it as the boundary for a membrane configuration on the original lattice. The same holds for $\mathcal{C}_m$ on the dual lattice.
Specifically, each site on the original lattice lies on an even number of the links in $\mathcal{C}_e$ (possibly zero) and similarly for the dual lattice and $\mathcal{C}_m$. Additionally the $\mathbb{Z}_2$ winding numbers of $\mathcal{C}_e$, $\mathcal{C}_m$ in each of the three directions are equal to zero modulo two (this is a requirement for a worldline configuration to be promoted to the boundary of a membrane configuration).} for details).
$\mathcal{C}_e$ may be identified with the membrane boundaries $\partial \mathcal{M}$  in the previous representation. 

The crucial feature in Eq.~\ref{eq:Zhybridrep} is the topological factor ${(-1)^X}$, which gives a factor of ${-1}$ for each linking between an $e$ worldline and an $m$ worldline. 
(That is, ${X(\mathcal{C}_e, \mathcal{C}_m)=0,1}$ is the $\mathbb{Z}_2$ linking number of the two worldline configurations. It can be computed, for example, by introducing an arbitrary membrane configuration $\widetilde{\mathcal{M}}$ such that ${\partial\widetilde{\mathcal{M}}=\mathcal{C}_e}$, and counting the number of intesections between $\widetilde{\mathcal{M}}$ and $\mathcal{C}_m$ modulo 2.)  The values of the dual link fugacities $y$ and $y'$ are defined in Eq.~\ref{eq:definexy}.
 
An interesting  model of $\mathrm{U}(1)$ (oriented) flux lines with a linking sign has been studied in \cite{geraedts2012monte} (see also variations in Refs.~\cite{geraedts2012phases,lee2016monte,geraedts2012line,geraedts2012monte}).
That model has many features  in common with Eq.~\ref{eq:Zhybridrep}, and also describes a problem of condensation of anyons with mutual statistics. 
However it  also has significant differences as a result of a global $\mathrm{U}(1)\times \mathrm{U}(1)$  symmetry. 
We expect the $\mathrm{U}(1)\times \mathrm{U}(1)$  model to be described, at least in principle, by a continuum Chern-Simons gauge theory. 
The ``$\mathbb{Z}_2\times \mathbb{Z}_2$'' loop model in Eq.~\ref{eq:Zhybridrep} is also closely related to a quantum wavefunction in 3+1D that sustains 2+1D $\mathbb{Z}_2$ topological order on its boundary \cite{wang2015topological,geraedtsunpublished,von2013three}.

Returning to Eq.~\ref{eq:Zhybridrep},
it is possible to sum over $\mathcal{C}_m$ exactly (App.~\ref{app:manifestlyselfdual}).
We then return to the membrane expression (\ref{eq:partitionfunctionmembranes}) for the partition function, with $\mathcal{C}_e=\partial\mathcal{M}$.
Similarly, integrating out $\mathcal{C}_e$ gives the dual membrane picture on the dual lattice.
(Note that the line tension $y'$ or $y$ of the species that is integrated out determines the surface tension of the membranes.) 

However the representation (\ref{eq:Zhybridrep}) makes the duality symmetry that exists when $y=y'$ (i.e. when $x=x'$) manifest.
We can think of the symmetry operation as a translation by ${(1/2, 1/2, 1/2)}$, where the lattice spacing of each cubic lattice is unity. This translation  exchanges the cubic lattice with its dual, so exchanges $e$ and $m$ worldlines.
Microscopically, this is not an \textit{internal} symmetry (since it involves translation) but we will argue in Sec.~\ref{sec:selfduality} that 
self-duality becomes an internal $\mathbb{Z}_2$ symmetry of the IR theory.

Sec.~\ref{sec:newmodel} presents an alternative loop model in which the $e$ and $m$ loops share the same lattice.

\subsection{Anyons and the toric code in a field}
\label{sec:quantummodel}

The 3D gauge theory is expected to capture the universal physics of a wide range of 2+1D quantum models with a $\mathbb{Z}_2$ spin liquid phase. 
An anisotropic limit of the 3D theory, where the $z$ direction becomes a continuous imaginary time coordinate, maps exactly to the partition function for such a Hamiltonian on the square lattice. See Refs.~\cite{kogut1979introduction, fradkin2013field} for details of such mappings. 
Here we review the basic excitations of the deconfined phase, and how they relate to the geometrical pictures above, 
in qualitative terms.  

It is convenient to start with the toric code \cite{kitaev2003fault},
a particularly simple  model lying in the deconfined phase.
The degrees of freedom are spin-1/2s on the links of the square lattice. 
The Hamiltonian includes a plaquette term and a vertex term:
\be\label{eq:tcH}
H =  - V  \bigg( \sum_\square XXXX + \sum_+ ZZZZ \bigg).
\ee
The first product is a shorthand for the Pauli-X operators on the four links making up a given plaquette, and the second for the four Pauli-Z operators on the  links touching a given site. 
Here $V$ is a coupling constant, which we have chosen equal for the two terms to ensure self-duality symmetry.
Noting that we can equally well view spins either as living at the midpoints of bonds on the original square lattice 
or at the midpoints of bonds on the dual square lattice, 
the duality symmetry operation may be viewed as a translation by $(1/2, 1/2)$, together with an exchange of $X$ and $Z$.\footnote{If a basis transformation is applied to the Hamiltonian the symmetry becomes a simple translation \cite{wen2003quantum}.}

\begin{figure}[t]
\centering
\includegraphics[width=0.95\linewidth]{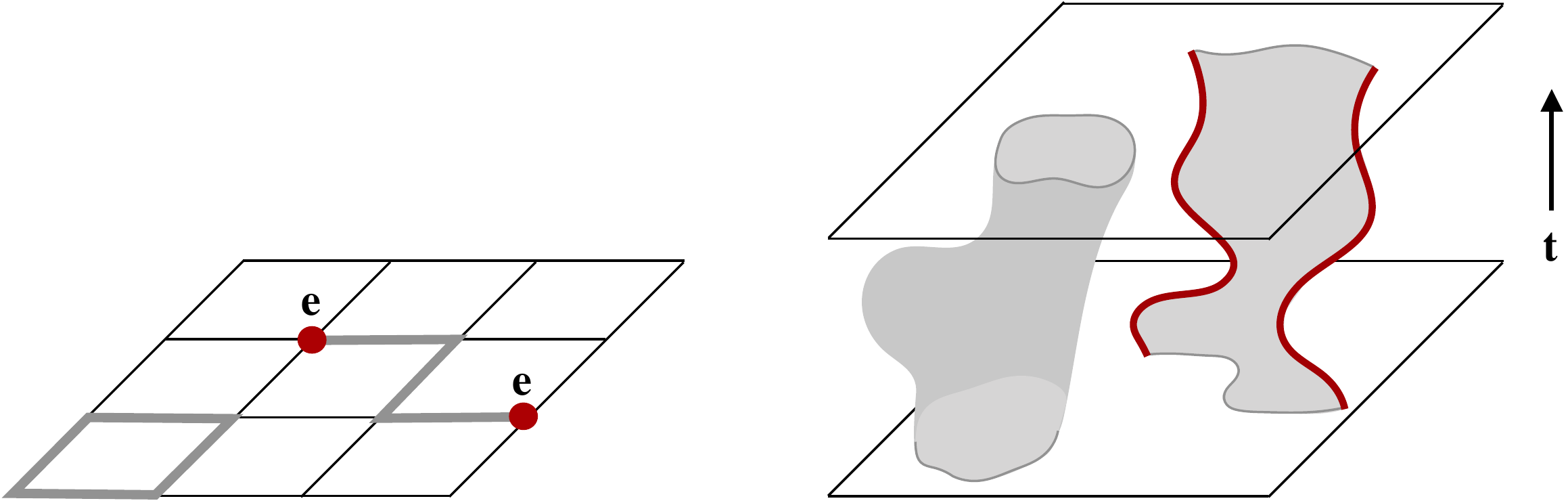}
\caption{The relation between $e$ particles and membranes. Left: in the $Z$ basis, the toric code wavefunction is a superposition of strings of occupied links (bold) representing $Z=-1$, and an $e$ excitation is a vertex where an odd number of occupied links meet. Right: constructing a path  integral (for a generic perturbed model) in this basis, worldsurfaces of strings become membranes $\mathcal{M}$ (grey), and worldlines of $e$ particles form the membrane boundary, or ``loops'', ${\partial\mathcal{M}=\mathcal{C}_e}$. 
(The figure on the right is schematic: in the model we study, membrane configurations look like Fig.~\ref{fig:plaquetteconfig}.)}
\label{fig:eworldline}
\end{figure}

Ground states\footnote{There is an order--1 degeneracy that depends on the system's topology.} have all the plaquette and  vertex terms in Eq.~\ref{eq:tcH} equal to 1,
and are superpositions of  ``strings'' --- either on the original lattice (if we use the $Z$ basis,  a link with $Z=-1$ being regarded as part of a loop) 
or the dual lattice (if we use the $X$ basis) \cite{kitaev2003fault}. 
There are two fundamental types of excitation, related by duality.
A vertex where $ZZZZ=-1$ is an ``$e$ particle'', and a plaquette where $XXXX=-1$ is an ``$m$ particle''.
These are distinct types of anyon. Each is a boson, but adiabatically braiding an $e$ around an $m$ changes the sign of the wavefunction.
(That is, they are mutual semions. The combination of an $e$ particle and an $m$ particle forms another type of anyon whose topological sector is denoted ``$\epsilon$'': this also has $-1$ statistics with $e$ and $m$, but is a fermion \cite{kitaev2003fault,nayak2008non}.) 
In the $Z$ basis, a $e$ excitation is the endpoint of a string (Fig.~\ref{fig:eworldline}).

The toric code is a fine-tuned limit in which the $e$ and $m$ particles are non-dynamical.
Critical phenomena are possible when the model is perturbed so that pair creation and annihilation of these particles becomes possible. 
This may be achieved by adding magnetic fields in both the $X$ and $Z$ directions \cite{vidal2009low,TupitsynTopological}. (For example, adding the operator $X$ to the Hamiltonian allows both hopping and pair creation/annihilation of bare $e$ particles on a given link.)   
The resulting model has been intensely studied \cite{vidal2009low,dusuel2011robustness,  TupitsynTopological, wu2012phase}. Duality exchanges these two magnetic fields, so that the line $h_X=h_Z$ preserves duality symmetry. 
  
The phase diagram of the toric code in $X$ and $Z$ fields is expected to be equivalent to that discussed in the previous sections, up to nonuniversal constants \cite{TupitsynTopological}.
The dimensionless field $h_X/V$,  which can induce condensation of the $e$ particle, plays the role of the vertical coordinate in Fig.~\ref{fig:schematic_phase_diag}, and $h_Z/V$ plays the role of the horizontal coordinate.

The connection with the geometrical pictures above arises from writing the imaginary-time partition function in various choices of basis. We describe this only in qualitative terms:

In the $Z$ basis, the wavefunction is a superposition of terms like that illustrated in Fig.~\ref{fig:eworldline} (Left) with $e$ particles (at sites of the square lattice) forming endpoints of strings. 
Constructing the sum over Feynman trajectories using this basis, the worldsurfaces of strings form a set of membranes $\mathcal{M}$, and the worldlines of $e$ particles form a set of loops that are the boundaries ${\mathcal{C}_e = \partial \mathcal{M}}$ of these membranes. 
(In the limit $h_X = 0$ there are no bare $e$ particles in the ground state, 
and correspondingly the membranes are closed surfaces.)
This picture is a continuous-time version of that in Sec.~\ref{sec:membranepicture}. 
The dual membrane picture is obtained by working in the $X$ basis, where the worldlines of $m$ particles are manifest. 

Alternately we may pick a basis in which \textit{both} the plaquette products $XXXX$ and the vertex products $ZZZZ$ are diagonal: this is possible since all these terms commute. The Feynman trajectory sum is then over worldline configurations $\mathcal{C}_e$ and $\mathcal{C}_m$ for both $e$ and $m$ particles, and is a continuous time version of the loop model in Eq.~\ref{eq:Zhybridrep}.

\subsection{Ising$^*$ and first-order lines}
\label{sec:phasetransitionsreview}

To conclude this overview of the model, we recap some features of the phase transition lines in Fig.~\ref{fig:schematic_phase_diag} or equivalently Fig.~\ref{fig:phase_diag}.

Starting in the deconfined phase (in the lower-right hand corner of Fig.~\ref{fig:phase_diag}) we may exit it in three ways, two of which are related by duality \cite{wegner1971duality,fradkin1979phase,TupitsynTopological}.
Condensing the $e$ particle while keeping the mass of $m$ finite corresponds to the upper boundary of the deconfined region in Fig.~\ref{fig:phase_diag} (main panel); this is the \emph{Higgs} transition in the lattice gauge theory.
Condensing $m$ while $e$ remains massive is equivalent by duality, and is the left-hand boundary of the deconfined region in Fig.~\ref{fig:phase_diag} (main panel). This is the \emph{confinement} transition in the lattice gauge theory.

These transitions are continuous with Ising exponents, at least sufficiently close to the boundaries of the phase diagram \cite{kogut1979introduction}, as we now rapidly review. 
These transitions, described by a weakly gauged Landau theory,
are sometimes referred to as ``Ising$^*$'' transitions (see for example Ref.~\cite{nandkishore2012orthogonal})
to denote the fact that, because of gauging, only the $\mathbb{Z}_2$--even operators of the Ising CFT survive as local operators.

Consider first the Higgs transition in the limit $x=1$, i.e. infinite gauge stiffness $K=\infty$. 
The freezing of gauge fluctuations in this limit gives an exact mapping to a standard cubic lattice Ising model.  
But Ising exponents are retained at least along some part of the phase transition line, for finite $K$. 
This can be argued by fixing the gauge and deriving an effective longer-range Ising Hamiltonian perturbatively in $e^{-K}$ \cite{wegner1971duality,fradkin1979phase}. 
In the quantum language, the point is that so long as the $m$ and $\epsilon$ anyons are gapped,  we can for many purposes neglect the fact that the $e$ particle which is condensing is an anyon, rather than a local excitation \cite{burnell2018anyon}. 
By duality, equivalent points hold for the confinement transition.

A more intuitive way to understand the relation to Ising is developed in Sec.~\ref{sec:percolation} and  Ref.~\cite{membranesforthcoming}.
Let us start with the membrane representation in the limit where membrane boundary is completely suppressed 
($y=0$ in Eq.~\ref{eq:partitionfunctionmembranes}, or $y'=0$ in the dual membrane picture). Again this corresponds to the transitions on the boundaries of the phase diagram.
In this limit the membranes form closed surfaces, so they may be mapped exactly to domain walls in a nearest-neighbor Ising model.\footnote{In this Ising model we must sum over both periodic and antiperiodic boundary conditions for each direction. This is to allow an odd number of domain walls to span the system perpendicular to each direction.}

Now when we increase $y$ slightly, the membranes acquire holes in them. 
This means that there is no longer an unambiguous mapping to an Ising model. 
But if the holes are sufficiently small, we might expect this ambiguity to be unimportant on large scales, so that we can again think in terms of an ordering transition for a fictitious Ising order parameter.

We make this idea of a fictitious Ising order parameter precise using an explicit construction,  based on the idea of ``repairing'' or ``patching'' the membranes in the representation (\ref{eq:partitionfunctionmembranes}) of the partition function 
(Sec.~\ref{sec:percolation} and Ref.~\cite{membranesforthcoming}).
We argue that this construction can be performed all the way along the Ising$^*$ critical lines, but not at the self-dual critical point, where a  different universality class takes over.

For completeness, let us note that we can also think of the Ising$^*$ transition in the loop model representation, Eq.~\ref{eq:Zhybridrep}. Loosely speaking, when one species of loops has a finite typical size, coarse-graining beyond this scale gives a loop model for a single species of unoriented loops. This is a standard representation of the Ising universality class, in terms of worldlines of the Ising quanta. 

The self-dual transition point \cite{fradkin1979phase,jongeward1980monte,TupitsynTopological,vidal2009low}, where the two Ising$^*$ lines meet, will be discussed in the rest of the text. 

The line of first order transitions occurs within the trivial phase, so is relatively conventional. It is also a line where self-duality symmetry is spontaneously broken.
The natural expectation is that the critical endpoint of this line is in the Ising universality class, with the Ising order parameter being the anti-self-dual operator defined below. 
Ising universality for this critical endpoint is consistent with a very rough estimate of  the universal crossing value of the Binder cumulant, as shown in~App.~\ref{app:coexistence}.

In addition to these thermodynamic transitions we may also define geometrical transitions \cite{huse1991sponge} using the geometry of the membranes in Eq.~\ref{eq:partitionfunctionmembranes} (Sec.~\ref{app:percolation}).

\section{Self-duality as a symmetry}
\label{sec:selfduality}

We anticipate that, for any scale-invariant critical point on the self-dual line, self-duality becomes an internal $\mathbb{Z}_2$ symmetry of the IR theory. 

One way to argue for this is via the manifestly self-dual representation of the partition function in Eq.~\ref{eq:Zhybridrep} with $y=y'$. 
This has a translation symmetry by $(1/2,1/2,1/2)$  which exchanges $e$ and $m$ worldlines.
Correspondingly, the 2D quantum model in Sec.~\ref{sec:quantummodel} has a symmetry involving translation by ${(1/2,1/2)}$ that exchanges $e$ and $m$ particles.

The simplest assumption is that,
at a scale-invariant critical point this microscopic symmetry gives rise to an internal  $\mathbb{Z}_2$ symmetry of the IR fixed point theory. 
Loosely speaking, the action of the translation on the rescaled spatial coordinate of the coarse-grained theory disappears in the IR limit, so that the microscopic symmetry transformation should map to a purely internal symmetry transformation on the operators of the IR theory.\footnote{See Ref.~\cite{metlitski2018intrinsic} for a careful discussion of the role of translation symmetries in the continuum in some other spin models.}

We can motivate this further by noting that alternative models for the deconfined phase can be constructed in which the duality symmetry, 
exchanging $e$ and $m$, 
is an internal symmetry even at the lattice level.
Refs.~\cite{heinrich2016symmetry, cheng2017exactly} give exactly solvable 2D string-net Hamiltonians for the deconfined phase, with this property.\footnote{The existence of microscopic models for the deconfined phase with an ``on-site'' self-duality symmetry \cite{heinrich2016symmetry, cheng2017exactly} suggests that this symmetry of the critical theory is not anomalous. In this respect it is different from the Kramers-Wannier duality symmetry of the 1+1D Ising model, which cannot be realized as an on-site symmetry \cite{aasen2016topological,karch2019web,jones20191d}.}
We may also define a variant of the 3D loop model  (\ref{eq:Zhybridrep}) 
in which the $e$ and $m$ loops live on the same lattice, with a $\mathbb{Z}_2$ symmetry exchanging them. This model is defined in Sec.~\ref{sec:newmodel}.
It is plausible that by varying the interactions in either of these models we could access the same self-dual fixed point, at the corner of the deconfined phase, as in the original model. 

\begin{figure}[t]
\includegraphics[width=0.95\linewidth]{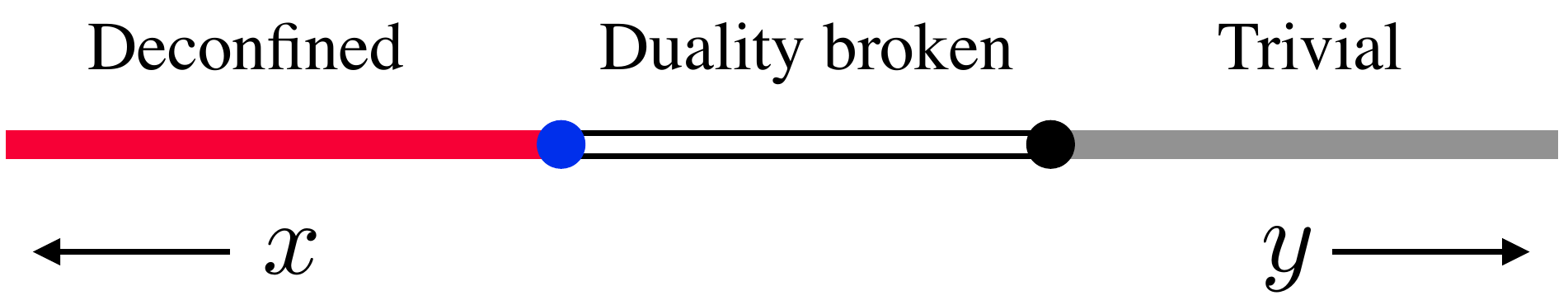}
\caption{The phase diagram on the self-dual line, i.e. on the line ${y=(1-x)/(1+x)}$ (where $x=x'$ and $y=y'$).}
\label{fig:1d_phase_diag}
\end{figure}

The phase diagram of the gauge-Higgs model, restricted to the self-dual line, is shown schematically in Fig.~\ref{fig:1d_phase_diag}. The first-order line in Fig.~\ref{fig:schematic_phase_diag} corresponds to spontaneous breaking of self-duality symmetry, as discussed below.

\subsection{Defining (anti)symmetric operators}
\label{sec:definingAandS}

Duality acts on the phase diagram as $y\leftrightarrow y'$.
We would now like to define lattice operators $S$ and $A$ that are conjugate to the self-dual and anti-self-dual couplings, namely $y+y'$ and $y-y'$ respectively.

We continue to use the language of membranes (\ref{sec:membranepicture}). 
First, define ``face'' and ``edge'' operators, ${\mathcal{F}(p)}$ and ${\mathcal{E}(\ell)}$ respectively, which are equal to either zero or one and which measure whether a given plaquette $p$ or link $\ell$ of the cubic lattice is occupied in membrane configuration $\mathcal{M}$. 
That is, $\mathcal{F}(p)=1$ if $p\in \mathcal{M}$ and $\mathcal{F}(p)=0$ if $p\notin \mathcal{M}$;
 $\mathcal{E}(\ell)=1$ if $\ell\in \partial \mathcal{M}$ and $\mathcal{E}(\ell)=0$ if $\ell\notin \partial \mathcal{M}$.

The duality transformation maps these operators to operators on the dual lattice. 
By extending the transformation 
 (App~\ref{app:membraneexpansion}) to the case of spatially varying couplings $x$ and $y$, we see that this mapping is
 \ba\label{eq:FEtransform1}
 \mathcal{F} 
 & \longrightarrow 
 - \f{2 x}{1-x^2} \, \mathcal{E} + \f{x}{1+x},
 \\ \label{eq:FEtransform2}
 \mathcal{E} & \longrightarrow
 -\f{2 y}{1-y^2} \, \mathcal{F} + \f{y}{1+y}.
 \end{align}
We have suppressed plaquette/link indices to avoid clutter. The transformed operator on the RHS is located at the link/plaquette that is dual to the plaquette/link of the operator on the LHS.

Next let us symmetrize these operators with respect to the lattice point group. 
This naturally leads to operators that are centred \textit{either} on a cube of the lattice, or on a vertex. 
We use $\mathcal{F}_\text{cube}(c)$ to denote the sum of $\mathcal{F}$ over the 6 plaquettes of a cube $c$, and 
$\mathcal{F}_\text{vertex}(v)$ to denote (one half times) the sum of $\mathcal{F}$ over the 12 plaquettes that touch a vertex $v$. 
Similarly $\mathcal{E}_\text{cube}(c)$ is (one half times) the sum over the  12 links in a cube  and $\mathcal{E}_\text{vertex}(v)$ is the sum over the 6 links touching a vertex. 
(We include the factors of $1/2$ so that the expectation values of $\mathcal{E}_\text{cube}$ and $\mathcal{E}_\text{vertex}$ are equal, and similarly for $\mathcal{F}_\text{cube}$ and $\mathcal{F}_\text{vertex}$.)

Finally, specializing to the self-dual line, we define
\ba
A_\text{cube} & = 
\mathcal{F}_\text{cube} + \f{2x}{1-x^2} \mathcal{E}_\text{cube} - \f{6 x}{1+x}
\label{eq:Acube}
\\
S_\text{cube} & = 
\mathcal{F}_\text{cube} - \f{2x}{1-x^2} \mathcal{E}_\text{cube} + \f{6 x}{1+x}
\label{eq:Scube}
\end{align}
and analogously for operators $A_\text{vertex}$ and $S_\text{vertex}$ at the vertices. 

These operators transform simply under duality:
\ba\label{eq:latticeAStransformation}
A_\text{cube} & \longleftrightarrow - \, A_\text{vertex}, 
\\
S_\text{cube} & \longleftrightarrow 
+ \,  S_\text{vertex}.
\end{align}
In addition, 
\ba\label{eq:Asum}
\sum_c A_\text{cube}(c) & =\sum_v A_\text{vertex}(v),
\\\label{eq:Ssum}
\sum_c S_\text{cube}(c) & =\sum_v S_\text{vertex}(v).
\end{align}
These ``integrated'' operators, which can be written either as sums over cubes or vertices, are the  anti-self-dual and self-dual perturbations of the self-dual line. 

Now we expand the lattice operators above in terms of continuum operators of a putative IR fixed point. 
Denote the leading $\mathbb{Z}_2$ odd and $\mathbb{Z}_2$ even  scalar continuum operators by $A(r)$ and $S(r)$  respectively, with no subscript. 
We will also write $A_\text{cube}(r)$, $A_\text{vertex}(r)$, etc., for lattice operators, where $r$ is the location of the appropriate cube/vertex. 

To be consistent with Eq.~\ref{eq:latticeAStransformation} and Eq.~\ref{eq:Ssum}, the operator $S_\text{cube}$  must be of the form 
\ba\notag
S_\text{cube}(r)  = &  (\text{self-dual operators}) + \\ \notag
& (\text{derivatives of anti-self-dual operators}),
\end{align}
and analogously for the other lattice $A$ and $S$ operators,
so that their integrated versions have well-defined symmetry under duality.
Taking into account point-group symmetry, some of the allowed terms in $S_\text{cube}$ and $S_\text{vertex}$ are:
\ba
S_\text{cube}(r) & = 
\alpha  S(r) + \beta \nabla^2 A(r) + \gamma \nabla^2 S(r) + \ldots \\
S_\text{vertex}(r) & = 
\alpha S(r) - \beta \nabla^2 A(r) + \gamma \nabla^2 S(r) + \ldots.
\end{align}
Here $\alpha$, $\beta$ and $\gamma$ are nonuniversal constants.
The sign of the $\mathbb{Z}_2$--odd term is reversed in the second line so that  mixed correlators of lattice operators are  consistent with Eq.~\ref{eq:latticeAStransformation}.
Equivalent formulae apply for the  lattice $A$ operators, with $A$ and $S$ exchanged, and separate nonuniversal constants.

We will use the operators $A_\text{cube}$ and $S_\text{cube}$ in our simulations.
We see that these lattice operators may be identified 
(up to derivative operators and other operators that are expected to be highly irrelevant)
with the leading self-dual and anti-self-dual continuum operators.

From now on we will denote the  lattice operators simply as $A(r)$ and $S(r)$, where $r$ is the coordinate of a cube.
We will use $A$ or $S$, without an argument, to denote the spatially averaged quantity, for example
\be\label{eq:spatialaverageA}
A = \f{1}{L^3}\sum_r A(r).
\ee
We write $x_A$ and $x_S$ for the scaling dimensions of the two operators.

\subsection{Spontaneous breaking of duality symmetry}
\label{sec:spontaneousbreakingduality}

\begin{figure}[t]
\centering
\includegraphics[width=0.9\linewidth]{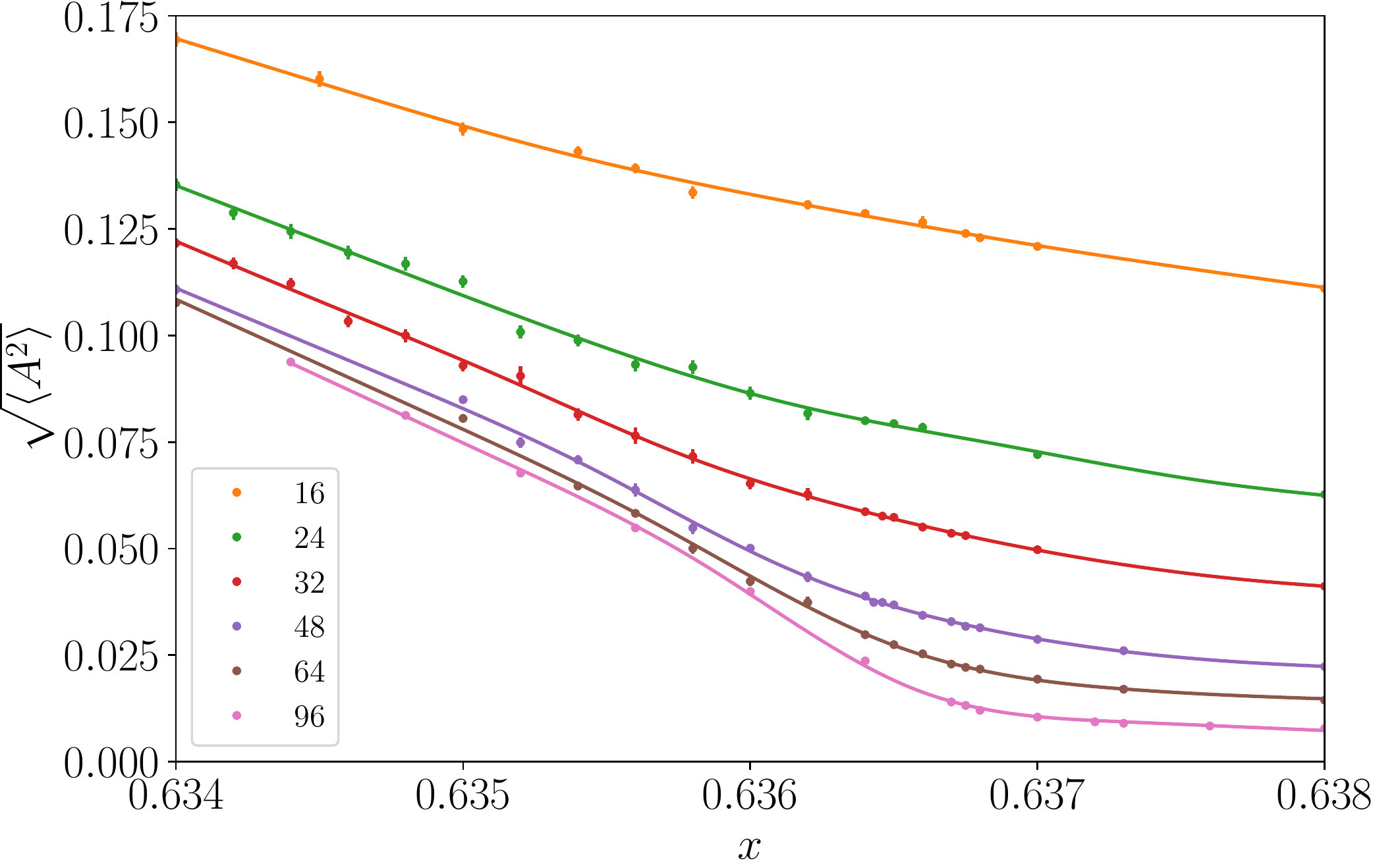}
\caption{Duality-breaking order parameter, 
$\sqrt{\langle A^2\rangle}$,
as a function of $x$ on the self-dual line,
for various system sizes (indicated in legend). 
Lines are just a guide to the eye. 
The deconfined phase is at larger $x$.}
\label{fig:A_x}
\end{figure}
\begin{figure}[t]
\centering
\includegraphics[width=0.9\linewidth]{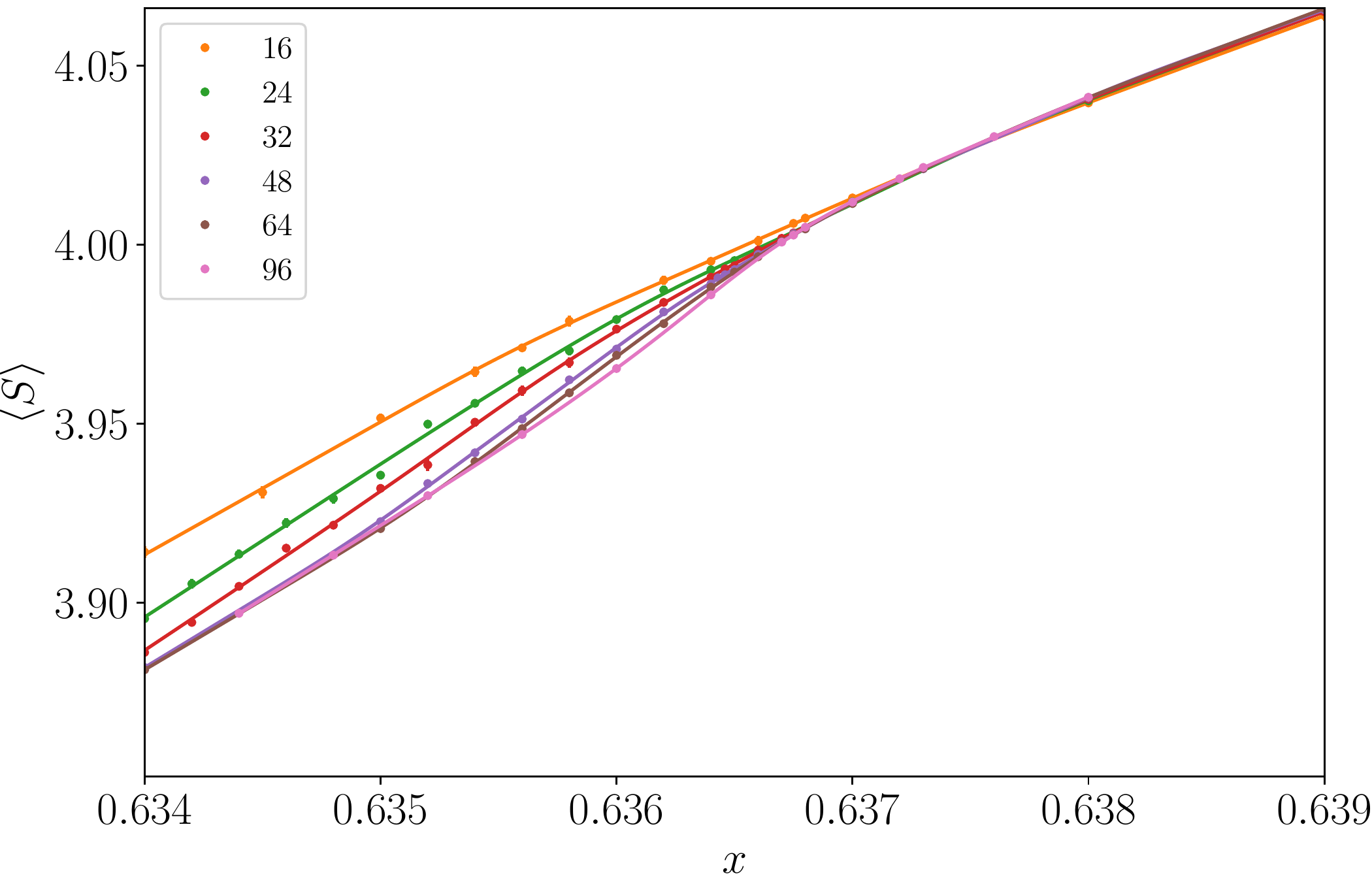}
\caption{$\langle S\rangle$ as a function of $x$ for different system sizes. The legend indicates system size. The lines are just a guide to the eye.}
\label{fig:S_x}
\end{figure}

The phase diagram on the self-dual line was shown in Fig.~\ref{fig:1d_phase_diag}.
$A$ in Eq.~\ref{eq:spatialaverageA} is an order parameter for the symmetry breaking that occurs when we exit the deconfined phase.
By self-duality symmetry, its average vanishes,
\be
\<A\> =0,
\ee
but in the duality-broken phase its magnitude $\sqrt{\<A^2\>}$ remains nonzero in the thermodynamic limit.

Raw data for this quantity are shown in Fig.~\ref{fig:A_x}, close to the critical point of interest.
In all plots we parameterize the position along the self-dual line with $x$, so the  deconfined phase corresponds to the right-hand-side of the figure.
At first glance, Fig.~\ref{fig:A_x} is consistent with the order parameter becoming nonzero in a continuous fashion 
below some $x_c$ 
(whose estimation we will discuss below).

The operator $S$, whose average is shown in Fig.~\ref{fig:S_x},
is analogous to the ``energy'' operator at a conventional classical transition, since it does not break symmetry: for a continuous transition, the correlation length exponent is 
\be
\nu = \f{1}{3-x_S}.
\ee

The  data in Figs.~\ref{fig:A_x},~\ref{fig:S_x} may be restated in terms of the average occupation of plaquettes and links in the membrane picture. On the section of the self-dual line where duality is spontaneously broken, there are two coexisting equilibria with different plaquette and link densities: we plot these two solutions explicitly in App.~\ref{app:coexistence}. 
In the critical regime of interest here, the average occupation number of links is relatively small {$\approx 2.5\%$}, but despite this they make up a scale-invariant ensemble of loops (Sec.~\ref{sec:percolation}).

We now discuss how to  establish the universal properties of the transition.

\section{Scale invariance}
\label{sec:scaleinvariance}

\subsection{Initial obstacles}

\begin{figure}[t]
\centering
\includegraphics[width=0.9\linewidth]{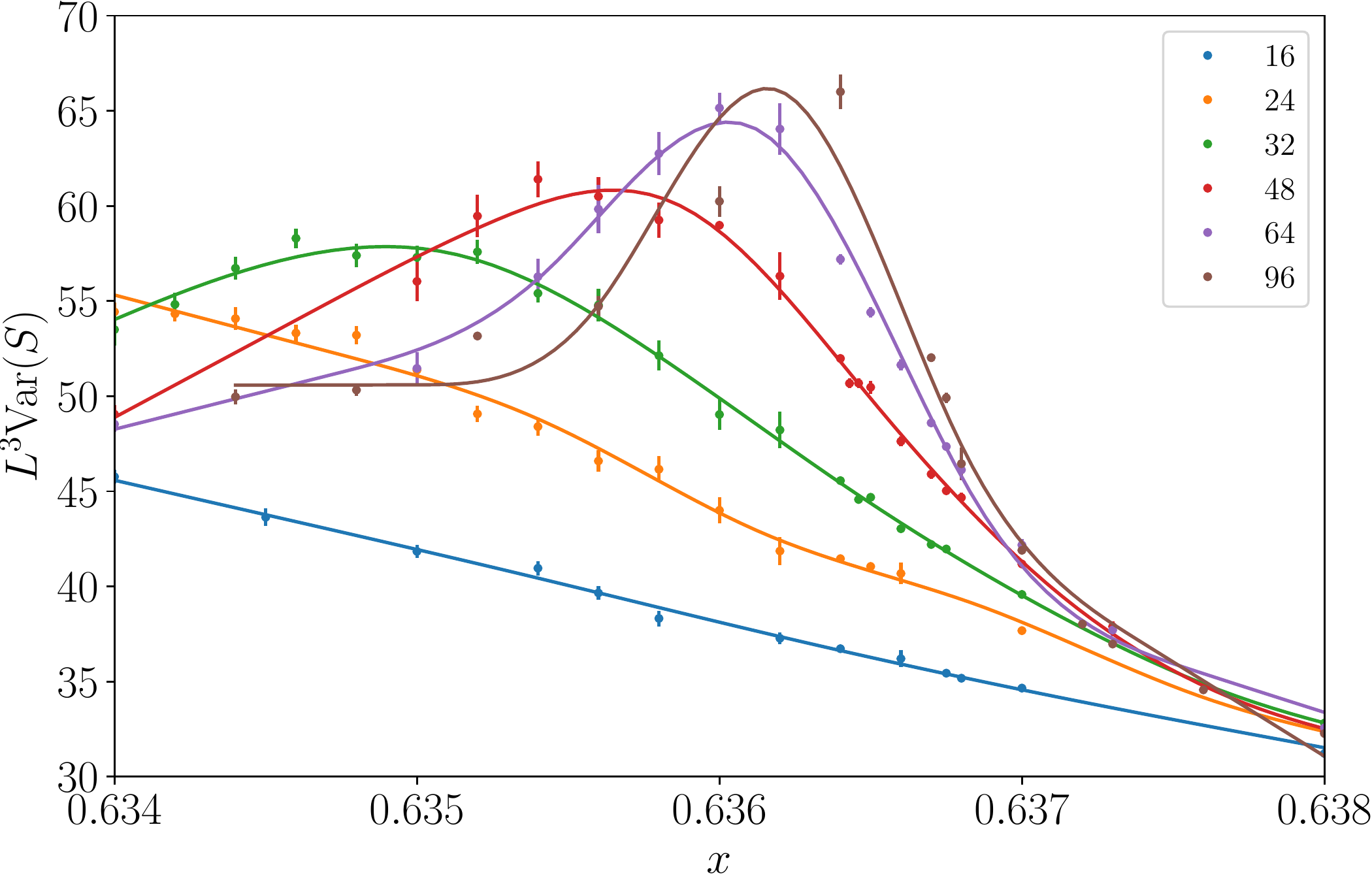}
\caption{The ``heat capacity'': Variance of $S$ (multiplied by $L^3$) as a function of $x$ for different system sizes. The legend indicates system size. The lines are b-spline fits and are just a guide to the eye.}
\label{fig:VS_x}
\end{figure}

One standard means of locating a phase transition is to analyze the specific heat, which for many simple ordering transitions diverges at the critical point. If so, data for different system sizes can typically be scaled, allowing the critical point and correlation length exponent to be determined.
Here the variable $S(r)$ is analogous to an energy, as discussed in the previous section, and ${L^3 \operatorname{var}(S)}$ is analogous to a specific heat. Values for different system sizes are shown in Fig.~\ref{fig:VS_x}.

At first sight the behaviour is the expected one: curves show a peak. 
But on  closer inspection it is unclear whether the peak diverges at large $L$ or tends to a constant.
It also becomes clear that variation of the width and height of the peaks does not follow the simple scaling form
\begin{equation}
\operatorname{Var}(S)=L^{-2 x_S} f(z),
\label{eq:scaling-VS}
\end{equation}
where $z=(x-x_c)L^{1/\nu}$ and $\nu=1/(3-x_S)$. 
At a first glance it looks like this transition will be plagued by large finite-size effects and it will be  difficult to see any sign of scale invariance. In fact this is not the case; this will become clear after analyzing the behaviour of the variable $A$. (We will return to the specific heat below.)

Another standard tool to determine the location of a critical point is the Binder cumulant for the order parameter \cite{binder1981critical}. 
Here $A$ is our order parameter and we define a rescaled version of the Binder parameter:
\begin{equation}
\operatorname{b}_4(A)=-\frac{\kappa_4(A)}{2 \kappa_2(A)^2},
\label{eq:b4A}
\end{equation}
where $\kappa_n(A)$ is the $n$-th order cumulant.\footnote{The standard definition of the Binder cumulant is $U_L=2 \operatorname{b}_4/3$.}
With this normalization, $\operatorname{b}_4(A)$ becomes zero in the deconfined phase (where $A$ is disordered and has a  Gaussian distribution)  and tends to one in the first-order coexistence region (where $A$ has a two-delta distribution).

\begin{figure}[t]
\centering
\includegraphics[width=0.9\linewidth]{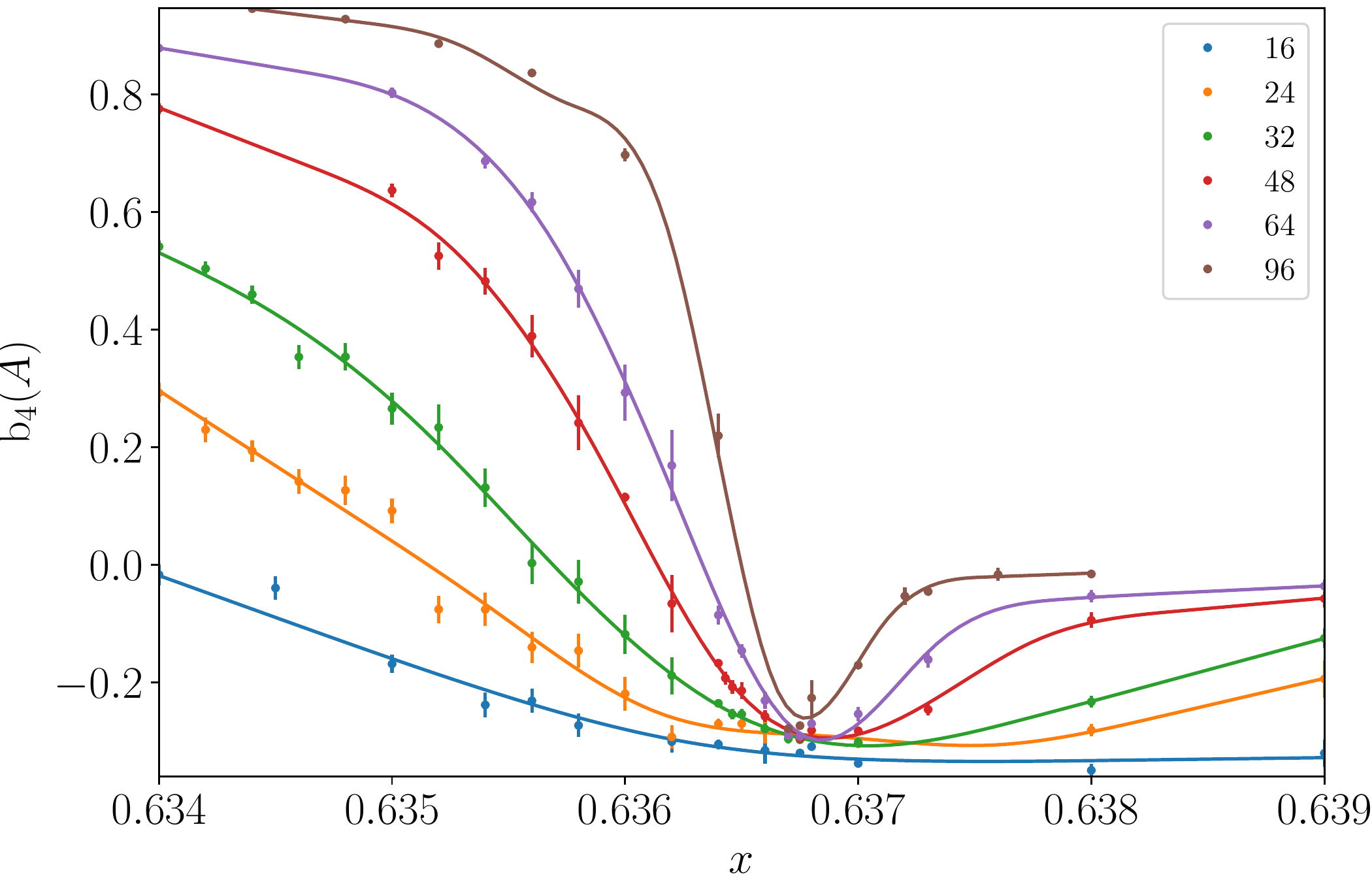}
\caption{$\operatorname{b}_4(A)=-(1/2)\kappa_4(A)/{\rm Var}(A)^2$  as a function of $x$ for different system sizes. The legend indicates system size. The lines are b-spline fits and are just a guide to the eye.}
\label{fig:b4a_x}
\end{figure}

At a conventional second-order symmetry-breaking transition 
(e.g. Ising), the Binder parameter varies monotonically
from zero to one, and  different system sizes  show a crossing that allows accurate location of the critical point. 
This is not the case here, as shown in Fig.~\ref{fig:b4a_x}. 
Rather than crossing, the curves present a minimum near $x=0.6367$. 
The curves do tend to touch here,
consistent with scale invariance 
($\operatorname{b}_4(A)$ is a dimensionless quantity, which should be asymptotically $L$-independent at a critical point),
although some finite-size effects can be appreciated. 
A previous estimate of $x_c\approx 0.6359$ \cite{TupitsynTopological} is not consistent with the location of the minimum (we will give a more accurate estimate below).

 So, after a first look at these two standard quantities it is hard to assess whether the data obeys scaling collapse, and it seems at first sight that accurate estimation of $x_c$ will be more troublesome than for other systems and plagued by finite-size effects.
 Having reached this point, a key step for our understanding was analysing a \textit{parameter-free} scaling collapse that we describe next.

\subsection{Parameter-free scaling collapse; RG trajectories}

We advocate using a parameter-free procedure to determine the quality of scaling collapse of the data near a critical point. 
We construct a parametric plot using as coordinates two dimensionless quantities, $\operatorname{b}_4(A)$ and $\operatorname{b}_1(A)$ (defined below).

In the scaling region, 
\begin{equation}
    \operatorname{b}_4(A)=f(z),
    \label{eq:b4a_scaling}
\end{equation}
where $z=(x-x_c)L^{1/\nu}$. 
Other dimensionless ratios of cumulants are candidates for the second dimensionless quantity.
Binder defined a ratio based on the sixth order cumulant: $V_L=\kappa_6(A)/(30 \kappa_2(A)^3)$ \cite{binder1981critical}.  
High order cumulants are sensitive to the tails of the distribution and can be difficult to estimate accurately. Therefore we instead advocate using the ratio $\<|A|\>/\<A^2\>^{1/2}$:
\begin{equation}
    \operatorname{b}_1(A)=\frac{1}{1-\sqrt{2/\pi}}\left(\frac{\<|A|\>}{\kappa_2(A)^{1/2}}
    - \sqrt{\frac{2}{\pi}} \,
    \right).
    \label{eq:b1a}
\end{equation}
The coefficients have again been chosen so that $\operatorname{b}_1(A)$ tends to zero in the deconfined phase and to one in the coexistence region. $\operatorname{b}_1(A)$ behaves qualitatively like $\operatorname{b}_4(A)$ in Fig.~\ref{fig:b4a_x}, 
and its expected scaling form is as in Eq.~\ref{eq:b4a_scaling}, with a different scaling function.
For a standard Ising transition $\operatorname{b}_1$ goes monotonically from 0 to 1 with a crossing for different system sizes: there, it can be used to determine the critical temperature with the advantage of being slightly easier to estimate than~$\operatorname{b}_4(A)$.

\begin{figure}[t]
\centering
\includegraphics[width=\linewidth]{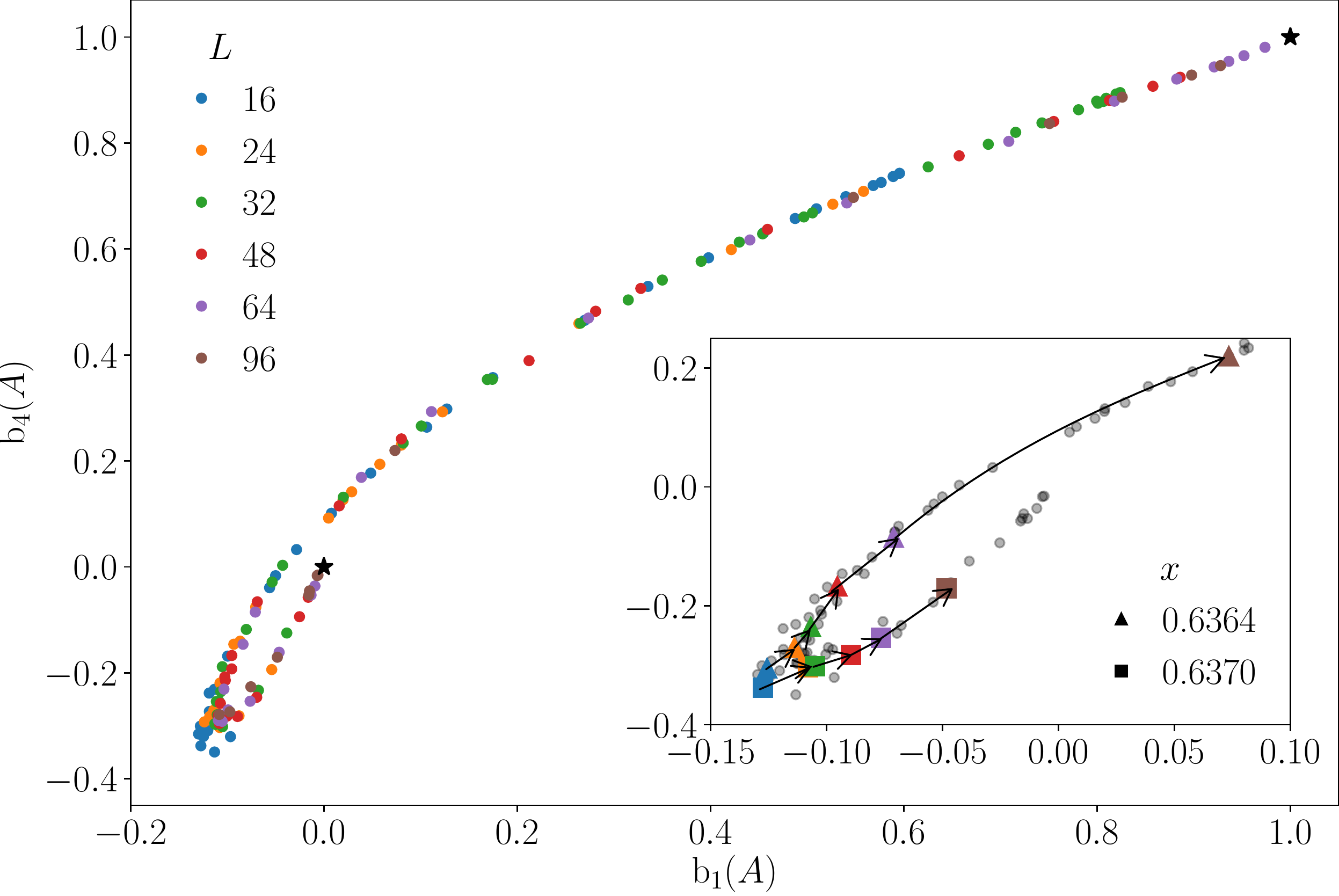}
\caption{Parameter-free scaling collapse for $\operatorname{b}_4(A)$ as a function of $\operatorname{b}_1(A)$ for several system sizes (colored). Black stars mark the two phases: the deconfined phase at $(0,0)$ and the broken-duality phase (i.e.  first-order coexistence) at $(1,1)$. Inset shows a zoom of the left lower corner of the main panel. Two selected $x$ values ($0.6364$ and $0.6370$) are highlighted and arrows are drawn in-between consecutive system sizes. Close to the critical point, $x \approx 0.6367$, values remain in a small region around $(\operatorname{b}_1(A), \operatorname{b}_4(A)) \approx (-0.109, -0.285)$.}
\label{fig:parameterfree}
\end{figure}

By plotting $\operatorname{b}_4(A)$ versus $\operatorname{b}_1(A)$ we obtain a parametric plot where $z$ is the parameter, see Fig.~\ref{fig:parameterfree}. 
If scaling is obeyed, points with different $x$ and $L$, but the same $z$, must overlap. This is a fair test of scale invariance because we do not have to fix or fit any parameters by hand and instead just plot raw data.  

The data traces a trajectory from the point (0,0) to the point (1,1), showing very good overlap,  except near the region $(\operatorname{b}_1(A), \operatorname{b}_4(A)) \approx (-0.109, -0.285)$ where we see some finite-size effects. However, these finite-size effects become small for $L>32$. 
This figure represents on its own strong evidence that the multi-critical point is a second-order phase transition.

This figure can be also used to estimate the critical point. We construct ``RG trajectories'' in the  ${(\operatorname{b}_1(A), \operatorname{b}_4(A))}$ plane by following the points for a fixed value of $x$ as $L$ is increased.
The points representing a system (for a fixed generic $x$) should flow along the universal line towards either the (0,0) or the (1,1) fixed point. 
The inset of Fig.~\ref{fig:parameterfree} shows this flow  for two different $x$ values. 
For the system sizes used this simple procedure already determines the critical point with four digits of precision. 
We observe that for $x> 0.6367$ the system flows towards (0,0), while for   $x<0.6366$ it flows to (1,1). The repulsive fixed point is located (within the precision of the procedure) at the lower left extreme of the universal curve.

We see that the data in Fig.~\ref{fig:b4a_x} should approximately scale for $L \gtrsim 32$ (fits  are given below). 
A similar figure using ratios involving $S$ (e.g. $\operatorname{b}_4(S)$ or $\kappa_3(S)/\kappa_2(S)^{3/2}$) does not show good overlap, 
as expected from the discussion of Fig.~\ref{fig:VS_x}. 
It would be strange to have very large finite-size effects in quantities depending on $S$  but not on those depending on $A$. The explanation turns out to be very simple. The exponent $x_S$ is very near
1.5, where the regular contribution to $\operatorname{Var}(S)$ cannot be neglected. When this is taken into account quantities depending on $S$ also obey scaling (Sec.~\ref{sec:scalingcollapseS}).

\section{Critical exponents}
\label{sec:criticalexponents}

We turn to scaling fits in order to determine $x_c$ and the scaling dimensions of $A$ and $S$ ($x_A$ and $x_S$ respectively). 
Details of how fits were constructed 
may be found in App.~\ref{app:MCscheme}.
The results of the various fits are summarized in Table~\ref{tab:exponents}.

\begin{table}[t]
\centering
\begin{tabular}{c|c|c|c|c|c}
\text{Variable} &    $x_c$     &  $x_S$    & $x_A$     & $\chi^2$ & \text{d.o.f.}  
\\ \hline
\hline
   $\operatorname{b}_1(A)$     & 0.636660(16) & 1.446(56) &           & 49.53    & 46     \\ 
\hline
   $\operatorname{b}_4(A)$     & 0.636670(14) & 1.445(62) &           & 65.8     & 46     \\ 
\hline
  $\sqrt{\<A^2\>}$& 0.636702(20) & 1.502(43) & 1.222(16) & 44.9     & 40     \\  
\hline
  $\<|A|\> $      & 0.636702(22) & 1.510(48) & 1.221(16) & 43.1     & 40     \\  
\hline
\hline
   $Var(S)$     & 0.636661(14) & 1.5(fixed)&           & 88.6     & 81     \\ 
\hline
 $\kappa_3(S)$  & 0.636651(18) &  1.506(9) &           & 68.6     & 66      \\ 
 \hline
\end{tabular}
 \caption{Results of fits.  Errors shown are purely statistical.}
  \label{tab:exponents}
\end{table}

\subsection{Scaling collapse for $A$}
\label{sec:scalingcollapseforA}

\begin{figure}[t]
\centering
\includegraphics[width=\linewidth]{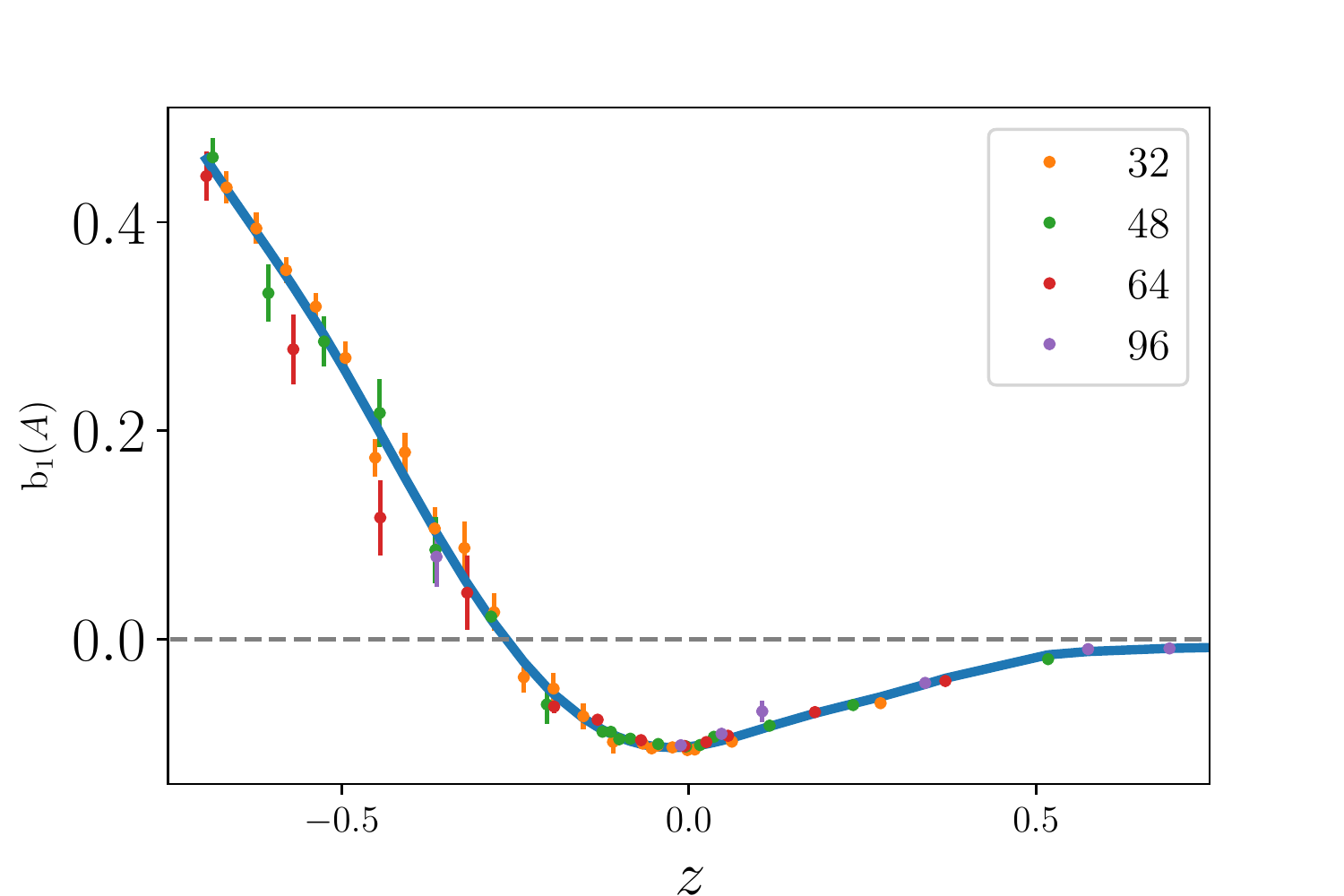}
\caption{Scaling collapse of $\operatorname{b}_1(A)$ versus scaling variable $z=(x-x_c)L^{1/\nu}$, where $1/\nu=3-x_S$. The blue line corresponds to the fitted scaling function using B-splines with 12 degrees of freedom. The legend indicates the different system sizes.}
\label{fig:fit-b1A}
\end{figure}

The scaling form for dimensionless quantities such as $\operatorname{b}_1(A)$ involves, in addition to the scaling function, the parameters $x_c$ and $\nu=1/(3-x_S)$. 
Fig.~\ref{fig:fit-b1A} shows the scaling collapse of the data for  $\operatorname{b}_1(A)$, 
and the fitted scaling function.
The critical coupling $x_c$ obtained (Table \ref{tab:exponents}) is very near the initial estimation made in section  \ref{sec:spontaneousbreakingduality} and $x_S$ is near 1.5, as noted above.
A fit of $\operatorname{b}_4(A)$ gives very similar results (Table \ref{tab:exponents}).

\begin{figure}[htb]
\centering
\includegraphics[width=0.87\linewidth]{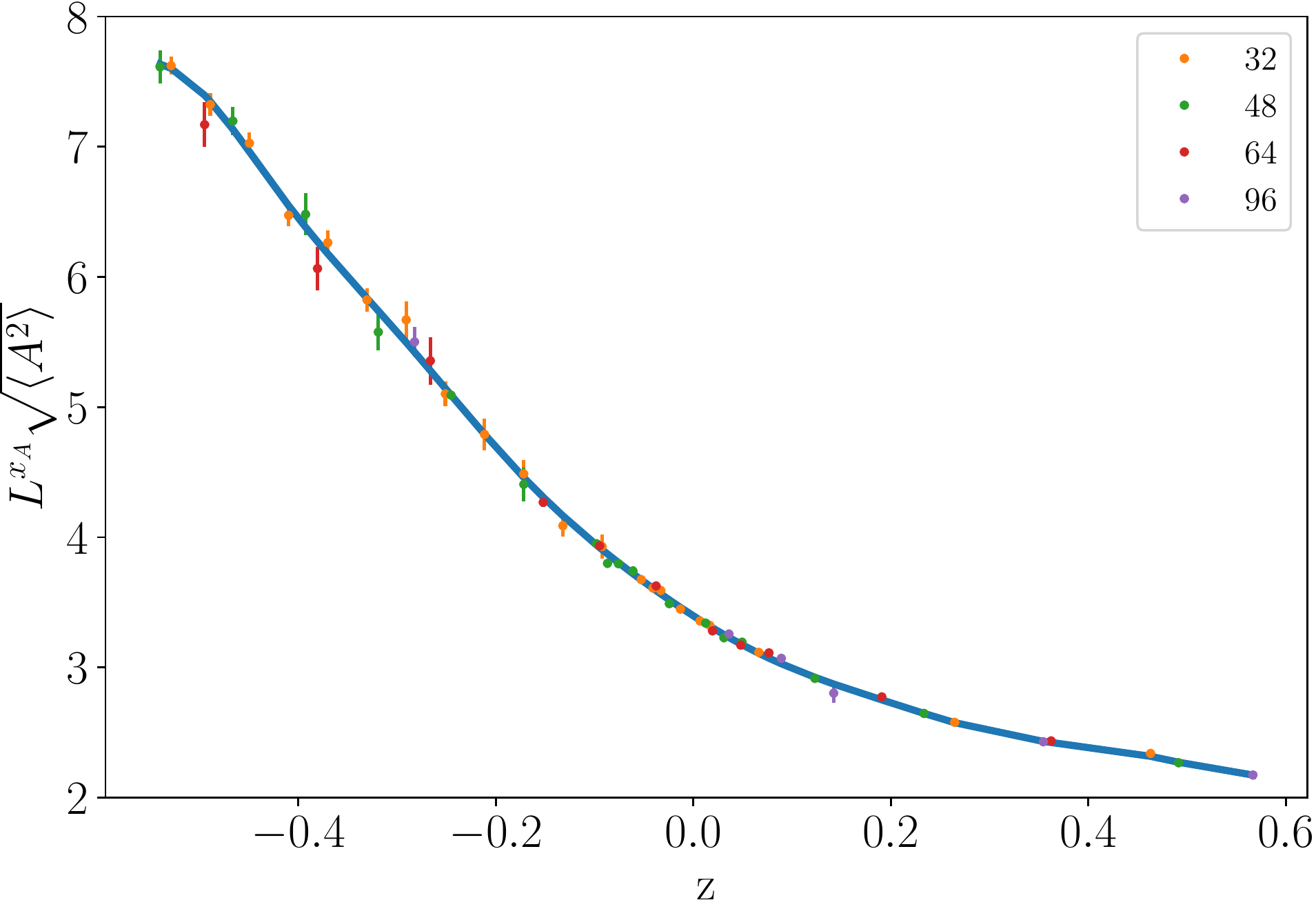}
\caption{Scaling collapse of $L^{x_A} \sqrt{\<A^2\>}$ versus scaling variable $z=(x-x_c)L^{1/\nu}$, where $1/\nu=3-x_S$.  Blue line is the fitted scaling function using B-splines with 10 degrees of freedom. Legend indicates  system sizes.}
\label{fig:fit-A}
\end{figure}

In order to obtain the exponent $x_A$ we fit
\begin{equation}
    \sqrt{\<A^2\>}=L^{-x_A} g(z).
\end{equation}
The resulting scaling function, $g(z)$, is shown in Fig.~\ref{fig:fit-A} and the fitted parameters are indicated in Table \ref{tab:exponents}. 
Fitting $\<|A|\>$ yields similar results.

No finite-size-scaling corrections are included in these fits, 
although for ${L=32}$ these corrections are still non-negligible (compared to the error bars). 
If data for ${L=32}$ is excluded from the fits for $\operatorname{b}_1(A)$ and $\operatorname{b}_4(A)$ the estimates of $x_S$ increase.

\subsection{Scaling collapse for $S$}
\label{sec:scalingcollapseS}

We have suggested above that the failure of a straightforward scaling collapse for $\operatorname{Var}(S)$ is due to $x_S$ being very close to $3/2$, the threshold where the regular contribution becomes comparable with the scaling contribution.
Fortunately, a simple modification of the scaling ansatz should be accurate when ${|x_S-3/2|\ll 1/\log L}$:
\begin{equation}\label{eq:varSmodifiedscaling}
    L^3 \operatorname{Var}(S) \simeq f(z) + 4\pi \alpha_S^2 \log(L).
\end{equation}
Here $\alpha_S^2$ is the normalization constant for the two-point function of $S$ (Sec.~\ref{sec:twopointcorrelations}).
The function $f(z)$ includes a $z$-independent constant contribution, which arises from nonuniversal short-distance correlations.\footnote{To see that Eq.~\ref{eq:varSmodifiedscaling} holds, recall that  $L^3\operatorname{Var}(S)$ is given by the integral over the connected two-point function of $S(r)$. 
If we neglect terms of size ${(x_S-3/2)\log L}$, then it is sufficient to replace the power-law $r^{-2 x_S}$ occuring in this integral with $r^{-3}$:
\be\label{eq:varSasintegral}
L^3 \operatorname{Var}(S)
\simeq
\int_{[0,L]^3} \dd^3 r \, r^{-3} H\lf \f{r}{L}, z \ri + B.
\ee
Here $H$ is a scaling function for the correlator in the finite system, and $H(0,0)=\alpha_S^2$.
The integral is cut off at $r$ of order 1, and the constant $B$ represents a short-distance contribution. This integral gives Eq.~\ref{eq:varSmodifiedscaling}, in which the nonuniversal constant $B$ has been absorbed into the function $f(z)$.}

\begin{figure}[t]
\centering
\includegraphics[width=\linewidth]{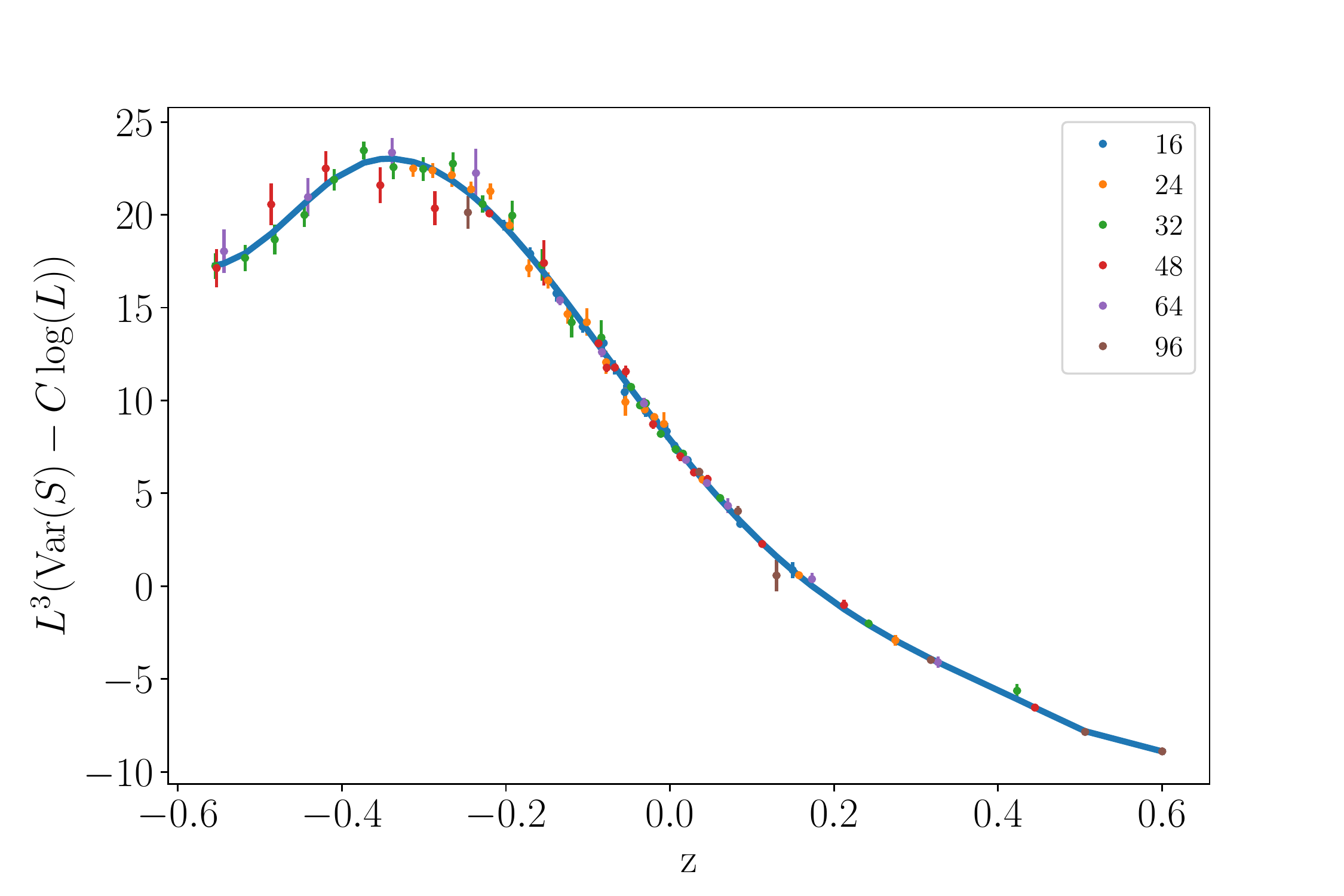}
\caption{Scaling collapse for $L^{3} (\operatorname{Var}(S)-C\log(L))$ versus scaling variable $z=(x-x_c)L^{1/\nu}$, where $1/\nu=3-x_S$. The blue line corresponds to the fitted scaling function using B-splines with 10 degrees of freedom. Legend indicates different system sizes. In this fit only $x_S$ has been fixed to 3/2, see text for explanation of the scaling ansatz.}
\label{fig:fit-VS}
\end{figure}
\begin{figure}[t]
\centering
\includegraphics[width=\linewidth]{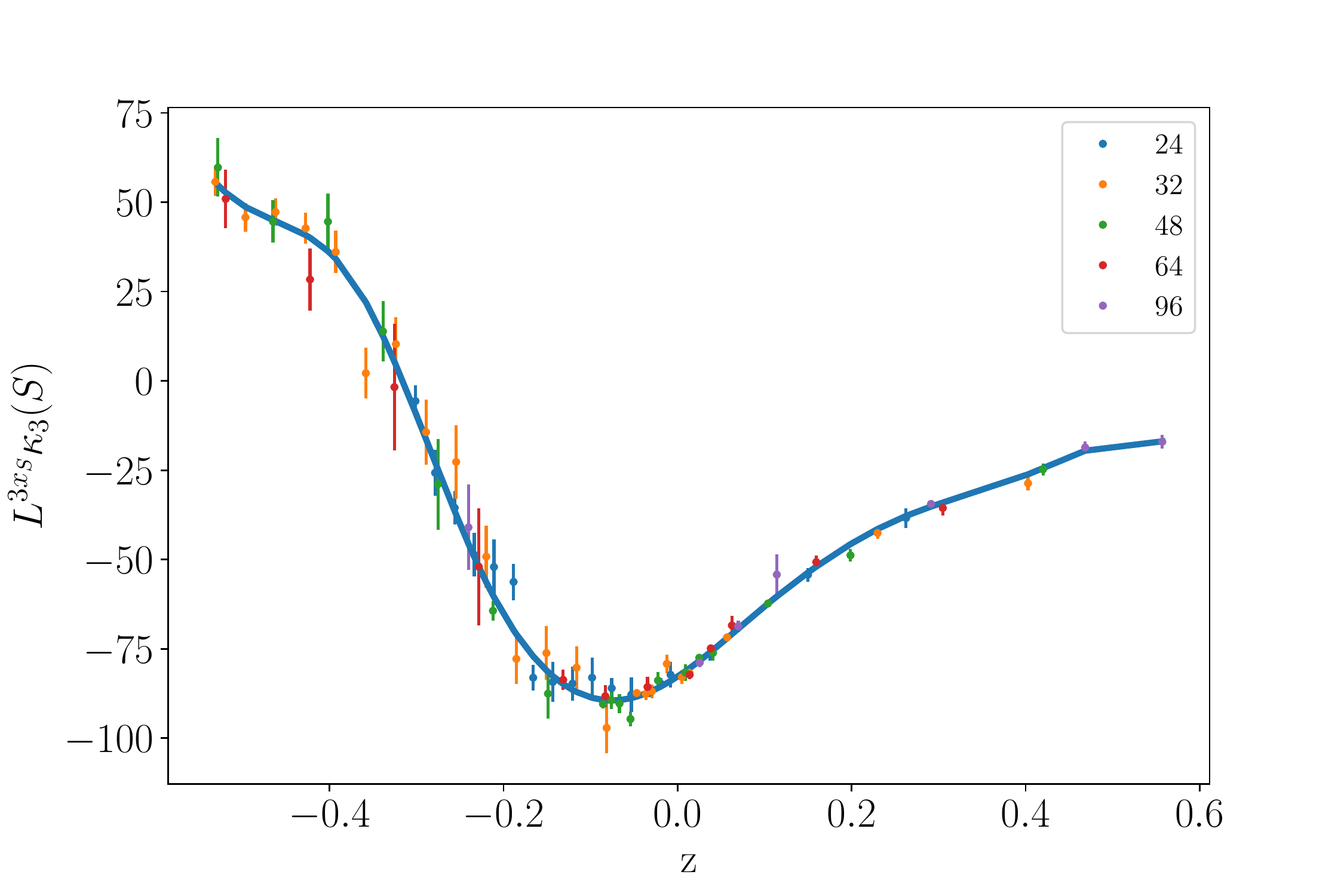}
\caption{Scaling collapse for $L^{3 x_S} \kappa_3(S)$ versus scaling variable $z=(x-x_c)L^{1/\nu}$, where $1/\nu=3-x_S$. The blue line corresponds to the fitted scaling function using B-splines with 10 degrees of freedom.}
\label{fig:fit-k3S}
\end{figure}

We have performed fits to this form keeping $\nu=2/3$ fixed (in line with the approximation above) so that only the critical coupling $x_c$ and the coefficient ${C=4\pi \alpha_S^2}$ of the logarithm can be adjusted to obtain scaling collapse. The scaling function is shown in Fig. \ref{fig:fit-VS}. We obtain a good fit, even when data from small system sizes are included. The estimated $x_c$ is again very similar to previous estimates. Also, the constant $C=10.05(23)$ obtained is consistent with our calculation of the correlation function in the next section. 
In summary, the fit to  $\operatorname{Var}(S)$ is consistent with the correlator of $S$ obeying scaling with $x_S$ very close to 3/2 (indeed, allowing $\nu$ to be free in this fit, instead of fixed to $2/3$, did not improve the fit quality).

An alternative way to avoid dealing with the regular contribution is to analyze higher-order cumulants. 
The  singular contribution near a critical point scales as ${\kappa_n(S)= L^{-n x_S}f(z)}$, while the regular contribution should scale as $\kappa_n(S)_\text{regular}\propto L^{-d(n-1)}$. 
For $x_S \approx 3/2$ and $n=2$ both contributions scale in the same way,
but for $n=3$ the singular contribution should dominate.
Indeed the data for $\kappa_3(S)$ can be collapsed, as shown in
Fig. \ref{fig:fit-k3S}. We obtain $x_c$ and $x_S$ values fully consistent with previous results (Tab.~\ref{tab:exponents}), although it is worth noting that the statistical error of $x_S$ is much smaller.

\subsection{Summary of exponents from the fits}
\label{sec:summaryexps}

We have provided clear evidence that the Ising gauge-Higgs model has a scale-invariant multicritical point. 
Simulations are inevitably restricted to finite lengthscales, 
so can never rigorously exclude an extremely weak first-order transition; 
but all of the observables we have examined exhibit good scaling collapse, with fairly modest finite-size effects. 

As there are some finite-size effects, we consider a reasonable confidence interval for the critical point to be  $x_c \in [0.63665,0.6367]$. 
For the study of correlation functions in the next sections we round to four digits and consider critical behaviour at $x_c\approx 0.6367$. 

For the exponent $x_S$, the value obtained from $\kappa_3(S)$ (Table~\ref{tab:exponents}) has the smallest statistical error and we take it as a reference in the following. Our statistical error bars do not take into account possible systematic errors related for example to finite size effects,
so a realistic confidence interval should be larger; however,
we have verified that dropping the smaller system sizes in the fit only very slightly increased $x_S$, remaining within the statistical error bars. 

Our estimate of $x_S$ leads to $\nu=0.669(4)$. 
This is not too far from estimates $\nu\approx  0.7$ \cite{vidal2009low} and $\nu\approx  0.69$ \cite{dusuel2011robustness} based on the calculation of the gap in the toric code using small-field series expansions (up to eighth order).
However, a basic issue with the series expansion method is that it cannot detect a first-order transition \cite{vidal2009low}, i.e. it must assume a continuous transition rather than demonstrating one. Ref.~\cite{dusuel2011robustness} attempts to rectify this by comparing estimates of the ground state energy from series expansion with a variational wavefunction, but we expect that the accuracy with which this method could detect a weak first order transition is severely limited by the accuracy of the variational wavefunction.

A realistic confidence interval for $x_A$ should again be larger than the statistical one in Tab.~\ref{tab:exponents}. We note that the $x_c$ estimates obtained from  $\sqrt{\<A^2\>}$ and $\<|A|\>$ are slightly larger than for the other fits; if we estimate $x_A$ keeping  $x_c=0.63666$ fixed, then the value  drops to 1.20, slightly below the statistical confidence interval.

Standard scaling relations \cite{cardy1996scaling} imply that the order parameter exponent $\beta$, defined by ${\sqrt{\<A^2\>}\sim (x_c-x)^\beta}$ in the infinite system, is ${\beta=x_A/(3-x_S)}$ (compare Fig.~\ref{fig:A_x}).
Asymptotically close to the self-dual critical point, the shape of the Higgs and confinement lines in the inset to Fig.~\ref{fig:phase_diag} should be   ${y_- \sim \pm |\delta y_+|^{(3-x_A)/(3-x_S)}}$, where ${y_\pm = (y\pm y')}$ gives the self dual and anti-self-dual couplings. 
Since  ${(3-x_A)/(3-x_S)}$ is a little larger than one, the Higgs and confinement lines are asymptotically parallel as they approach the critical point.

The values obtained for the critical exponents clearly differ from Ising values, but they are surprisingly close to certain  exponents in the XY model. This point will be discussed in Sec.~\ref{sec:relatedmodels}.

\section{Two-point correlators}
\label{sec:twopointcorrelations}

\begin{figure}[t]
\centering
\includegraphics[width=0.95\linewidth]{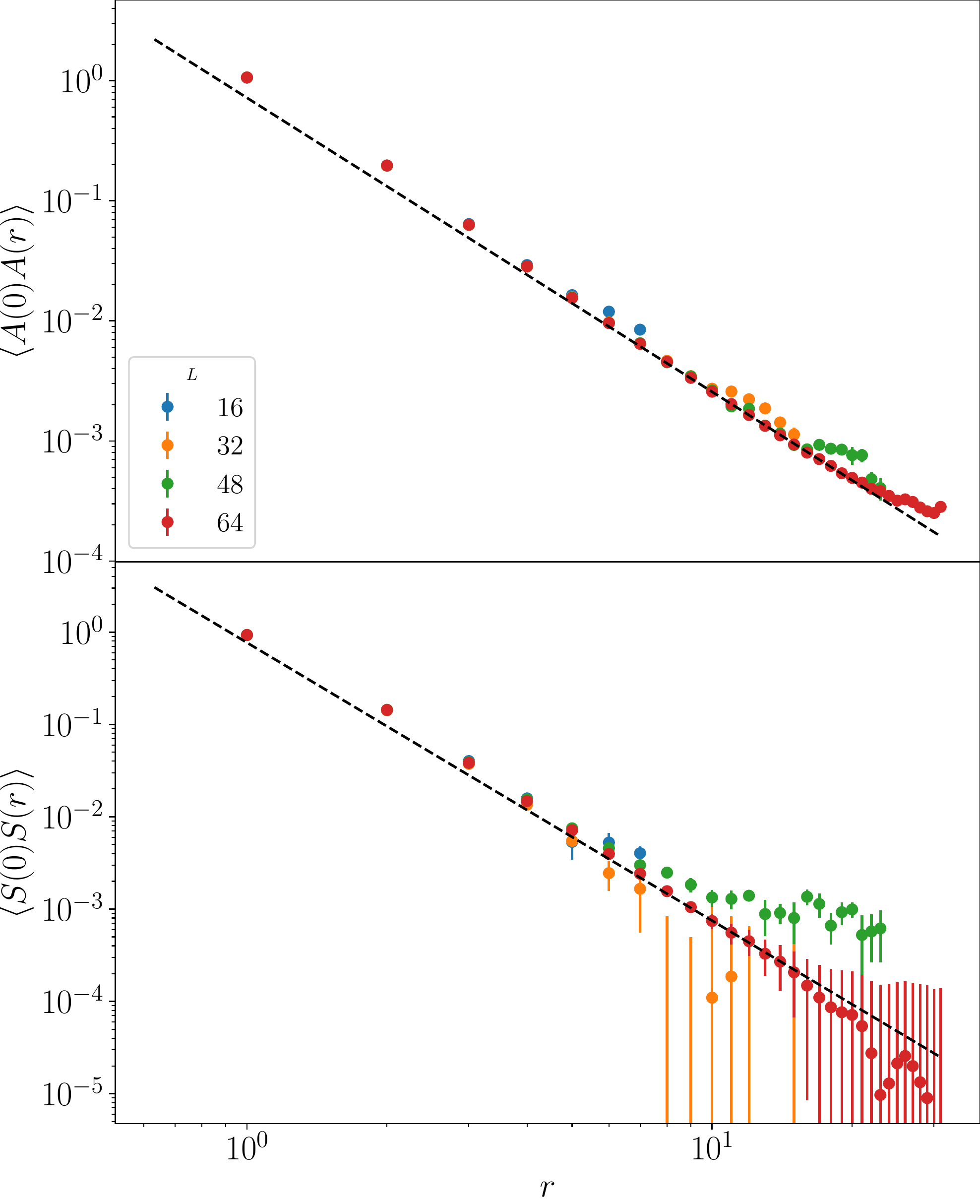}
\caption{Two-point correlators for the operators $A$ (top) and $S$ (bottom). 
The displacement between the two operators is taken parallel to a lattice direction.  
Dashed lines are fits of the $L=64$ data in the range $r\in [10, 15]$ 
to the forms in Eq.~\ref{eq:2pointfnforms}
with $x_A=1.224$, $x_S=1.506$ fixed and $\alpha_{A,S}$ free, giving $\alpha_A^2 = 0.72 $ and $\alpha_S^2 = 0.77$.
}
\label{fig:C2}
\end{figure}

We now show that two-point functions of the local operators $A(r)$ and $S(r)$ are consistent with scale invariance,
\ba\label{eq:2pointfnforms}
\<A(0)A(r)\> & = \f{\alpha_A^2}{r^{2x_A}},
&
\<S(0)S(r)\>_\text{conn} & = \f{\alpha_S^2}{r^{2x_S}}.
\end{align}
Fig.~\ref{fig:C2} compares data for the critical two-point functions to such power-law fits, giving good agreement at larger separations. 
The exponents $x_A$ and $x_S$ in the fits have been fixed to the values $1.224$ and $1.506$,   respectively (see Tab.~\ref{tab:exponents}), 
while the nonuniversal constants  $\alpha_{A,S}$, which we will require in Sec.~\ref{sec:threepointfunction}, have been left free.

The simulations also give access to dynamical correlation functions in Monte-Carlo time, which we analyze in Sec.~\ref{sec:stochasticdynamics}. These also encode the exponents $x_A$ and $x_S$, together with a dynamical exponent $z$.

\section{Three-point function and conformal invariance}
\label{sec:threepointfunction}

\begin{figure}[t]
\centering
\includegraphics[width=0.95\linewidth]{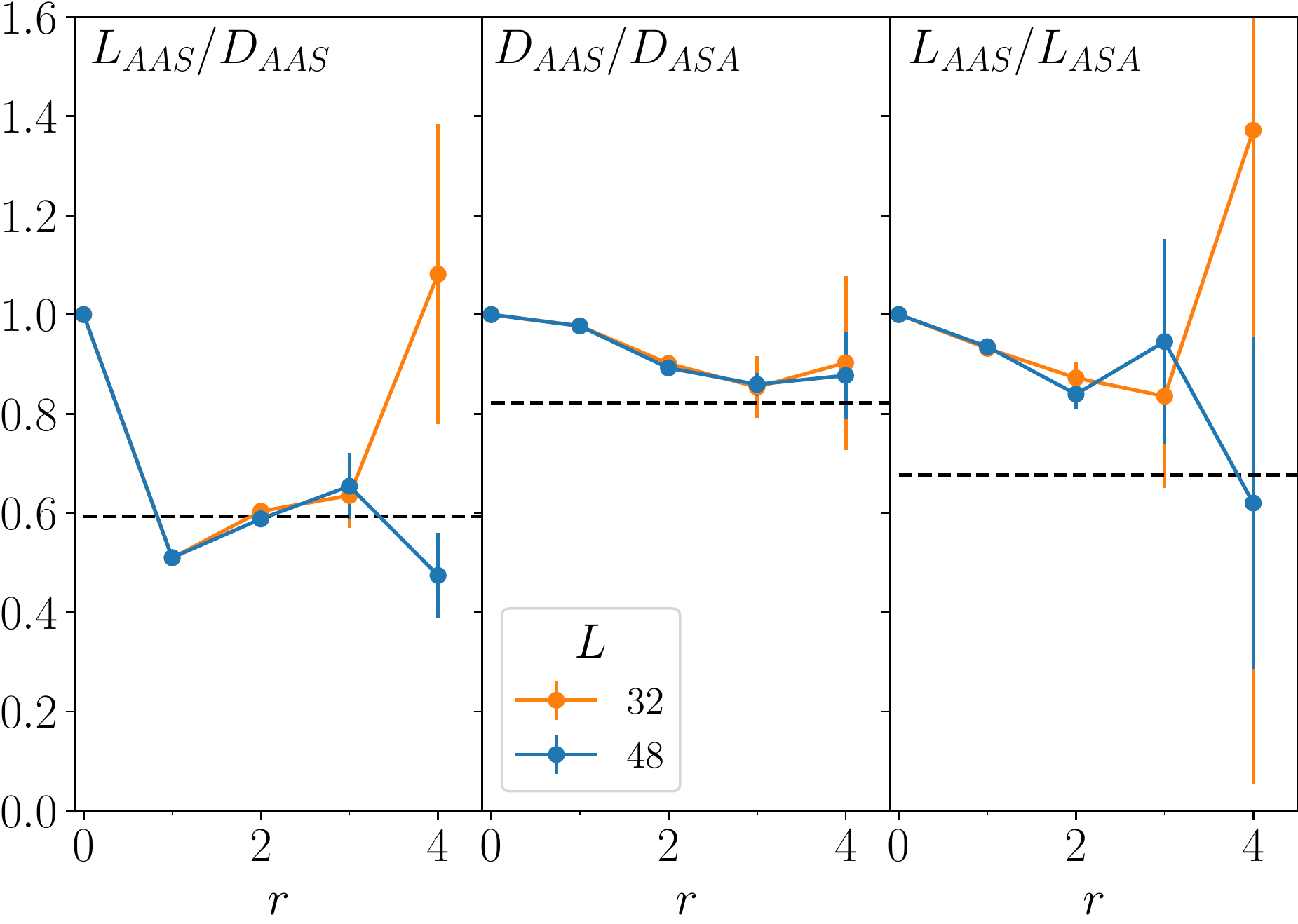}
\caption{Test of conformal invariance: 
ratios of the three-point correlators defined in the text. 
Assuming exponent  values $x_A= 1.224$, $x_S = 1.506$, 
conformal invariance requires these ratios to converge at large $r$ to the values  indicated by dashed lines  (error bars are from variation between 6 samples). 
We find agreement with the predicted value, within error bars, once $r\gtrsim 3$. Error bars become too large for a useful comparison once $r>4$.}
\label{fig:C3-ratios}
\end{figure}

Conformal invariance fixes the three point functions in terms of the fields' scaling dimensions and  operator product expansion (OPE) coefficients \cite{cardy1996scaling}.
Conversely, data for three-point functions allow a direct numerical test of conformal invariance.

The OPE coefficients for the fields $A$ and $S$ that are allowed by duality symmetry to be nonzero are $C_{AAS}$ and $C_{SSS}$.
Here we examine the three-point function $\< A(0) A(\vec  r) S(\vec r') \>_\text{conn}$ and give a very rough estimate of the corresponding OPE coefficient $C_{AAS}$. 
Data for $\< S(0) S(\vec  r) S(\vec r') \>_\text{conn}$ was too noisy for a similar analysis.

The form dictated by conformal invariance for the three-point function is 
\be\label{eq:threepointgeneralform}
\< A(0) A(\vec  r) S(\vec r') \>_\text{conn} = \f{C_{AAS} \times  \alpha_A^2 \alpha_S  }{|\vec r|^{2x_A - x_S} |\vec r'|^{x_S} |\vec r - \vec r'|^{x_S}}.
\ee
where $\alpha_{A,S}$ are the same  operator normalization constants that appear in the two-point functions (\ref{eq:2pointfnforms}). 

We consider four possible spatial arrangements for the three points in the correlator, lying either on a line ($L$) or on the vertices of a right triangle ($D$): 
\ba
L_{AAS}(r) & \equiv \< A(0,0,0)A(r,0,0)S(2r,0,0)\>_\text{conn},\\
L_{ASA}(r) & \equiv \< A(0,0,0)S(r,0,0)A(2r,0,0)\>_\text{conn},\\
D_{AAS}(r) & \equiv \< A(0,0,0)A(r,0,0)S(r,r,0)\>_\text{conn},\\
D_{ASA}(r) & \equiv \< A(0,0,0)S(r,0,0)A(r,r,0)\>_\text{conn}.
\end{align}
We use ratios of these three-point functions to test for conformal invariance. 
The CFT  prediction depends only on  $x_A$ and $x_S$ (and on the arrangement of points in the correlator), 
so this test does not require an independent estimate of the nonuniversal constants $\alpha_{A,S}$ in Eq.~\ref{eq:threepointgeneralform}.

Fig.~\ref{fig:C3-ratios} compares each of the three independent 3-point function ratios with the CFT prediction (for $x_A=1.224$, $x_S=1.506$), 
which is marked with a dashed line.
Modulo uncertainty in the exponent estimates, the data should converge to these lines at large $r$.
Statistical errors limit us to small $r$, because of the rapid decay of the 3-point functions with $r$.
Despite this, 
there is agreement, within errors, with the CFT prediction once $r\gtrsim 3$.

\begin{figure}[t]
\centering
\includegraphics[width=0.9\linewidth]{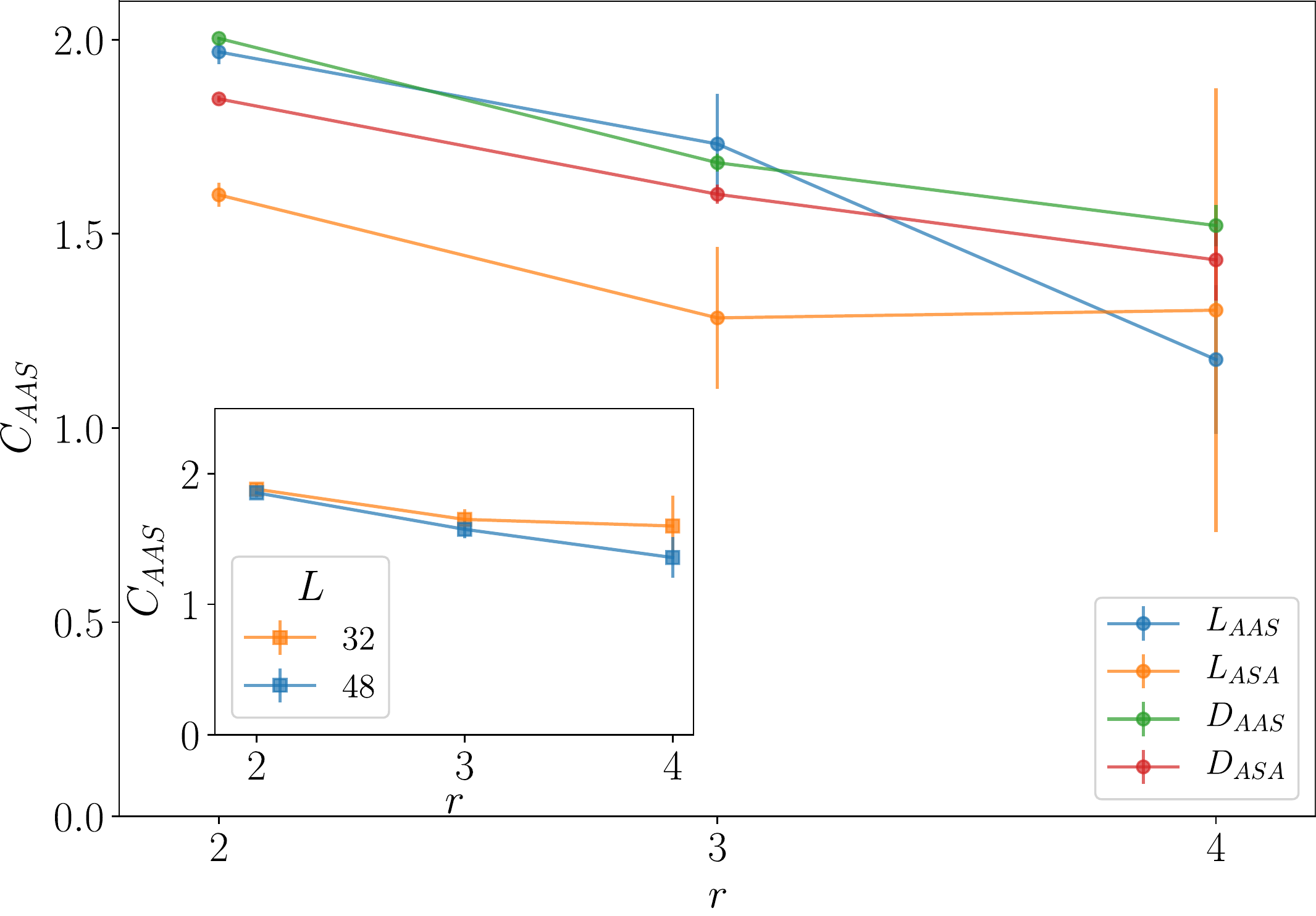}
\caption{
Main panel: Finite-$r$ estimates of the OPE coefficient $C_{AAS}$ using 3-point functions with four different geometries, using data from system size $L=48$ (error bars are from variation between 6 samples). Inset: Average of the four estimates for  $L=48$ and also for $L=32$.}
\label{fig:CAAS}
\end{figure}

Motivated by this consistency, we make a very preliminary estimate of the universal constant $C_{AAS}$.
Fig.~\ref{fig:CAAS} shows finite-$r$ estimates obtained from Eq.~\ref{eq:threepointgeneralform} (for each geometry of 3-point function). 
The data suggest that ${C_{AAS}\sim 1.5}$.
The uncertainty is large, because of the very small range of $r$,
and because the uncertainty in  $\alpha_{A,S}$ (obtained from the 2-point function in Sec.~\ref{sec:twopointcorrelations}) is hard to estimate. 
It would be worthwhile to improve this estimate.  Refs.~\cite{caselle2015numerical} and \cite{hasenbusch2020two} discuss methods for numerical estimation of OPE coefficents.

\section{Stochastic dynamics of membranes}
\label{sec:stochasticdynamics}

So far we have discussed the gauge theory as a problem of equilibrium statistical mechanics in either 2+1 or 3+0 dimensions.
But our simulations involve a fourth coordinate, which is Monte Carlo time  (denoted $t$). 
The Monte Carlo dynamics may be interpreted physically as a model for stochastic thermal motion of classical membranes (Sec.~\ref{sec:membranepicture}), or alternately of classical spins in 3D (Sec.~\ref{sec:modelasspins}). These dynamics contain additional universal data beyond the data in static correlations:
most importantly, the dynamical exponent $z$ that dictates how the typical relaxational timescale $\tau$ scales with system size $L$ at the critical point, ${\tau \sim L^z}$.\footnote{The dynamical exponent $z$  of the 3+1D stochastic dynamics should not be confused with the dynamical exponent $z_\text{QM}=1$ of the 2+1D quantum system.}
Two-time correlation functions in this dynamics are also an alternative means of determining the  exponents $x_A$ and $x_S$, as shown below.

\subsection{Universal dynamics and duality}
\label{sec:universaldynamicsduality}

There is great freedom in 
the microscopic definition of the stochastic dynamics,
i.e. the choice of update for our Monte Carlo Markov chain.
But we expect to find a dynamical fixed point that embraces a large class of microscopic updates that are local and preserve detailed balance
(our updates are local and are described in Sec.~\ref{sec:MCshortsummary} below).
This is analogous to, say, the critical 3D Ising model which shows a robust universality class for spin-flip dynamics with no conservation laws (the universality class of ``Model A'' \cite{glauber1963time,hohenberg1977theory,wansleben1987dynamical,wansleben1991monte,munkel1993dynamical,ito1993non,grassberger1995damage,jaster1999short,ito2000nonequilibrium,murase2007dynamic,collura2010off,niermann2015critical,
mesterhazy2015quantum,duclut2017frequency,adzhemyan2018diagram,hasenbusch2020dynamic}).

As in the Ising model, the dynamical universality class may  change if we introduce conservation laws \cite{hohenberg1977theory}. For example, dynamics that conserve the total membrane area (the total number of occupied plaquettes) could  be relevant to some experimental settings.
The dynamical universality class may also change if we include nonlocal updates in the Monte Carlo simulations: finding a nonlocal update  that speeds up simulations by reducing $z$ is a challenging open problem (Sec.~\ref{sec:outlook}).

The present dynamical critical point has one subtlety that arises from self-duality.
We have argued that self-duality is a $\mathbb{Z}_2$ symmetry of the 3D fixed point, allowing us to classify scaling operators as $\mathbb{Z}_2$ even or odd ($S$ and $A$ respectively). 
The mixed correlator $\<AS\>$ therefore vanishes in the equilibrium ensemble.\footnote{As discussed in Sec.~\ref{sec:definingAandS}, the lattice operators are only self-dual or anti-self-dual up to derivative terms.}
But the Monte-Carlo dynamics itself is not $\mathbb{Z}_2$-symmetric  \cite{jongeward1980monte}.
To define the dynamics we had to choose one of the two dual representations, either on the original cubic lattice or on its dual, breaking the symmetry between them.
As a result, the mixed correlator $\<AS\>$ can be nonzero for non-equal times.

Assuming that the scaling operators ${A(r)}$ and ${S(r)}$ of the three-dimensional theory are lifted to scaling operators ${A(r,t)}$ and ${S(r,t)}$ in the dynamical theory, 
 standard dynamical scaling  \cite{hohenberg1977theory} at large $|r|$ and $t$ gives:\footnote{Detailed balance implies that the correlator is invariant under $t\rightarrow-t$ so we take $t>0$.}
\begin{align}\label{eq:dynamicalscalingform1}
\< A(r,t) A(0,0) \> & = t^{-{2 x_A}/{z}} F_{AA} (t/|r|^z, t/L^z),
\\ \label{eq:dynamicalscalingform2}
\< S(r,t) S(0,0) \> & = t^{-{2 x_S}/{z}} F_{SS} (t/|r|^z, t/L^z), \\
\label{eq:dynamicalscalingform3}
\< A(r,t) S(0,0) \> & = t^{-{(x_A+x_S)}/{z}} F_{AS} (t/|r|^z, t/L^z).
\end{align}
The $\mathbb{Z}_2$ symmetry of the equilibrium critical point ensures that the last line vanishes at equal time.

\begin{figure}[t]
\includegraphics[width=0.98\linewidth]{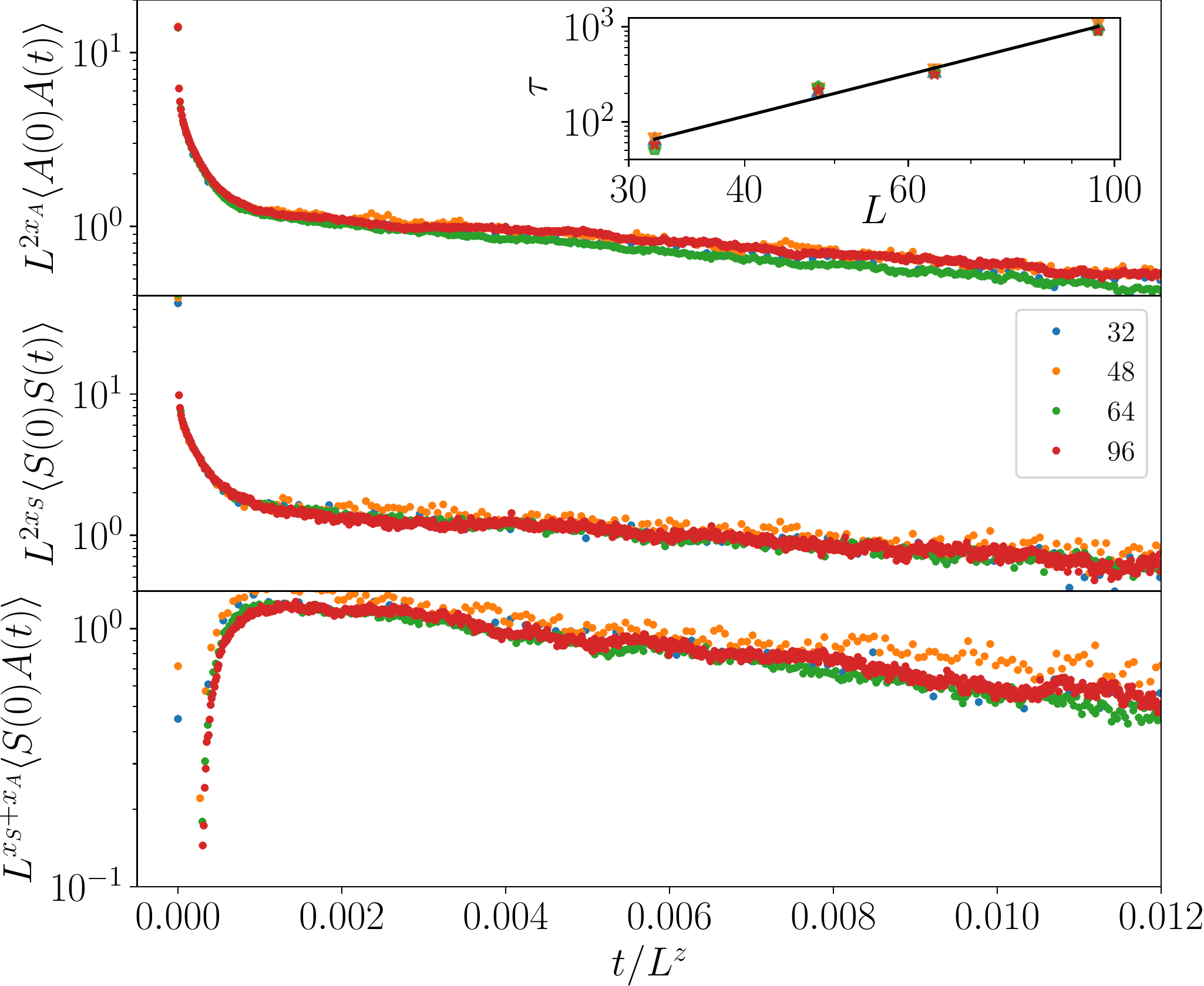}
\caption{Main panels: Scaling collapse of autocorrelation functions for   $\langle A(0) A(t) \rangle$ (top),  $\langle S(0) S(t) \rangle$ (center) and  $\langle S(0) A(t) \rangle$ (bottom) as a function of $t/L^z$ (using $x_A=1.224$ and $x_S=1.506$). Inset: Autocorrelation time as a function of system size for 4 different correlators: $\langle E(0) E(t) \rangle$ (blue triangle), $\langle A(0) A(t) \rangle$ (orange triangle), $\langle S(0) S(t) \rangle$ (green pentagon), $\langle S(0) A(t) \rangle$ (red star). Straight line fits all the data points to a power-law $\tau = A L^z$ with $z = 2.48$.
}
\label{fig:dynamics}
\end{figure}

\subsection{Dynamical scaling collapse}

First we obtain the dynamical exponent  from the typical relaxation timescale $\tau(L)$ of a sample of size $L$. 
We estimate this timescale  from the exponential decay of  various two-time correlators, in particular those of $S$ and $A$ (inset to Fig.~\ref{fig:dynamics}).
The various estimates are consistent with each other,\footnote{At first glance we might have expected the relaxation times for duality-odd and duality-even operators to differ by a nontrivial order-one factor, due to coupling to eigenstates of the Markov matrix with different symmetry under duality. 
But the Monte-Carlo dynamics is not in fact symmetric under duality, so all operators can couple to the lowest excited state, whose eigenvalue determines $\tau$.} and fitting $\tau(L)$ to a power law gives:
\be
z \simeq 2.48.
\ee
For comparison, this is larger than the dynamical exponent for spin flip dynamics in the 3D Ising model \cite{wansleben1987dynamical,wansleben1991monte,munkel1993dynamical,ito1993non,grassberger1995damage,jaster1999short,ito2000nonequilibrium,murase2007dynamic,collura2010off,niermann2015critical,
mesterhazy2015quantum,duclut2017frequency,adzhemyan2018diagram,hasenbusch2020dynamic}, for which a recent estimate is $z=2.0245(15)$  \cite{hasenbusch2020dynamic}.

The main panel of Fig.~\ref{fig:dynamics} demonstrates scaling collapse for the temporal correlators of the spatially averaged operators $A$ and $S$, using this exponent.
(The relevant scaling forms are given by integrating {Eqs.~\ref{eq:dynamicalscalingform1}--\ref{eq:dynamicalscalingform3}}.)
Results are consistent with expectations from Sec.~\ref{sec:universaldynamicsduality}, including the continuous vanishing of the scaling function for $\<S(0)A(t)\>$  as $t\rightarrow 0$.

\subsection{Monte-Carlo updates}
\label{sec:MCshortsummary}

The simplest Monte Carlo update is one that flips the state of a single plaquette with the appropriate Metropolis probability.
However, a notable feature of configurations close to the multi-critical point is that only a very small fraction ($\approx 2.5 \%$) of \textit{links} are occupied.
When occupied links are rare,  an attempted plaquette update has a high chance of creating four new occupied links, significantly increasing the energy, and therefore a high chance of being rejected.

This suggests that while plaquette updates are necessary for allowing occupied links to move, they are inefficient at moving surfaces around. 
To speed up the equilibration of surfaces we therefore combine plaquette updates with a second update that flips the state of all six surfaces of a cube.  Since this move never changes the number of occupied links, it does not face the problem above. Since this move is still a local update we do not expect it to change $z$, as we have confirmed numerically. See App.~\ref{app:MCscheme} for further details including the scheme for parallelization.

\section{Membrane patching, emergent one-form symmetry, and worldline percolation}
\label{sec:percolation}

In this section we widen our focus to  transitions out of the deconfined phase generally. 
We give a construction for the ``fictitious'' Ising order parameters 
(that are the key feature of Ising$^*$ transitions) on the Higgs and confinement lines. 
We find that these fictitious order parameters can be constructed all the way along the Ising$^*$ lines, but not at the self-dual critical point. 
The disappearance of the fictitious order parameter at the self-dual critical point is associated with the emergence of a scale-invariant ensemble of loops there.

Studying this ensemble of loops with percolation-like observables \cite{huse1991sponge,cardy1996scaling} 
allows another numerical test of scale-invariance at the self-dual  critical point,
and gives another critical exponent with which to characterize it (Secs.~\ref{sec:percolationsummary},~\ref{sec:percolationsubsec}).

\subsection{Patching membranes}

The fact that the $e$ and $m$ condensation transitions  have Ising exponents
(away from the self-dual line)
is easy to understand at the boundaries of the  phase diagram (Fig.~\ref{fig:phase_diag}), 
as reviewed in Sec.~\ref{sec:phasetransitionsreview} \cite{wegner1971duality,fradkin1979phase}. 
In these limits the partition function can be written as a sum over {closed} membrane configurations. 
Mapping these closed membranes to Ising domain walls gives the relation to Ising.

Moving away from this extreme limit, the membranes acquire ``holes'' \cite{huse1991sponge}. 
(We use the term ``hole'' loosely: more precisely, we mean any connected cluster of links in $\partial\mathcal{M}$, as defined in Sec.~\ref{sec:membranepicture}.)
But it is natural to think that,
if these holes have a finite typical size, 
coarse-graining beyond this size may restore a picture in terms of closed membranes that can be interpreted as Ising domain walls.\footnote{Analogs of this phenomenon may also be found in 2D \cite{shi2011boson}.} This is a heuristic explanation for why an effective Landau theory is useful even some distance away from the boundary  of the phase diagram.

Here we wish, first, to give an explicit construction of these emergent degrees of freedom.  Our approach is simply to ``repair'' the membranes  $\mathcal{M}$ in a given configuration.
We will be schematic, deferring further details and a numerical demonstration to Ref.~\cite{membranesforthcoming}.
Second, we wish to understand what happens to the fictitious order parameter when move along the Ising$^*$ transition line towards the self-dual critical point.
This issue is closely connected to the question of where in the phase diagram emergent ``one-form'' symmetries \cite{kapustin2017higher,gaiotto2015generalized, wen2019emergent} exist.

We consider the membrane ensemble in Eq.~\ref{eq:partitionfunctionmembranes},
which is convenient for describing one of the two dual one-form symmetries. 
By duality, analogous considerations apply for the dual symmetry.
(The Ising$^*$ transition that we discuss below is the $m$ condensation line.)

Let us briefly make the connection with a formal point of view. 
The fictitious Ising order parameter will make sense if
(perhaps after coarse-graining) 
we can consistently define  string operators $V_P=\pm 1$, supported on arbitrary paths $P$ in spacetime, that count the parity of the number of membranes that intersect $P$.
Let us assume that we can define such operators which are functions of the membrane configuration $\mathcal{M}$, and whose value is unchanged if the shape of the path $P$ is deformed 
(while preserving the locations of the  endpoints, if $P$ is not a closed loop). 
We can then define a coarse-grained Ising variable $\phi_r$, up to a global $\mathbb{Z}_2$ ambiguity,
by identifying $\phi_r \phi_{r'}$ with $V_P$ for any path $P$ between $r$ and $r'$. (Here for simplicity we consider an infinite system.\footnote{If the original system is finite with periodic boundary conditions, then we must sum over periodic and antiperiodic boundary conditions for $\phi_{r}$.})

Such string operators, obeying an invariance under deformations, define a $\mathbb{Z}_2$ one-form symmetry 
(see Ref.~\cite{gaiotto2015generalized,wen2019emergent} for definitions).\footnote{Formally, the relation between the models with $\mathbb{Z}_2$ global and $\mathbb{Z}_2$ one-form symmetries is via gauging of these symmetries with flat gauge fields \cite{gaiotto2015generalized}.}
Switching briefly to the language of 2D quantum states, 
the analogous quantum operators in the toric code
are simply the familiar topological string operators \cite{kitaev2003fault} which can be used to create pairs of $m$  anyons at their endpoints (a similar dual operator creates pairs of $e$ anyons).
Perturbing away from the solvable limit of the toric code, 
dressed versions of these string operators are expected to exist in principle so long as the other anyons, which braid nontrivially with $m$, remain gapped  \cite{hastings2005quasiadiabatic, wen2019emergent}.\footnote{Ref.~\cite{hastings2005quasiadiabatic} gives a rigorous result for the case when all excitations are gapped.}\footnote{The various string operators can be connected to the Fredenhagen Marcu order parameter \cite{fredenhagen1986confinement} and the Huse-Liebler horseshoe \cite{huse1991sponge, gregor2011diagnosing}, long used as diagnostics for deconfinement. The dressed string operators above obey the property of invariance under deformations. ``Bare'' string or Wilson line operators do not, and their correlators generically include a product of local contributions from along the length of the string. However, universal data can still be extracted by dividing an appropriate open expectation value for an open Wilson line by the expectation value of a closed Wilson line, in order to cancel the UV contributions \cite{fredenhagen1986confinement,huse1991sponge}.}

\begin{figure}[t]
\includegraphics[width=0.95\linewidth]{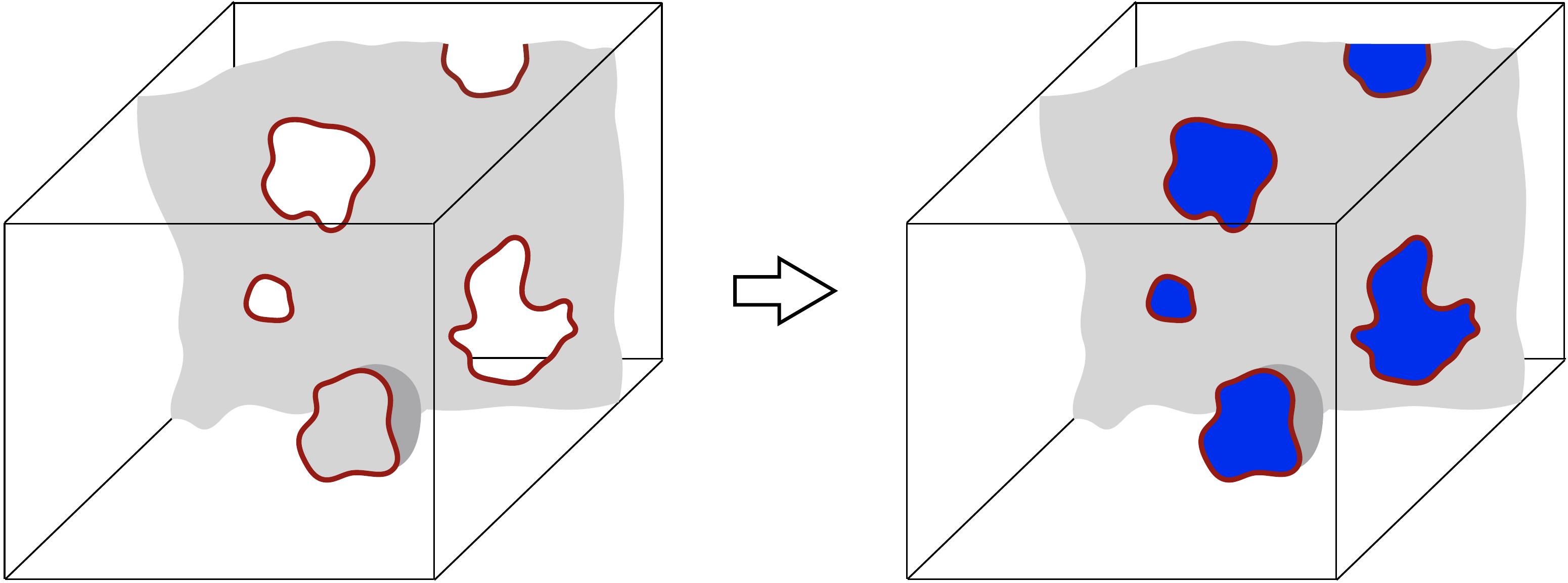}
	\caption{Membrane patching process (schematic). A configuration $\mathcal{M}$ of membranes, with a non-empty boundary $\partial M$ made up of finite clusters of occupied links (shown red) is ``patched'' by attaching finite surfaces (blue) to the components of $\partial \mathcal{M}$. In the resulting ensemble of \textit{closed} membranes,  line operators may be defined, specifying an emergent one-form symmetry.
	A ``fictitious'' scalar field ${\phi=\pm 1}$ may also be introduced, whose domain walls are the patched membranes. The Ising$^*$ transition is the ordering transition for this fictitious field.
}
\label{fig:membranepatchingschematic}
\end{figure}

Returning to the membrane picture, how would we explicitly define such string operators, or the $\phi_r$ configuration, in a simulation? 
A natural approach is to start with the membrane configuration, and try to ``patch up'' the holes, to give closed membranes. 
This is not a strictly local process, since holes can be of any size.
We also have some freedom in the convention, or algorithm, for constructing the patching surfaces.\footnote{We defer a discussion of numerical implementation to Ref.~\cite{membranesforthcoming}. To have a simple convention in mind, we can define the surface associated with a given connected cluster of links in ${\partial \mathcal{M}}$ as the (possibly self-intersecting) surface traced out when the cluster is shrunk down onto its centre of mass.}
But, \textit{if} holes have a finite typical size, and large holes are exponentially rare, we expect the nonlocality in the patching operation to be mild.
Each finite  ``loop'' (cluster of links) in $\partial \mathcal{M}$ may be patched by attaching a finite surface of comparable size. This is illustrated in Fig.~\ref{fig:membranepatchingschematic}.

Having done this, we may define $\phi_r$ (again with a global $\mathbb{Z}_2$ freedom). 
Because of the nonlocality of the patching operation, this effective field only really makes sense on lengthscales larger than the typical size of a loop. 
(It will therefore be most useful as an effective field in a Lagrangian when it has a correlation length that is parametrically larger than this loop size, or infinite.)
This construction allows us in principle to compute the ``two-point'' correlation function of $\phi_r$ in a simulation, and extract the corresponding anomalous dimension, despite the fact that $\phi_r$ is not a local gauge invariant quantity~\cite{membranesforthcoming}.

We may also define thickened string operators.
Let $P$ be, say,  a straight path of length $\ell\gg 1$. In order to determine how many domain walls $P$ passes through after patching, 
we must check how many  loops  $P$  threads in the unpatched configuration $\mathcal{M}$. 
This requires us to examine a cigar-shaped region around $P$, wide enough to contain (with probability close to one) all the loops which $P$ threads.
The operator $V_P$ is therefore a function of the degrees of freedom within this cigar-shaped region.
When large loops are exponentially suppressed, the largest loop that $P$ threads will typically be of size ${\sim \ln \ell}$ (due to rare large loops), 
so the typical width of the cigar should be of this order. 
However,  close to the endpoints of $P$, it is sufficient for the width to be only somewhat larger than the typical loop size (i.e. $\ell$-independent).

The above pertains to the case where the ``holes'' have a finite typical size $\xi_h$.  If on the other hand $\xi_h$ diverges, so that samples of arbitrarily large size $L$ contain holes of size comparable with $L$, then this procedure for defining the string operator and effective Ising order parameter is liable to fail (since $\phi_r$ and $V_P$ can become highly nonlocal).
That is, a sufficient\footnote{This is only a sufficient condition: see e.g. the comment on doubled strands in Sec.~\ref{sec:percolationsummary} below.} condition for this procedure to work is that the appropriate set of worldlines, $\partial \mathcal{M}$, is in the non-percolating phase when viewed as a bond percolation configuration.

For this reason it is interesting to revisit the question of where in the phase diagram these worldlines percolate \cite{huse1991sponge}, which we do next.
For example, as we move along the confinement transition line, starting at $y=0$ (where there are no worldlines) and moving towards the self-dual critical point, where does the fictitious order parameter $\phi_r$ --- as defined by the above simple algorithm --- stop making sense? The results below indicate that it makes sense all the way along the confinement line, but not at the self-dual critical point where that line terminates.
They are consistent with the simplest expectation: that the  one-form symmetry which exists at $y=0$ persists as an emergent symmetry all the way along the confinement transition line, 
but disappears at the self-dual critical point. 
Therefore, the RG flow from the self-dual fixed point to the Ising$^*$ fixed point involves the emergence of the one-form symmetry.
Similar considerations apply for the dual one-form symmetry along the Higgs line.

\begin{figure}[t]
\includegraphics[width=0.98\linewidth]{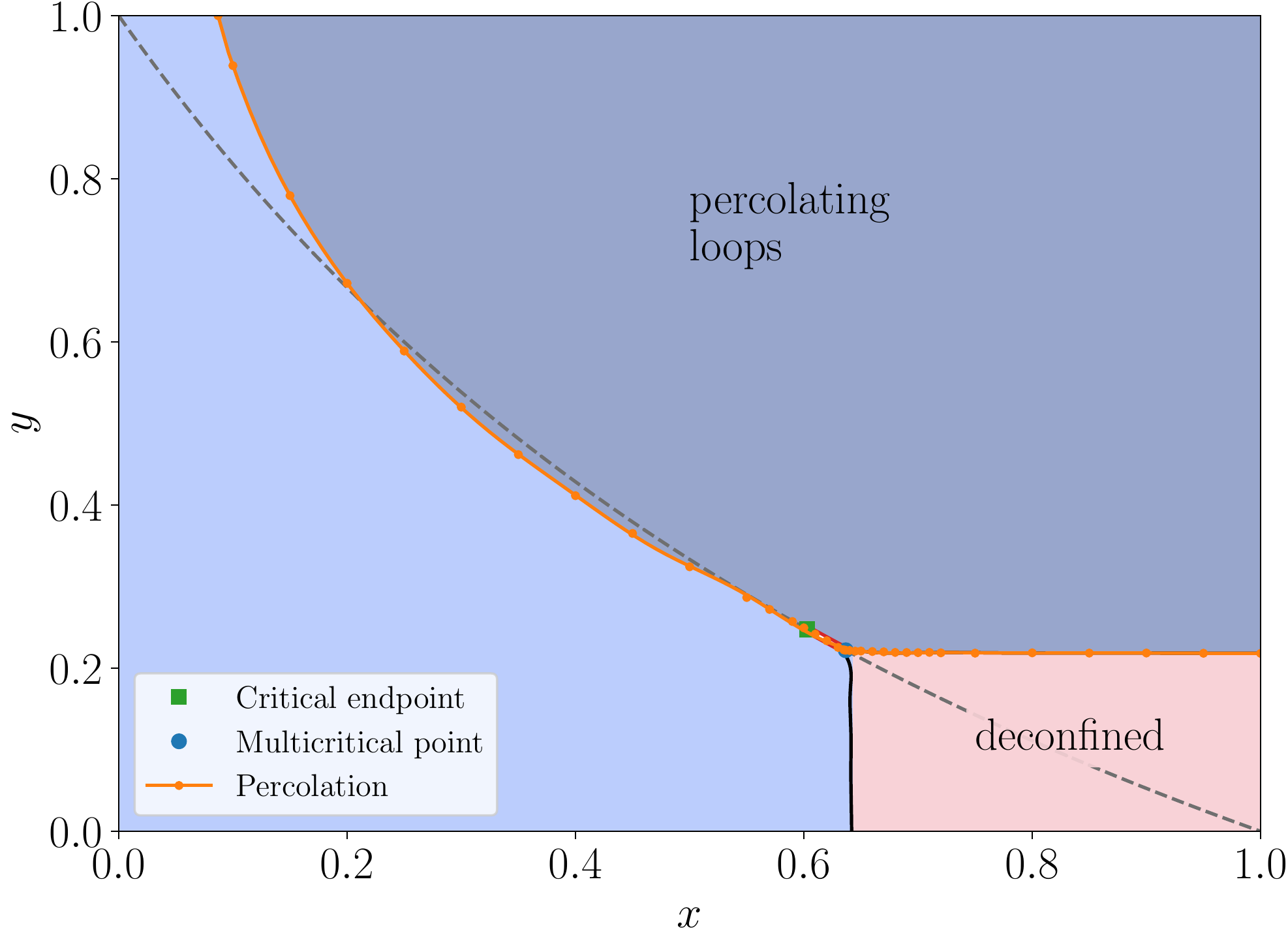}
	\caption{Phase diagram showing the phase boundary for percolation of $e$ worldlines (defined as clusters of links in $\partial \mathcal{M}$, see Eq.~\ref{eq:partitionfunctionmembranes}). The deconfined phase lies within the non-percolating phase.  The self-dual line is shown dashed.
}
\label{fig:phase_diag_perco}
\end{figure}

\subsection{Percolation summary}
\label{sec:percolationsummary}

Our result for the percolation phase diagram is shown in Fig.~\ref{fig:phase_diag_perco} and   explained in Sec.~\ref{sec:percolationsubsec}.
Within our numerical precision, 
the percolation phase boundary matches the thermodynamic boundary of the deconfined phase along the entire Higgs transition line 
(to the right of the self-dual line),
and passes through the self-dual critical point.
(The percolation transition line also lies very close to the first order line, though closer examination indicates that these two lines do not entirely coincide, see App.~\ref{app:percolation}.)

The fact that the percolation line passes through the self-dual critical point agrees with the scenario in  Ref.~\cite{huse1991sponge}. A far as we are aware, however, this result is not guaranteed a priori:  the geometrical percolation transion could have  separated from the Higgs transition at some point along the Higgs line, with the multicritical point lying in the interior of the percolating phase (see footnote\footnote{This is because there are in principle two ways for the loops $\partial \mathcal{M}$ to undergo a percolation transition. Heuristically these correspond to proliferation either of single strands or of doubled strands (thin ribbons). 
The former results in a Higgs transition, but the latter does not. 
This can be made slightly more precise by extending the lattice gauge-Higgs theory to include a replica index $\alpha=1,\ldots, n$ on the matter field $\tau$, with the limit $n\rightarrow 1$ recovering the initial problem. Condensation of $\tau_\alpha$ results both in percolation and in a Higgs transition, while condensation of the composite field $\tau_\alpha\tau_\beta$, which is gauge neutral but charged under replica symmetry, results in percolation but not a Higgs transition~\cite{nahum2012universal} (in this case the percolation transition has no thermodynamic effect \cite{cardy1996scaling}).
The ``double-line'' percolation transition does not necessarily obstruct the membrane-patching procedure, since there is no large-scale ambiguity about patching the surface between two nearby lines.}).

\begin{figure}[t]
\includegraphics[width=0.8\linewidth]{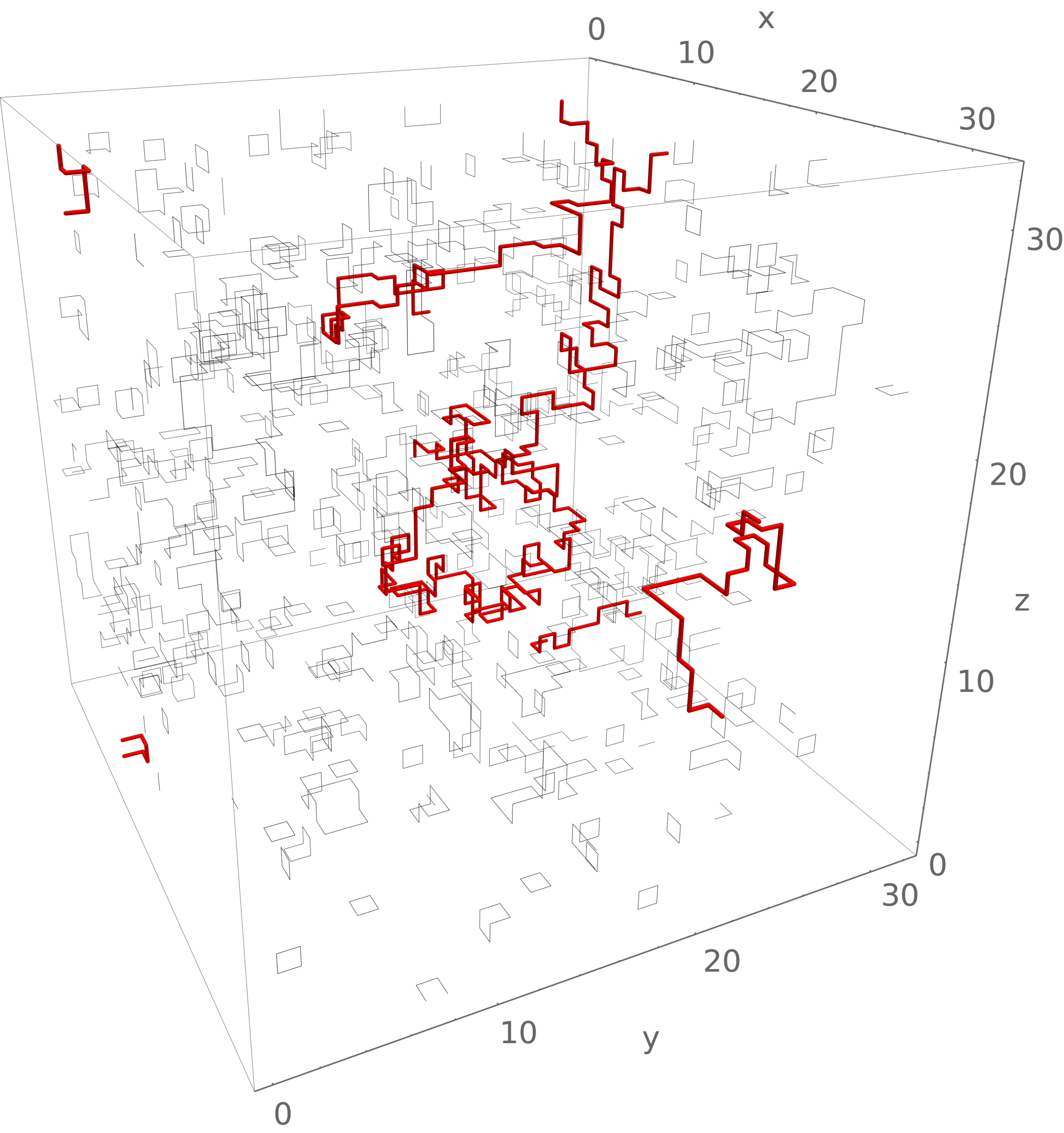}
\caption{Loops in a sample of system size ${L=32}$ at ${x=0.6367}$. A spanning loop is highlighted in red (note periodic B.C.s).  Our definition of a ``loop'' allows for junctions where 4 or 6 occupied links meet, but  as can be seen they are relatively rare at the critical point.}
\label{fig:sampleloops}
\end{figure}

It is also striking that the self-dual critical point lies on the percolation phase boundary despite having a very low fraction of occupied links, around $2.5\%$.
Despite their low density, these links make up a scale-invariant ensemble of clusters.  
Fig.~\ref{fig:sampleloops} shows the loops $\partial\mathcal{M}$ in an example configuration.

This scale invariance allows us to define a new exponent at the self-dual critical point, namely the fractal dimension $d_f$ of the critical loops. 
A priori, this exponent is independent of the scaling dimensions of local operators discussed above:
${d_f=3-x_\text{conn}}$ is determined by the scaling dimension $x_\text{conn}$ of a nonlocal geometrical operator of the type familiar from percolation \cite{cardy1996scaling}.\footnote{The probability that two links separated by a distance $r\gg 1$ lie on the same cluster decays as $r^{-2x_\text{conn}}$.}
Interestingly, though, our numerical result for $d_f$ below (Sec.~\ref{sec:percolationsubsec}),
\be\label{eq:dfresult}
d_f = 1.77(2),
\ee
is consistent with $x_\text{conn} \overset{?}{=} x_A$, perhaps hinting at additional hidden symmetry structure at this critical point. 
(See Sec.~\ref{sec:newmodel} for an argument that ${x_A\leq x_\text{conn}}$.)

\subsection{Percolation observables}
\label{sec:percolationsubsec}

We locate the boundary between percolating and  short-loop (non-percolating) phases using the spanning probability, $P_s$. 
This is  the probability that the sample contains a  
loop which spans the sample in a given axis direction.\footnote{A loop is defined to span the system in (say) the $x$ direction if it visits each of the $L$ distinct planes of $x$-directed links.}
In the thermodynamic limit, this   quantity converges to zero and to one in the nonpercolating and percolating phases respectively, 
and it is expected to take a universal value in between 0 and 1 at a continuous transition between the two phases.

We estimate the percolation phase boundary from crossings in $P_s$, plotted as a function of $y$,  using small system sizes $L=8$, $12$ and $16$ (data not shown).  
We use larger sizes to analyse the transitions at $x=1$, at  $y=1$, at the multicritical point and in the region around it.
We may also obtain the correlation length exponent from a scaling collapse of $P_s$.

The phase diagram Fig.~\ref{fig:phase_diag_perco} shows three different phases.
The deconfined phase has short loops, while the thermodynamically trivial phase splits into a percolating and a non-percolating phase 
(this is possible because the percolation transition need not have any thermodynamic signature).
Note that here we are considering  percolation of $e$ worldlines:
the phase diagram for  percolation of $m$ worldlines (in the dual membrane representation) may be obtained by duality.

As a check, we first examine the percolation transitions on the boundary of the phase diagram, where we expect to see standard universality classes (data in App.~\ref{app:percolation}).
At $x=1$ results are as expected from the Ising mapping, with a fractal dimension consistent with the known value for critical Ising worldlines \cite{winter2008geometric, kompaniets2020fractal}, and correlation length exponent consistent with the Ising value. 
At $y=1$, where the percolation transition 
is purely geometrical
(has no thermodynamic signature) 
 exponents are consistent with the standard 3D percolation universality class.

\begin{figure}[t]
\includegraphics[width=0.98\linewidth]{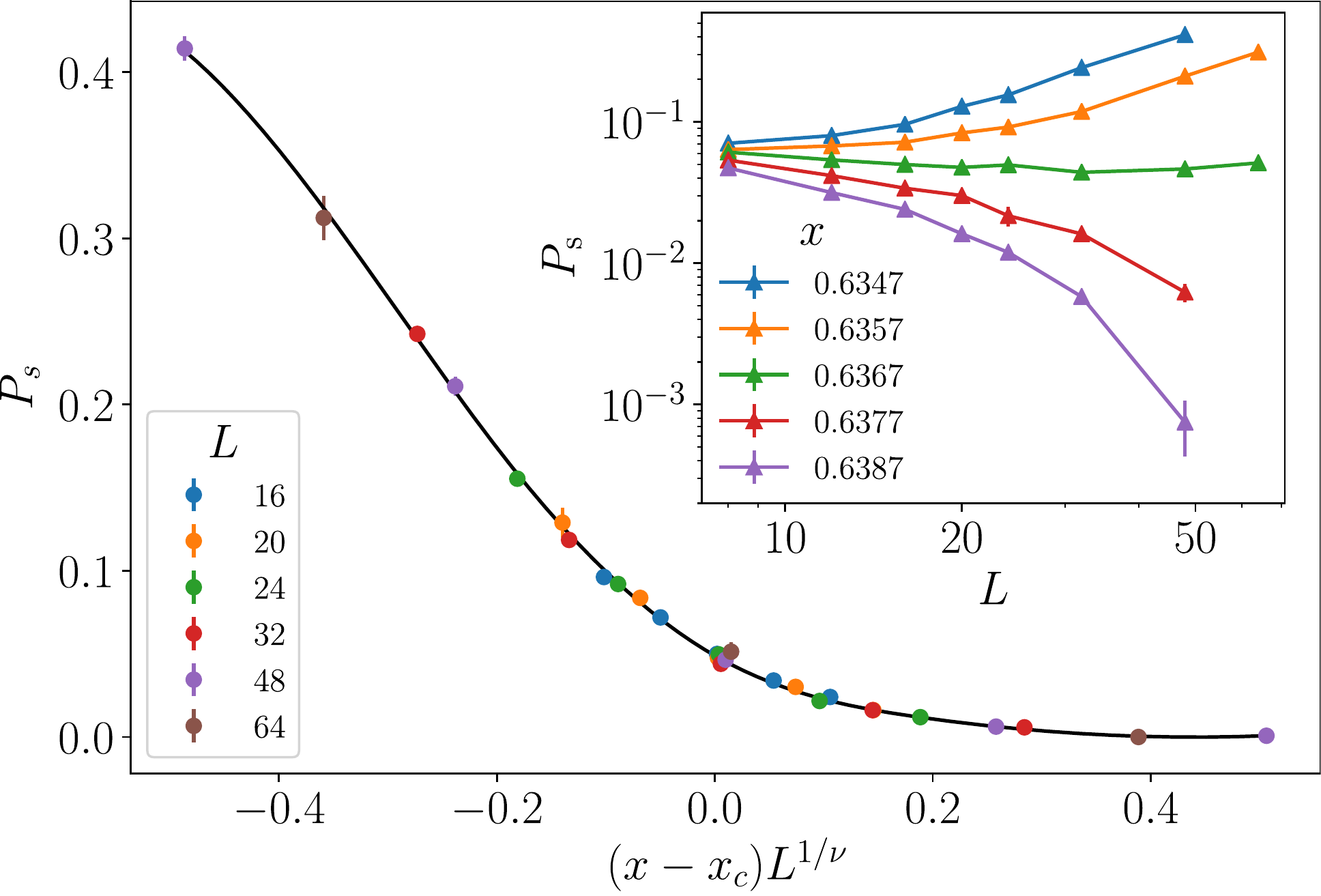}
	\caption{Scaling collapse of $P_s$ as a function of $(x-x_c)L^{1/\nu}$ with $x_c = 0.63666$ fixed. Inset: flow of $P_s$ as a function of the system size for several $x$ values close to the multicritical point, showing approximate scale invariance at $x=0.6367$.}
\label{fig:scalingPs}
\end{figure}

\begin{figure}[t]
\includegraphics[width=0.8\linewidth]{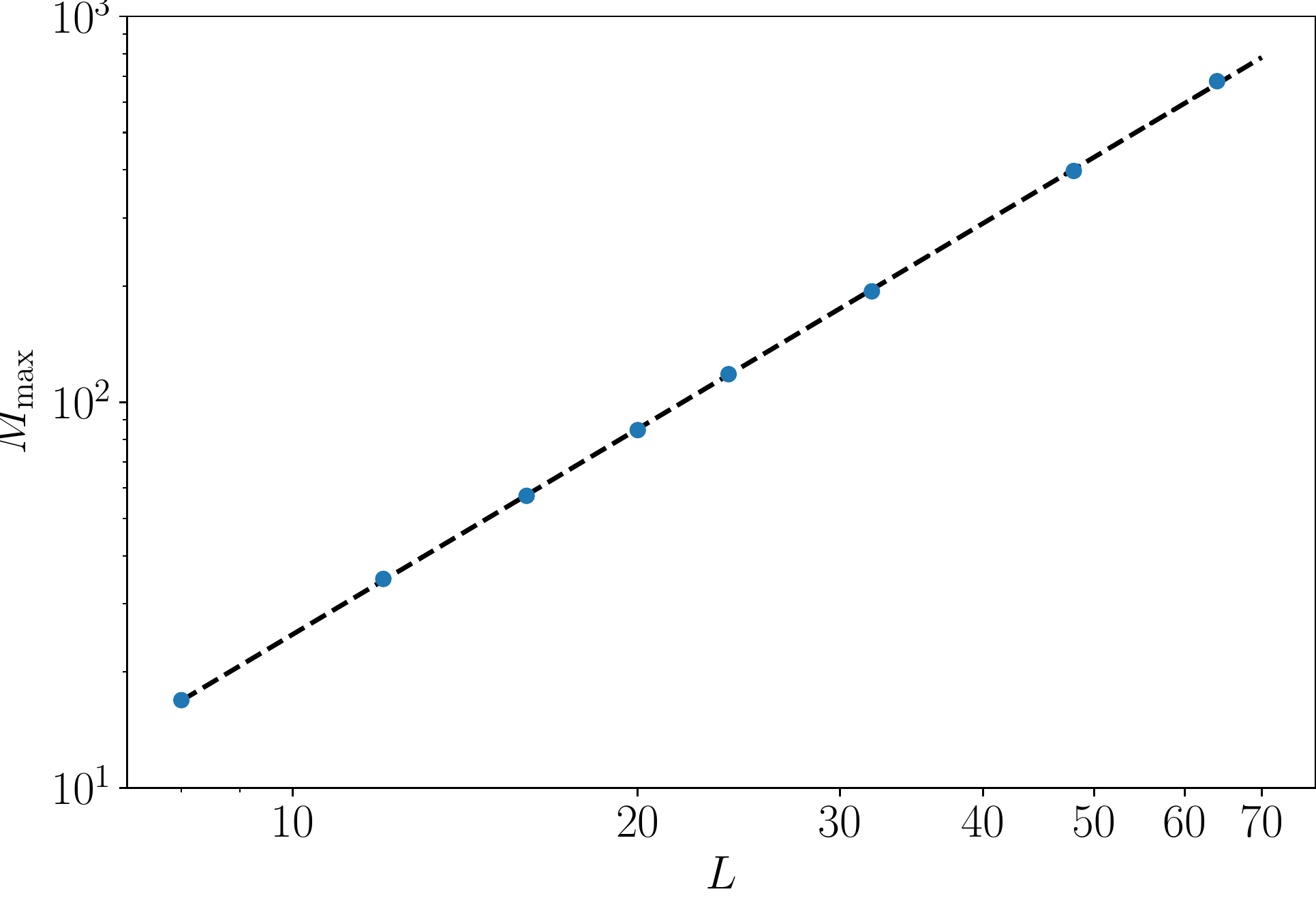}
	\caption{Mass of the largest cluster (number of links) as a function of the system size at $x= 0.6367$. The dashed line fits system sizes as a power-law $A L^{d_f}$, with $d_f=1.77(2)$.}
\label{fig:df}
\end{figure}

Fig.~\ref{fig:scalingPs} shows data on the self-dual line, close to the self-dual critical point. 
The data  are compatible with a critical point very close to ${x = 0.6367}$ 
(inset of Fig.~\ref{fig:scalingPs}), 
i.e. with geometrical criticality coinciding with the
self-dual critical point at the corner of the deconfined phase.
A scaling collapse of  $P_s$ as a function of $(x-x_c)L^{1/\nu}$, leaving $x_c$ and $\nu$ free (not shown) gives $x_c = 0.63664(10)$ and $\nu = 0.69(6)$ compatible with our best estimates for the self-dual multicritical point. 
In Fig.~\ref{fig:scalingPs}, we show the scaling collapse when $x_c$ is fixed to our previous best estimate $x_c = 0.636660$ (Sec.~\ref{sec:scalingcollapseforA}). 
We use B-splines with 5 knots and obtain $\nu = 0.70(6)$ for a fit that gives $\chi^2 = 27$ for 24 degrees of freedom.
Using $\nu^{-1} = 3-x_S$, this result for $\nu$ is consistent with our previous estimate of $x_S$, though with lower precision.

At the self-dual critical point, loops are fractal, and exist on all scales (Fig.~\ref{fig:sampleloops}).
The fractal dimension $d_f$  can be estimated from fitting the total mass of the  largest loop to a power-law in $L$, Fig.~\ref{fig:df}. 
The straight line fits the whole range of system sizes from 8 to 64, providing the estimate $d_f = 1.77(2)$ quoted above.

\section{Related models}
\label{sec:relatedmodels}

The previous section concludes our analysis of numerical data. We now consider some variations of the model, and relations to other models. Sec.~\ref{sec:newmodel} connects the self-dual critical point to another partition function, for a  ``topologically constrained'' ensemble of loops, which it may be interesting to study further.
Sec.~\ref{sec:perturbations} and Sec.~\ref{sec:dimcrossover} discuss perturbations and crossovers in the gauge-Higgs model.  Sec.~\ref{sec:XY} discusses our numerical observation that the exponents $x_A$ and $x_S$ are close to exponents in the XY model.

\subsection{An unusual self-dual loop model}
\label{sec:newmodel}

In Sec.~\ref{sec:manifestlyselfdual} we considered a representation of the gauge-Higgs partition function as a ``loop model'', for two species of ``loops'',\footnote{Recall that the ``loops'' in Eq.~\ref{eq:Zhybridrep} are really clusters, since any even number of occupied links can meet at a node.} with a topological sign factor $(-1)^{\text{linking}}$ in the Boltzmann weight. 
In that model the two species of loops live on distinct cubic lattices. 
Here we consider a modified loop model in which the loops live on the same cubic lattice. 
This allows the partition function to be re-expressed in a form involving a topological constraint rather than a topological sign factor.

Let $\mathcal{C}_e$ and $\mathcal{C}_m$ be two species of loops on the cubic lattice. Here we define the allowed loop configurations differently to those in  Sec.~\ref{sec:manifestlyselfdual}: now we insist that the loops are strictly self-avoiding and mutually avoiding (a  loop may visit a given site at most once, for example). With this definition the linking number $\hat X$ is well-defined.

The partition function is
\be\label{eq:Zmod}
Z_\text{mod} = 4 \sum_{\mathcal{C}_e, \mathcal{C}_m} y^{|\mathcal{C}_e|} y'^{|\mathcal{C}_m|}(-1)^{\hat X(\mathcal{C}_e, \mathcal{C}_m)}.
\ee
($|\mathcal{C}_e|$ is the number of occupied links in $\mathcal{C}_e$, etc., and the loops in $\mathcal{C}_e \cup \mathcal{C}_m$ are mutually avoiding.) 
We do not yet know the full phase diagram of this new model, but it is plausible that it may also show a self-dual critical point, in the same universality class as the lattice gauge theory studied above.

As an aside, we note that the model could be varied in many ways. 
We could allow the ``loops'' to be clusters (as in the previous model, Eq.~\ref{eq:Zhybridrep}), by allowing the number of occupied links adjacent to a site to be any even number (rather than just 0 or 2 as in Eq.~\ref{eq:Zmod}).
With this choice the model maps on to the original gauge-Higgs model in the limit $y\rightarrow 0$ and in the limit $y'\rightarrow 0$. 
This choice may have advantages for simulations (as may other choices of lattice as noted below). 
These changes do not affect the points we make here, so we consider the more easily-visualized ensemble of strictly self-avoiding loops.

Let $\mathcal{C}$ denote the full loop configuration, without regard to species labels, and let us specialize to the self-dual line where ${y=y'}$:
\be
Z_\text{mod} = 
4 \sum_{\mathcal{C}}
y^{|\mathcal{C}|}
\sum_{
\substack{
{\text{species}}
\\
{\text{labels}}
}
}
(-1)^{
\hat X(\mathcal{C}_e, \mathcal{C}_m)
}.
\ee
The final sum is over assignments of the loops in $\mathcal{C}$ to species $e$ or $m$, 
i.e. over splittings of $\mathcal{C}$ into $\mathcal{C}_e$ and $\mathcal{C}_m$.
For simplicity, let us choose (non-periodic) boundary conditions such that loops cannot end on the boundary or wind around the system.

We can sum over the species assignments explicitly, for a fixed $\mathcal{C}$. The result is simple:\footnote{To see this, let $i, j$ be indices running over the distinct loops in $\mathcal{C}$. 
Let $s_i$ be a species index, with $s_i=1$ for an $e$ worldline and $s_i=-1$ for an $m$ worldline. Finally let $n_{i,j}=0,1$ be the $\mathbb{Z}_2$ linking number of loops $i$ and $j$, which is straightforwardly defined since all loops contract to a point. The linking sign for $\mathcal{C}_e$ and $\mathcal{C}_m$ may be written $ (-1)^{\hat X(\mathcal{C}_e,\mathcal{C}_m)}= e^{\f{i\pi}{2} \sum_{i<j} n_{i,j}(2-s_i-s_j)}$. Summing over the $s_i$ gives the result in the text. The sum vanishes unless $\sum_{j(\neq i)} n_{i,j}$ is even for every $i$.}
\be\label{eq:Zmodfactorsof2}
Z_\text{mod} = 
4 \sum_{\mathcal{C}}
y^{|\mathcal{C}|} \times
2^{(\text{\# loops in $\mathcal{C}$})} \times \chi_\mathcal{C} .
\ee
Here $\chi_\mathcal{C}=0,1$ 
depends only on the topology of $\mathcal{C}$, 
and simply imposes a restriction (constraint) on the allowed topologies.
${\chi_\mathcal{C}=1}$ so long as every loop 
in the configuration links with an \textit{even} number of other loops, and ${\chi_\mathcal{C}=0}$ otherwise.\footnote{{If the topological constraint in Eq.~\ref{eq:Zmodfactorsof2} is relaxed (by removing the factor $\chi_\mathcal{C}$) we have the partition function for a version of the XY model \cite{nienhuis1982exact}.}}

Strikingly, this expression is sign-free, and  could be sampled with Monte-Carlo, using a local update that preserved the mod 2 total linking number of each loop.
It would be interesting to know the phase diagram of this model or variants of it. (For an efficient numerical study, it might be useful to modify the lattice geometry of the model so that loops can form nontrivial links on a shorter lengthscale.\footnote{In the model (\ref{eq:Zmod}) as it stands, the smallest loop that can be nontrivially linked by another loop is a square of side length 2.})

This model also allows an interesting topological interpretation for  correlation functions of the anti-self-dual operator.

In the ensemble (\ref{eq:Zmod}), let us define the operator $\widetilde A(r)$ at a site $r$ to take the value $0$ if the site is not visited by a loop, $1$ if the site is visited by an $e$ loop, and $-1$ if the site is visited by an $m$ loop.  This operator is odd under duality, so analogous to the operator $A(r)$ defined for the gauge-Higgs model in Sec.~\ref{sec:definingAandS}.

By again explicitly summing over the loops' species labels, 
we may write correlators of $\widetilde A$ in the formulation of Eq.~\ref{eq:Zmodfactorsof2}. 
First, an insertion of $\widetilde A(r)$ forces a loop to pass through $r$.
Second, the $\widetilde A(r)$  insertion forces  the total linking number of this anchored loop (with other loops) to reverse its parity. In the original ensemble (\ref{eq:Zmodfactorsof2}) every loop has even linking. 
In the presence of $\widetilde A$ insertions, the linking number of a loop that passes through an odd number of $\widetilde A$ operators must instead be odd (while the linking number of a loop that passes through an even number of $\widetilde A$ operators remains even).

\begin{figure}[t]
\includegraphics[width=\linewidth]{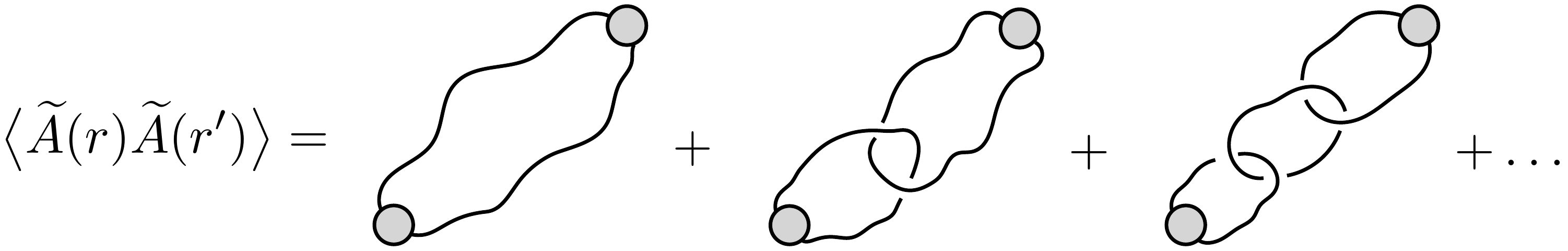}
\caption{Schematic: some of the configurations contributing to the correlator $\langle\widetilde A(r)\widetilde A(r')\rangle$ in the model (\ref{eq:Zmodfactorsof2}). Circles indicate locations {$r$,~$r'$}.}
\label{fig:loopchain}
\end{figure}

In Fig.~\ref{fig:loopchain} we illustrate some of the configurations that contribute to  ${\langle \widetilde A(r) \widetilde A(r')\rangle}$.
For simplicity, we show the schematic situation at small $y$, where loop length is suppressed.
The first term involves a single loop  (with zero linking) that passes through both insertions. 
The other terms involve a \textit{chain of linked loops}, with the loops at the two ends of the chain,  passing through $r$ and $r'$, having odd linking.
(This picture shows that in the limit of small $y$ the correlation length $\xi$ for this correlator is proportional to  ${|\ln y|^{-1}}$, since we must pay a factor of $y$ for every unit of loop length.)
Moving away from the regime of small $y$, it remains true that there are two types of terms: those where the two $\widetilde A$ insertions lie on the same loop, and those where they lie on distinct loops which (unlike all the other loops in the configuration) have odd linking.

Assuming this model has a self-dual critical point, then this topological picture for ${\langle\widetilde A\widetilde A\rangle}$ yields an inequality for the fractal dimension $d_f$ of loops.
This sheds some light on the coincidence of exponents that we found numerically in Sec.~\ref{sec:percolation}.
Recall that the connectivity correlator, ${P_\text{conn}(r,r')\sim r^{-2x_\text{conn}}}$, is the probability that two distant sites are connected by a loop. 
Above we have shown that these connected configurations are a subset of the configurations that contribute to  ${\langle\widetilde A\widetilde A\rangle}$. That is, ${\langle\widetilde A\widetilde A\rangle=P_\text{conn}+R}$, 
where $R$ is the sum over the remaining configurations and is positive.
Therefore 
\be
x_A \leq x_\text{conn},
\ee
or equivalently $d_f \leq 3-x_A$.

We expect that when the model above is perturbed away from the self-dual line, an Ising$^*$ transition can take place, as in the original gauge-Higgs model. 
In the language of Eq.~\ref{eq:Zmod} this occurs in the same manner discussed in Sec.~\ref{sec:phasetransitionsreview}: if one of the loop species has a finite typical size, it can be integrated out at large scales, leaving a simple ensemble of Ising-like worldlines (topologically unconstrained loops with a fugacity 1 per loop).
It is also possible to see this crossover in the language of 
Eq.~\ref{eq:Zmodfactorsof2}: adding the
duality-breaking perturbation $\widetilde A$ relieves the 
topological linking constraint in Eq.~\ref{eq:Zmodfactorsof2} and leads to an ensemble where large loops have a fugacity of 1.\footnote{We can see this by summing over all possible spatial patterns of insertions of ${\widetilde A}$ (obtained by expanding in this perturbation) for a given configuration of loops. Since a large loop may lie on either an even or an odd number of $\widetilde A$ insertions, its linking number may be either even or odd. Further, summing over patterns of insertions on an asymptotically large loop, with a fixed parity for the number of insertions, yields a factor of $1/2$ that cancels the
fugacity 2 in Eq.~\ref{eq:Zmodfactorsof2}.}

\subsection{Perturbations of the gauge-Higgs model}
\label{sec:perturbations}

After this detour, we return to the standard gauge-Higgs model to discuss some remaining questions.

We have characterized the leading self-dual and anti-self-dual scalar (spin-zero) operators at the self-dual critical point numerically,
but it remains to characterize the subleading operators in these sectors, as well as operators with higher spin.
One motivation for this is to formally determine the number of relevant scaling parameters once duality is broken, as we explain below.

On the appropriate line in parameter space, 
self-duality is an exact property of the standard gauge-Higgs model.
But in many settings where the gauge-Higgs model is a useful effective theory,
exact self-duality will be broken in the ultraviolet by additional interactions.
It is natural to conjecture that the phase diagram structure in Fig.~\ref{fig:schematic_phase_diag} can nevertheless survive, 
with self-duality appearing as an emergent symmetry at the corner of the deconfined phase, where Higgs and confinement transitions meet.
In order for this to be the case, $A$ and $S$ should be the only relevant scalar operators at the self-dual critical point.

At first glance this is demonstrated by the fact that we only had to tune two parameters to reach this critical point. 
However this is not quite correct: the microscopic self-duality symmetry of the self-dual line forces all anti-self-dual perturbations to vanish there (not only the leading $A$ perturbation). 
Therefore, in principle we should separately check whether the \textit{subleading} duality-odd scalar operator is relevant or irrelevant.
Since $A$ itself has a large scaling dimension, we might expect that this subleading operator will be irrelevant, but this should be checked. 

The subleading duality-even operator is irrelevant, but a sufficiently large duality-even perturbation may yield a ``self-dual tricritical'' point with an additional relevant direction.

In Ref.~\cite{dusuel2011robustness} it was argued, using series expansions, that the toric code with $X$, $Y$ and $Z$ fields had a critical line, with varying exponents, in the $h_x=h_z$ plane. 
This will be interesting to investigate further, as continuously-varying exponents in 3D are rare. 
However, it should be noted that, in the present language, the
perturbation $h_y$ breaks both internal and spatiotemporal symmetry.
The toric code Hamiltonian with $h_x=h_z$, discussed in Sec.~\ref{sec:quantummodel}, has a duality symmetry $D$ that we may take to be $X\rightarrow T(Z)$, $Z\rightarrow T(X)$, $Y\rightarrow-T(Y)$, where $T$ represents a translation by  $(1/2,1/2)$. It also has an antiunitary time-reversal symmetry which we may take to act as ${X\rightarrow X}$, ${Z\rightarrow Z}$, ${Y\rightarrow -Y}$, ${i\rightarrow -i}$. Adding the $h_Y$ field breaks both of these symmetries. (It preserves their product.) It would be interesting to identify the leading continuum perturbation of the self-dual critical point that is induced by the $h_y$ coupling.

Recent work has demonstrated infinite-randomness scaling for a range of Higgs transitions in ${\text{2+1D}}$ quantum gauge theories with quenched disorder in the couplings \cite{kang2020superuniversality}. It would be interesting to study the effect of disorder on the self-dual topological phase transition.
The exponents $x_S$ and $x_A$ imply that spatially uncorrelated quenched disorder is a strongly relevant perturbation of the self-dual critical point in its 2+1D quantum manifestation, regardless of whether this disorder preserves duality or not.\footnote{In the classical interpretation of the critical point 
(where disorder is uncorrelated in all three directions, rather than being translationally invariant in the imaginary time direction) disorder that breaks self-duality is relevant (even if it preserves self-duality on average) while since $x_S\simeq 1.5$ disorder that preserves self-duality is close to being marginal \cite{cardy1996scaling}.}

\subsection{Dimensional crossovers}
\label{sec:dimcrossover}

Various dimensional crossover effects may also be worth studying.
By making one of the three lattice directions finite and of width ${1/T \gg 1}$ (with periodic boundary conditions) we may study the effect of a low but nonzero temperature in the quantum problem  \cite{cardy1996scaling}.
Standard considerations show that on the boundaries of the phase diagram (at $y=0$ or $y'=0$), the 3D Ising$^*$ transitions give way to 2D Ising transitions, but that in the interior of the phase diagram these transitions become crossovers, with a finite correlation length \cite{svetitsky1986symmetry}.
This correlation length is exponentially large in $1/T$ at small $T$. In the worldline representation (\ref{eq:Zhybridrep}) this scaling is associated with closed worldlines of the massive anyon which are of length $1/T$ and wrap around the temporal cycle.\footnote{Such configurations are suppressed by a factor ${\sim \exp(\text{const.}/T)}$ due to the line tension of the worldlines.
This factor can be mapped to an exponentially weak magnetic field ${h\sim \exp(\text{const.}/T)}$ in an effective 2D Ising model \cite{svetitsky1986symmetry}.} (This exponential scaling may be why a numerical study of the $\mathbb{Z}_2$ gauge-Higgs model instead reported  finite-temperature transitions \cite{genovese2003phase}.)
A line of first-order transitions on the self-dual line will remain at small finite temperature.
What happens to the self-dual critical point at nonzero temperature is less clear.
The simplest possibility is that it becomes a conventional critical endpoint, so that the interior of the phase diagram contains only a first-order line, bounded by two conventional critical endpoints. 

Other boundary condition choices for the finite dimension give other phase diagrams.
For example, consider the loop model (\ref{eq:Zhybridrep}) in a slab geometry of thickness $\ell$, with open boundary conditions in the finite direction, with loop strands forbidden from terminating on the boundary. 
Physically this can be obtained by taking a 2D quantum system deep in the deconfined phase (corresponding to $y,y'\ll 1$) and then varying the couplings inside a strip of width $\ell$ to allow anyons to proliferate there. 
Isolated strands that span the finite direction are now forbidden, so the mechanism that  rendered the correlation length finite in the previous quasi-2D geometry is removed. Instead, after coarse-graining to scales $\gg \ell$  we may argue for an effective 2D loop model with two species of loops (and with no nontrivial topological sign factor).
Away from the self-dual line, 2D Ising transitions are likely, associated with proliferation of a single loop species (a single anyon type). 
It may also be possible to have a gapless Ashkin-Teller-like regime on part of the self-dual line, where both species are critical. 

If we think of the length-$\ell$ direction here as imaginary time instead of space, this setup may be related to the interesting finding of 2D Ashkin-Teller criticality in a deformed toric code wavefunction, for which equal-time correlators map to correlators in a 2D classical model \cite{zhu2019gapless}. 
This deformed wavefunction is given by a finite-depth non-unitary circuit acting on the toric code wavefunction. This can be visualized in a path integral representation. 
In the zero-temperature path integral for the deconfined phase, the evolution for imaginary times $(-\infty,0)$ can be viewed as preparing the ground state ket and the evolution for times $(0,\infty)$ as preparing the corresponding bra. 
To obtain equal-time correlators in the deformed wavefunction, we insert a ``slab'' of finite temporal extent in between these two pieces, representing the action of the nonunitary circuit on bra and ket. This is reminiscent of the setup for the quasi-2D loop model above (since in the spactime region outside the slab we set $y=y'=0$, meaning that, in the loop model picture, worldlines are forbidden except inside the slab).

It will also be interesting to characterize boundary critical phenomena, and conformally invariant boundary conditions, for the self-dual topological transition.

\subsection{Comparison with XY exponents}
\label{sec:XY}

A striking feature of our numerical results is that the values for scaling dimensions are close to certain values for the 3D XY universality class. Below we discuss why this is surprising.

At first sight (however, see below) a relationship with XY appears a natural guess \cite{huse1991sponge}, by analogy with  conventional ordering transitions,\footnote{As well as ``starred'' (weakly gauged) versions of such transitions.} where two Ising critical lines (together with a first order line) can meet at an XY critical point. 
Given two conventional Ising-like order parameters  $\varphi_x$ and $\varphi_y$, and an additional $\mathbb{Z}_2$ symmetry that exchanges them, 
XY criticality for ${{\vec{\varphi}} = (\varphi_x,\varphi_y)}$ can arise by tuning one parameter because the symmetry-allowed ``cubic'' anisotropy ${\varphi_x^4 + \varphi_y^4 - 6 \varphi_x^2 \varphi_y^2}$ is  a (weakly)  irrelevant operator at the XY fixed point \cite{pujari2015transitions,shao2020monte}.

In the present model, we know that the Ising$^*$ lines can be understood as ordering transitions for ``fictitious'' (non-gauge invariant) Ising-like order parameters. 
Therefore at first sight it is tempting to make the above analogy. 
We would then identify the operator $S$ with the thermal operator ${{\vec \varphi}^2}$, 
and the operator $A$ with the symmetry-breaking mass operator ${\varphi_x^2- \varphi_y^2}$.
The scaling dimensions of these operators in the XY model are 
${x_{\vec\varphi^2}=1.51136(22)}$)
and ${x_{\varphi_x^2- \varphi_y^2}=1.23629(11)}$ 
\cite{chester2020carving,hasenbusch2011anisotropic,hasenbusch2019monte}. 
Strikingly, the differences between these values and our results for $x_S$ and $x_A$ in Tab.~\ref{tab:exponents} are small, comparable in size with the (statistical) error bars quoted in the table.\footnote{Our result for the fractal dimension of an $e$ worldline is also consistent with the value for an XY worldline, but it is unclear at present whether this is an independent  exponent or whether ${d_f = 3-x_A}$
(see the discussion below Eq.~\ref{eq:dfresult} and in Sec.~\ref{sec:newmodel}). In the XY model, ${d_f^\text{XY} = 3- x_{\varphi_x^2 - \varphi_y^2}}$ by virtue of known scaling relations.}\footnote{Our data for the OPE coefficient $C_{AAS}$ in Fig.~\ref{fig:CAAS} is not sufficient to make a very useful comparison, but the known XY OPE coefficient value $C_{{\varphi_x^2- \varphi_y^2},{\varphi_x^2- \varphi_y^2},{\vec \phi}^2}=1.25213(14)$ \cite{chester2020carving,hasenbusch2020two} cannot be ruled out.}  

The problem with this analogy is that it ignores the nontrivial mutual statistics between $e$ and $m$ excitations \cite{vidal2009low,TupitsynTopological, geraedts2012monte} that are the key feature of the transition. 
These mutual statistics do not affect critical exponents on the Ising$^*$ lines, because only one of the two excitations is massless on these lines.
But both excitations become massless at the self-dual critical point.

For example, any consistent description of the fixed point should correctly reproduce  the spectrum of low-lying anyonic quasiparticles that exists when we perturb slightly away from the self-dual critical point, into the deconfined phase. It is hard to see how this could be consistent with a mapping that related the present fixed point and the XY fixed point.

The obstacle to making a connection with XY can also be seen in the geometrical pictures. In the membrane picture, the possibility of defining a fictitious Ising order parameter
is associated with the membranes being effectively closed on large scales, as discussed in Sec.~\ref{sec:percolation}. But at the self-dual critical point, we have ``holes'' in these membranes on all scales, as we have demonstrated explicitly. Therefore the attempt to make a connection with a simple Landau theory, at least in this manner, fails at this critical point.

It therefore seems likely that the exponents $x_A$ and $x_S$ at the self-dual critical point are numerically close to XY exponents, but distinct from them. 
If on the other hand the exponents are the same as those of the XY fixed point,  this relationship between a topological phase transition and a simple ordering transition would have to be of a fundamentally new kind. 
We plan to return to these questions elsewhere.

\section{Outlook}
\label{sec:outlook}

The three-dimensional $\mathbb{Z}_2$ gauge-Higgs model is the simplest nontrivial lattice gauge theory \cite{wegner1971duality, wilson1974confinement,balian1974gauge,fradkin1979phase,TupitsynTopological,wu2012phase,vidal2009low}.
Its remarkable duality property allows for a self-dual topological phase transition whose properties have long been unresolved.
We have given direct evidence for scale invariance at this transition, 
exploring system sizes up to two orders of magnitude larger than the lattice spacing. Exciting directions remain open, on the computational, experimental, and theoretical fronts.

First, there are many intriguing questions that could be addressed using further simulations.
At the basic level, armed with the accurate estimate of $x_c$, further characterization of the critical point will be possible, examining the scaling dimensions of a wider range of operators (Sec.~\ref{sec:perturbations}), and pinning down OPE coefficients more precisely (Sec.~\ref{sec:threepointfunction}).

We have also proposed new models that could be simulated. The loop model in Sec.~\ref{sec:newmodel} has a simplified action of self-duality. 
It has a sign-free reformulation of a nonstandard kind, 
as an ensemble of loops with a simple ``topological constraint''. 
(This connects, heuristically, to the longstanding question from polymer physics of how to think about the renormalization group for  models with topological constraints \cite{edwards1977theory,
de1979scaling,
des1981ring,
khokhlov1985polymer,
rubinstein1986dynamics,
cates1986j,
halverson2014melt,
imakaev2015effects,
serna2015topological,dai2020quantum}.)
This sign-free formulation could be exploited to determine the model's phase diagram, and may suggest a more general strategy for obtaining sign-free lattice models for topological transitions.

In the context of the standard lattice gauge theory, a range of  perturbations and crossovers may be studied (Sec.~\ref{sec:perturbations} and Sec.~\ref{sec:dimcrossover}), for example to search for self-dual tricriticality.

The self-dual topological phase transition can be viewed as a paradigmatic challenge for  Monte Carlo algorithm design.
Although it is Monte-Carlo sign-free (unlike many other lattice gauge theories \cite{de1995progress,
peardon2002progress,
schaefer2011algorithms,
ishikawa2009recent}),
the lack of a nonlocal cluster update \cite{newman1999monte} for ensembles of membranes, and the large dynamical exponent (Sec.~\ref{sec:stochasticdynamics}), make it expensive to simulate. 
Creative algorithmic improvements would be valuable. 
We might consider updates acting on larger finite clusters, perhaps optimized using machine learning \cite{liu2017selflearningmontecarlomethod,huang2017accelerated}.

If we are in the deconfined phase, but  close to the self-dual critical point, various features of the spectrum of massive quasiparticles \cite{vidal2009low, dusuel2011robustness} will be universal and could perhaps be examined using Monte Carlo \cite{suwa2016level}, series expansion \cite{vidal2009low, dusuel2011robustness} or tensor network techniques \cite{vanderstraeten2015excitations,vanderstraeten2019simulating}.
For example, does the fermionic $\epsilon$ excitation exist as a stable bound state in this regime, or inevitably decay into an $e$ and an~$m$?

Even away from the self-dual point, interesting questions remain.
The existence of ``fictitious'' Ising order parameters on the Higgs and confinement transition lines is the key to the theoretical understanding of these transitions \cite{wegner1971duality, fradkin1979phase, huse1991sponge,gregor2011diagnosing}.
We have argued that we can construct these field configurations explicitly by a quasi-local patching process in the  membrane representation of the partition function, so that for example the Ising ``two-point function'' $G(r,r')$
can be computed numerically. 
Formally, this is the expectation value of a dressed string operator that extends from $r$ to $r'$ (Sec.~\ref{sec:percolation}).
In separate work we will analyze the emergence of this structure in more detail \cite{membranesforthcoming}.

The self-dual critical point may be accessible experimentally, either in its 3D classical or its ${2+1\mathrm{D}}$ quantum manifestation.
It would be exciting to see the full structure of the gauge-Higgs phase diagram, with the meeting of the two Ising$^*$ lines, in experiments on amphiphilic membranes (verifying a longstanding conjecture \cite{huse1991sponge}).
In order to access this point, the membranes must have free edges, i.e. a nonempty membrane boundary $\partial \mathcal{M}$. However, by analogy with results in App.~\ref{app:coexistence}, the required density of free edges may be relatively small.

Strategies for \textit{quantum} simulation of lattice gauge theories are under intensive development  \cite{dalmonte2016lattice,zohar2017digital,zohar2017digital2,schweizer2019floquet,barbiero2019coupling,banuls2020review,davoudi2020towards,cui2020circuit,paulson2020towards}, so it may one day be possible to explore the self-dual critical point and its real-time quantum dynamics experimentally.

Perplexing theoretical questions remain. Why are our estimates for $x_A$ and $x_S$ so close to XY values (Sec.~\ref{sec:XY})? 
Further numerical characterizations of the critical point mentioned above may shed light on this. 
Significant input may also come from the conformal bootstrap \cite{poland2019conformal,el2012solving,kos2014bootstrapping,kos2016precision,poland2019conformal}, by exploring the space of theories with the requisite $\mathbb{Z}_2$ symmetry.

There remains the fundamental question that we started with: can we formulate a useful continuum field theory for the self-dual topological transition? Criteria for ``usefulness'' could include the possibility of calculating exponents in a systematic expansion, as well as the possibility of deriving the structure of phase diagram analytically. More generally, the time seems ripe for a numerical and theoretical attack on phase transitions where multiple species of anyons, with nontrivial statistics, simultaneously condense \cite{vidal2009low,TupitsynTopological, geraedts2012monte,burnell2018anyon}.

\acknowledgments 
We thank Hans Evertz, Paul Fendley, David Huse,
Jack Kemp, Miguel Ortu\~no,
Siddharth Parameswaran,
 Yang Qi, T. Senthil, and
Sagar Vijay for useful discussions. A.S. acknowledges support by AEI(Spain)/FEDER(EU) grant PID2019-104272RB-C52. A.S. and P.S acknowledge support by Fundación Séneca grant 19907/GERM/15.
AN acknowledges support from the Gordon and Betty Moore Foundation under the EPiQS initiative (grant No.~GBMF4303), from EPSRC Grant No.~EP/N028678/1, and from a Royal Society University Research Fellowship.

\appendix

\section{Membrane representation of $Z$}
\label{app:membraneexpansion}

We now review the standard relationship between the Ising gauge theory partition function and partition functions for membranes on either the original cubic lattice or its dual \cite{huse1991sponge}.
In the interpretation as a 2D quantum system in imaginary time, these membranes are worldsurfaces of either electric or magnetic strings (cf. Fig.~\ref{fig:eworldline}), depending on whether we use the original lattice or the dual lattice.  
In general the strings (which live in a 2D spatial plane) can be open lines, terminating at $e$ or $m$ particles in the respective cases.
Therefore the membranes (which live in 3D spacetime) are not closed in general, but rather have boundaries, which are the wordlines of $e$ or $m$ particles respectively. 
The ``action'' of a given membrane configuration (the logarithm of the Boltzmann weight) is, up to constants, the area of the worldsurfaces plus the length of the worldlines.

\subsection{Membranes on the original lattice}

Using the fact that the variables take only the values $\pm 1$, $Z$ in Eq.~\ref{eq:Z3Dgaugerep} can be rewritten in a form convenient for a standard graphical expansion:
\ba\label{eq:Z(x,y)sigmatau}
Z(x,y) \equiv 
\f{1}{2^{4L^3}} 
\sum_{\{\sigma\}, \{\tau\}}
\prod_{\square} \lf 1 + x \prod \sigma \ri 
\prod_\ell \lf 1 + y \sigma\tau \tau \ri,
\end{align}
with
${K = \f{1}{2} \ln  \f{1+x}{1-x}}$ and ${J = \f{1}{2} \ln \f{1+y}{1-y}}$.
We expand out the products over (1) plaquettes $\square$
and (2) links $\ell$
in Eq.~\ref{eq:Z(x,y)sigmatau}, and represent a given term by colouring plaquettes of the lattice and highlighting links in bold, as in Fig.~\ref{fig:plaquetteconfig}. A plaquette is coloured (``occupied'')  iff  we pick the ``$x \prod \sigma$'' term for that plaquette and similarly a link is bold if we pick the ``$y\sigma\tau\tau$'' term. 
For a given term in the expansion, the collection of occupied plaquettes constitutes the membrane configuration $\mathcal{M}$.

Now for each term we must sum over $\sigma$ and $\tau$. The term will vanish if there is any link $\ell$ where the terms we have chosen contribute $\sigma_\ell$ an odd number of times in total. 
This means that the set of bold links must coincide with the membrane boundary $\partial \mathcal{M}$ to have a nonvanishing term ($\partial \mathcal{M}$ is defined as the set of links where an odd number of coloured plaquettes meet).
If this is satisfied, then the sums over $\sigma$ and $\tau$ are both nonvanishing, giving trivial factors $2^{3L^3}$ and $2^{L^3}$ respectively that cancel the  normalization term chosen in Eq.~\ref{eq:Z(x,y)sigmatau}. We are left with the partition function as sum over membrane configurations weighted by $x^{|\mathcal{M}|}y^{|\partial\mathcal{M}|}$, Eq.~\ref{eq:partitionfunctionmembranes}.

From this expansion we also see that the face and edge operators defined in Sec.~\ref{sec:selfduality} may be written as 
\begin{align}
\mathcal{F}(p) & =\f{x}{1-x^2}\left( \prod\sigma-x \right), 
& 
\mathcal{E}(\ell) & =\f{y}{1-y^2} \left( \sigma\tau\tau-y \right),
\end{align}
for a plaquette $p$ and link $\ell$ respectively.\footnote{For example,  ${\mathcal{F}(p)(1-x \prod_{\ell\in p} \sigma_\ell)=x \prod_{\ell\in p}\sigma_\ell}$, so that
inserting $\mathcal{F}(p)$ in a correlator has the effect of restricting the expansion to terms where plaquette $p$ is occupied. Equivalently, $\mathcal{F}(p)$ is 1 if $p$ is occupied and 0 otherwise.}

The expansion above is the standard high-temperature expansion, meaning that terms are weighted by powers of the ``fugacities''  $x$ and $y$ which are small when $K$ and $J$ are small. Since the lattice is finite the expansion may be done exactly, to all orders: i.e. one may think of it as a reformulation of the partition function and not as a perturbative series. It is a generalization of the high-temperature expansion of the Ising model, which would obtain if the $\sigma$ field was absent and we just had loops associated with $y\tau\tau$.

\subsection{Membranes on the dual lattice}

Eq.~\ref{eq:Z(x,y)sigmatau} can be related to membranes on the \textit{dual} lattice even more directly. 

Let us choose the gauge $\tau=1$ so that the partition function is a sum over only the $\sigma=\pm 1$ on each link. 
We can represent a given term by a collection of occupied links, where a link $\ell$ is occupied iff ${\sigma_\ell=-1}$. 
(Note that this notion of a link being occupied is unrelated to the one in the previous subsection.) 
Next, recall that plaquettes of the dual lattice are in 1:1 correspondence with links of the original lattice, so a configuration of occupied links is equivalent to a configuration of occupied plaquettes $\widetilde{\mathcal{M}}$ on the dual lattice. 
What is the Boltzmann weight of $\widetilde{\mathcal{M}}$? 
Each occupied plaquette costs ${x'\equiv(1-y)/(1+y)}$ (from the ratio of the ${1+y\sigma}$ term in Eq.~\ref{eq:Z(x,y)sigmatau} with ${\sigma=-1}$ to that with ${\sigma=+1}$). 
Further a link of the dual lattice where an odd number of occupied plaquettes meet means a square on the original lattice where $\prod \sigma=-1$. So each link in $\partial\widetilde{\mathcal{M}}$ contributes ${y'\equiv (1-x)/(1+x)}$. Including the normalization,
\be\label{eq:Z(x,y)sigmataudual}
Z(x,y) = 
\f{(1+x)^{3L^3} (1+y)^{3L^3}}{2^{3 L^3}}
\sum_{\widetilde{\mathcal{M}}} \, {x'}^{|\widetilde{\mathcal{M}}|} \, {y'}^{|\partial\widetilde{\mathcal{M}}|}.
\ee

\subsection{Manifestly self-dual representation}
\label{app:manifestlyselfdual}

Next we demonstrate the reformulation in terms of two species of loops (or more precisely, clusters), cf. Fig.~\ref{fig:linkedlattices}.

In addition to the degrees of freedom $\sigma$ and $\tau$ on the links and sites (respectively) of the original lattice, let us add degrees of freedom $\tilde \sigma$ and $\tilde \tau$ on the links and sites of the dual lattice. Let us denote the links of the original lattice by $\mathcal{L}$ and those of the dual lattice by $\widetilde{\mathcal{L}}$.
Define
\ba\label{eq:apphybrid}
Z'  = &
\sum_{\sigma,\tau,\tilde\sigma,\tilde\tau}
e^{- S_\text{top}[ \sigma, \widetilde \sigma]}
\prod_{\ell\in \mathcal{L}} \lf 1 + y \sigma\tau \tau \ri
\prod_{\tilde \ell\in \widetilde{\mathcal{L}}} \lf 1 + y' \tilde\sigma\tilde\tau \tilde\tau \ri.
\end{align}
The ``topological'' action $S_\text{top}[ \sigma, \widetilde \sigma]$ is both gauge invariant and symmetric between $\sigma$ and $\tilde \sigma$: $e^{-S_\text{top}}=(-1)^{\hat X}$,
where $\hat X$ is the $\mathbb{Z}_2$ linking number of the flux lines of $\sigma$ with those of $\tilde \sigma$. 
However it is convenient here to define it as
\be\label{eq:appStop}
e^{- S_\text{top}}=
\prod_{\tilde \ell \in \widetilde{\mathcal{L}}}
\lf 
\delta_{\tilde\sigma_{\tilde\ell} , 1}
+ \delta_{\tilde\sigma_{\tilde\ell} , -1} \prod \sigma
\ri
\ee
where these properties are not manifest.

To see the equivalence to the original Ising gauge theory (\ref{eq:Z(x,y)sigmatau}) we simply pick the gauge $\tilde \tau=1$ and do the sum on $\tilde \sigma$ separately for each link, 
\be
\sum_{\tilde\sigma}
\lf 
\delta_{\tilde\sigma_{\tilde\ell} , 1}
+ \delta_{\tilde\sigma_{\tilde\ell} , -1} \prod \sigma
\ri
(1+ y' \tilde \sigma)=
(1+y') \lf 1+ x \prod \sigma \ri,
\ee
so that 
\be
Z' = 2^{L^3} (1+y')^{3L^3} 
\sum_{\{\sigma\}, \{\tau\}}
\prod_{\square} \lf 1 + x \prod \sigma \ri 
\prod_\ell \lf 1 + y \sigma\tau \tau \ri.
\ee

To obtain the expression in terms of $\mathcal{C}_e$ and $\mathcal{C}_m$ in Eq.~\ref{eq:Zhybridrep} we first perform the graphical expansion of the two products in (\ref{eq:apphybrid}), giving the sum over  ``loop'' configurations $\mathcal{C}_e$ and $\mathcal{C}_m$ (Fig.~\ref{fig:linkedlattices}). In addition to the fugacities $y$ and $y'$ these are weighted by 
\be\label{eq:Stoploops}
\sum_{\sigma,\tilde\sigma}
e^{- S_\text{top}}
\lf \prod_{\ell \in \mathcal{C}_e} \sigma_\ell \ri 
\lf \prod_{\tilde \ell \in \mathcal{C}_m} \tilde\sigma_{\tilde \ell} \ri
= 
2^{(4L^3+2)} (-1)^{X(\mathcal{C}_e,\mathcal{C}_m)}.
\ee
We can see this by using (\ref{eq:appStop}) to make a graphical expansion of the left hand side above, in terms of a membrane configuration $\mathcal{M}$ on the original lattice with boundary ${\partial \mathcal{M}= \mathcal{C}_e}$. For a given term in the expansion, the $\tilde \sigma$ are fixed by the Kronecker deltas, which dictates the sign of the product $\prod_{\tilde \ell \in \mathcal{C}_m} \tilde\sigma$ on the LHS of (\ref{eq:Stoploops}). There are, for periodic boundary conditions, $2^{L^3+2}$ choices of $\mathcal{M}$ for fixed ${\partial \mathcal{M}=\mathcal{C}_e}$, but they all give the same sign.
Altogether,
\be
Z' = 
2^{(6L^3+2)}
\sum_{\mathcal{C}_e, \, \mathcal{C}_m}
(-1)^{X(\mathcal{C}_e, \mathcal{C}_m)} 
y^{|\mathcal{C}_e|}
{y'}^{|\mathcal{C}_m|}.
\ee

Summarizing,  $Z'$ can be related to $Z$ in Eqs.~\ref{eq:partitionfunctionmembranes},~\ref{eq:Z(x,y)sigmatau} by
\be
Z= 4\, c \sum_{\mathcal{C}_e, \, \mathcal{C}_m}
(-1)^{X(\mathcal{C}_e, \mathcal{C}_m)} 
y^{|\mathcal{C}_e|}
{y'}^{|\mathcal{C}_m|},
\ee
where ${c=2^{L^3} \left(1+y'\right)^{-3L^3}=2^{-2L^3} \left(1+x\right)^{3L^3}}$.

\section{First-order coexistence}
\label{app:coexistence}

\begin{figure}[t]
\centering
\includegraphics[width=0.9\linewidth]{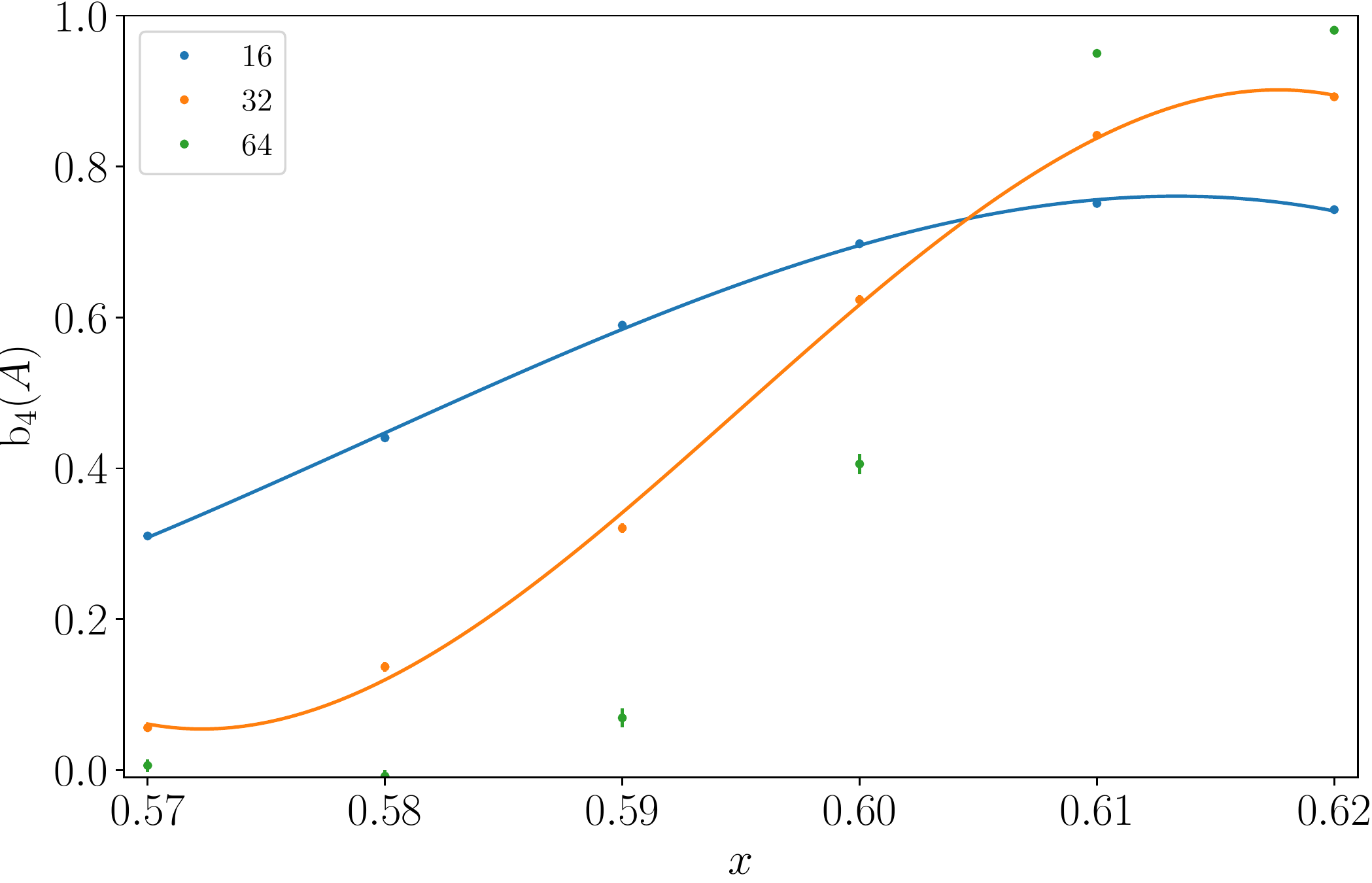}
\caption{$\operatorname{b}_4(A)=-(1/2)\kappa_4(A)/{\rm Var}(A)^2$  as a function of $x$ for three system sizes. The lines are cubic polynomial fits.}
\label{fig:b4a_x_cep}
\end{figure}

Although we have concentrated our study on the vicinity of the multicritical point, we can extract from the data some information related to the first-order coexistence region along the self-dual line. Starting from the deconfined phase (large $x$) this region starts at the multicritical point, $x_c$, and ends at a critical endpoint, $x_\mathrm{cep}$. The estimate $x_c\approx 0.6367$ was obtained in Secs.~\ref{sec:scaleinvariance} and \ref{sec:criticalexponents}. 
The location of $x_\mathrm{cep}$ is, in principle, easier to determine because in this region $\operatorname{b}_4(A)$ (defined in Eq.~\ref{eq:b4A}) behaves monotonically and presents a crossing, as  shown in Fig.~\ref{fig:b4a_x_cep}. From the figure we roughly estimate $x_\mathrm{cep}\approx 0.605$. 
Though this is a rough estimate, it is worth noting that the value of $\operatorname{b}_4(A)$ at the crossing point is consistent with standard Ising universality 
(as for the liquid-gas critical endpoint), for which ${\operatorname{b}_4(A)\approx 0.7}$ \cite{ferrenberg2018pushing}.

\begin{figure}[t]
\centering
\includegraphics[width=\linewidth]{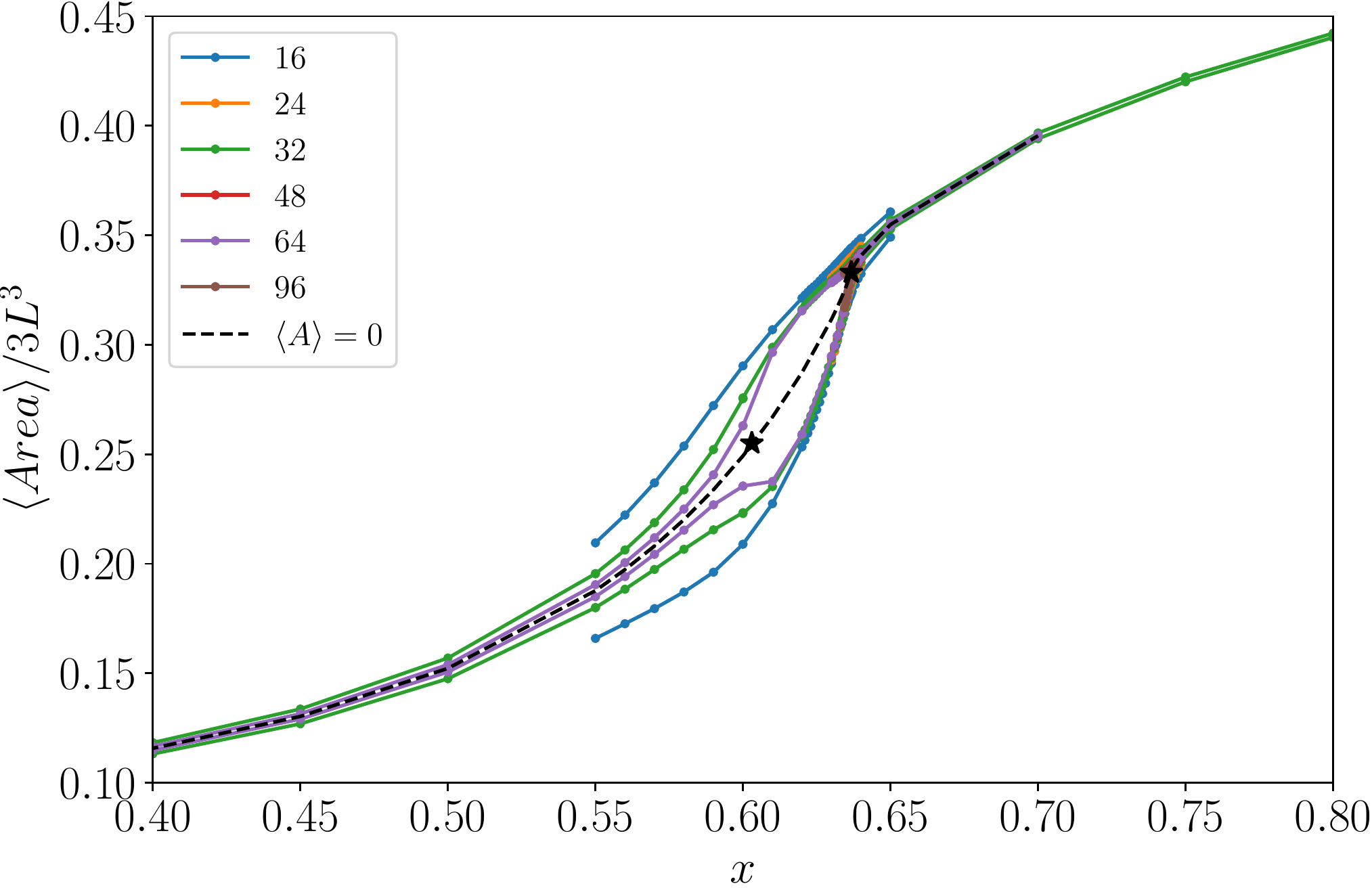}
\includegraphics[width=\linewidth]{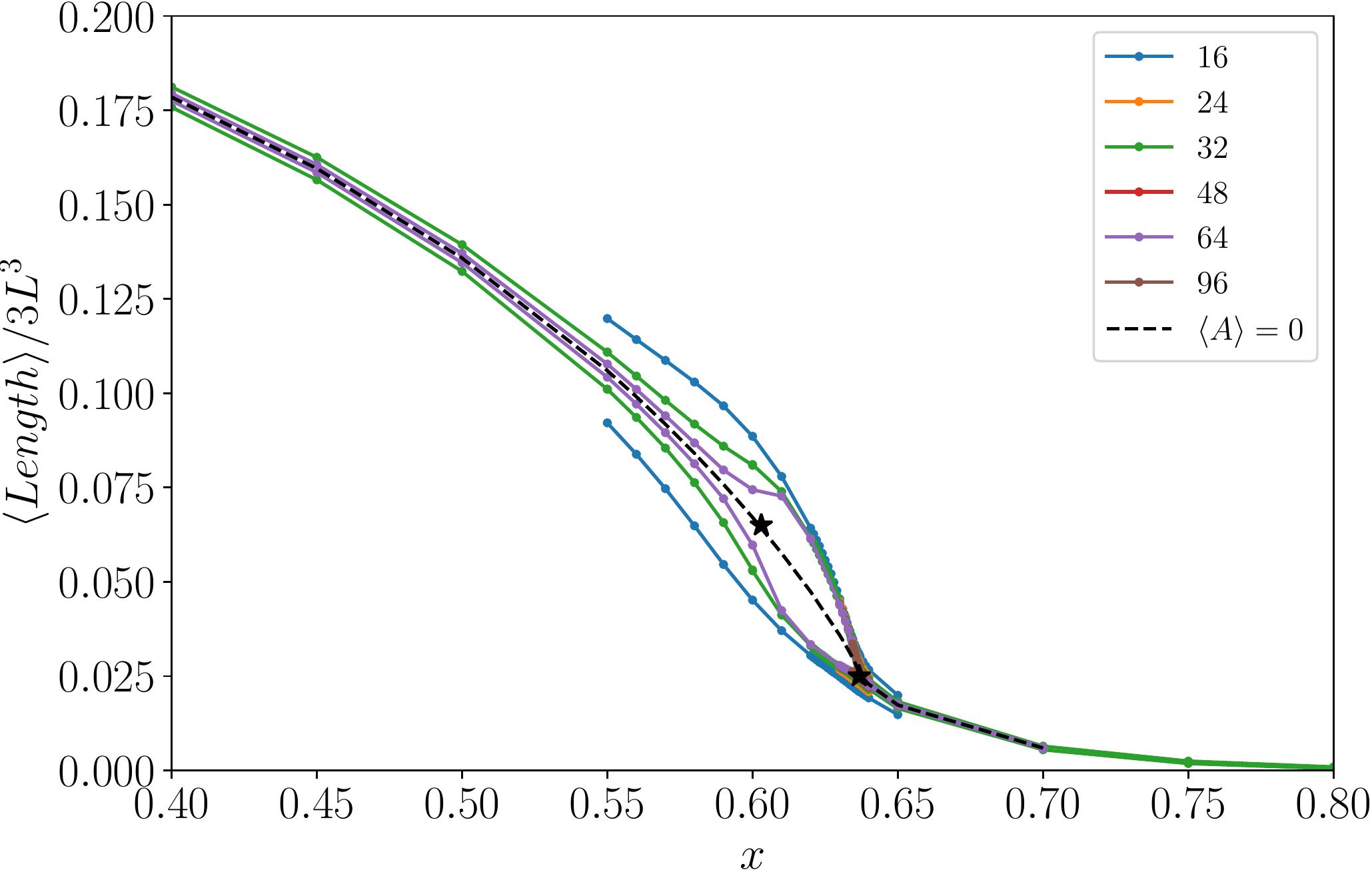}
\caption{Top: Average plaquette occupation number versus $x$ obtained from $\<|A|\>$ and $\<S\>$ on the self-dual line (see text). The limits of these curves as $L\rightarrow\infty$ give the two equations of state for the two coexisting phases.
The colors in the legend indicate different system sizes. The dashed black line is the average of the two coexisting phases, determined by $\<A\>=0$, and the black stars indicate the locations of $x_\mathrm{cep}$ and $x_c$ (${x_\mathrm{cep}<x_c}$). Lines are just a guide to the eye. Bottom: Similarly for the average link occupation number.}
\label{fig:Area_x}
\end{figure}

In between $x_c$ and $x_\mathrm{cep}$, histograms of $A$ or of the total membrane Area or membrane boundary Length  have two peaks, corresponding to the two coexisting phases. 
For large system sizes our MC scheme will not properly sample both minima, so it could become hard to obtain equations of state for each phase. However, we can exploit the symmetry properties of $A$ and $S$.
Denoting expectation values in the two equilibria by ${\<\ldots\>_\pm}$, 
in the thermodynamic limit we have ${\<A\>_\pm = \pm \< |A|\>}$.
Therefore by Eqs.~\ref{eq:Acube} and \ref{eq:Scube},
\ba
\frac{\<\mathrm{Area}\>_\pm}{3 L^3} & = \frac{ \pm \<|A|\>+\<S\>}{12} 
\label{eq:Area}
\\
\frac{\<\mathrm{Length}\>_\pm}{3 L^3} & =\frac{1-x^2}{2x} \frac{ \pm \<|A|\>-\<S\>}{12} +\frac{1-x}{2},
\label{eq:Length}
\end{align}
on the coexistence region of the self-dual line and in the thermodynamic limit.
These equations give equations of state for each phase.
The results are shown in Figs.~\ref{fig:Area_x}.

\label{app:firstorder}

\section{Details of MC scheme and of fits}
\label{app:MCscheme}

For most of our simulations, each MC step consists in updating all of the plaquettes (taking each of the three orientations in turn) and then updating all of the cubes.
To allow parallelization we divide plaquettes parallel to the $(x,y)$ plane into 2 sublattices, and  similarly for plaquettes in the $(y,z)$ and $(x,z)$ planes. We also divide cubes into two sublattices. We used one MC step as our unit of time.
We studied system sizes up to $L=96$ and  our longest simulations had $4\times 10^9$ MC steps.
Error bars for cumulants of $A$ and $S$ are calculated using bootstrap methods \cite{newman1999monte} (for this purpose the correlation time is estimated as the time for the correlation to decay by a factor of 10). 

For the fits in Sec.~\ref{sec:criticalexponents} the scaling functions were described using B-splines with 8  to 12 degrees of freedom. The data used for the fits were restricted to ${x \in [0.633,0.640]}$ and to scaling variable $z\in [-0.5,0.5]$, although the particular intervals could slightly change from fit to fit.  The system sizes included correspond to the best statistical fit, in the sense that the $p$-value (probability of getting a $\chi^2$ value below the one obtained from the fit, for the degrees of freedom used) was maximized.

\section{Further percolation data}
\label{app:percolation}

\subsection{Percolation at  the Ising$^*$ transition ($x=1$)}

\begin{figure}[htb]
\includegraphics[width=0.98\linewidth]{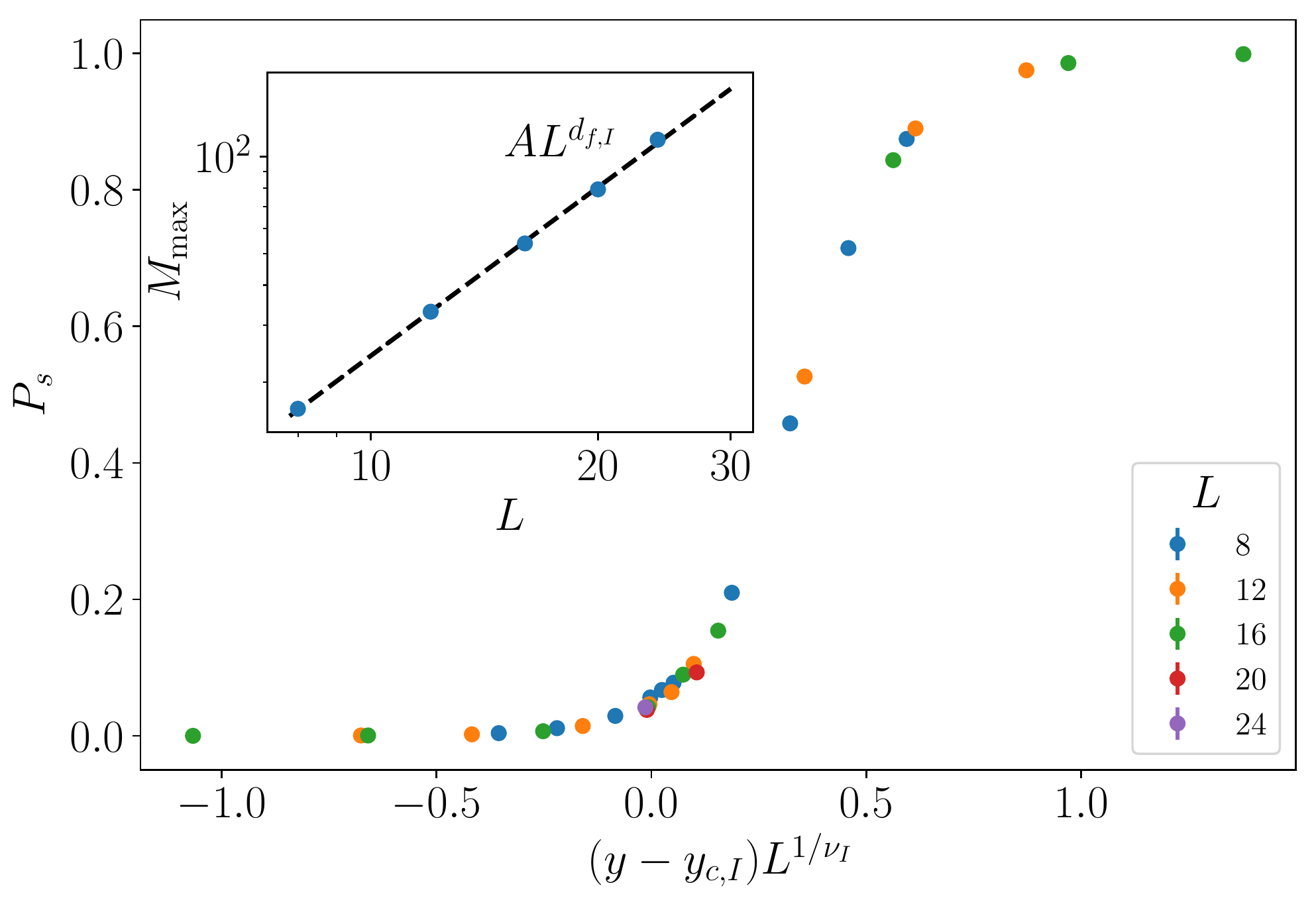}
\caption{Main Panel: scaling collapse of $P_s$ as a function of $(y-y_{c,I})L^{1/\nu_I}$, for $x=1$, with ${y_{c,I} = 0.21809}$ and ${\nu_I = 0.63012}$. Inset: $M_\mathrm{max}$ as function of the system size $L$ at $x = 1,\, y=0.218$. Straight-line shows a power-law using the fractal dimensions of Ising worldlines ${d_{f,I} = 1.7349}$. }
\label{fig:PsIsing}
\end{figure}

The transition at $x=1$ maps to 3D Ising.
Up to a difference in boundary conditions, the  wordlines are those in a standard high-temperature expansion of the Ising model, and the percolation transition happens precisely at the Ising critical point for the cubic lattice, $y = 0.21809$.
We indeed find that curves for the spanning probability  $P_s$ cross close to this value,
and can be collapsed by plotting as a function of $(y-y_c)L^{1/\nu}$,
 using known Ising critical values, $y_{c,I}=0.21809$ and  $\nu_I = 0.63012$ \cite{winter2008geometric}, Fig.~\ref{fig:PsIsing}.
We also check that the mass of the largest loop (number of links, $M_\mathrm{max}$) follows a power-law with a fractal dimension  consistent with the known value $d_{f,I}=1.7349(65)$ for Ising worldlines \cite{winter2008geometric}. The inset of Fig.~\ref{fig:PsIsing} shows $M_\mathrm{max}$ as a function of the system size at $y=0.218$.

\subsection{Percolation on the $y=1$ boundary}

\begin{figure}[t]
\includegraphics[width=0.98\linewidth]{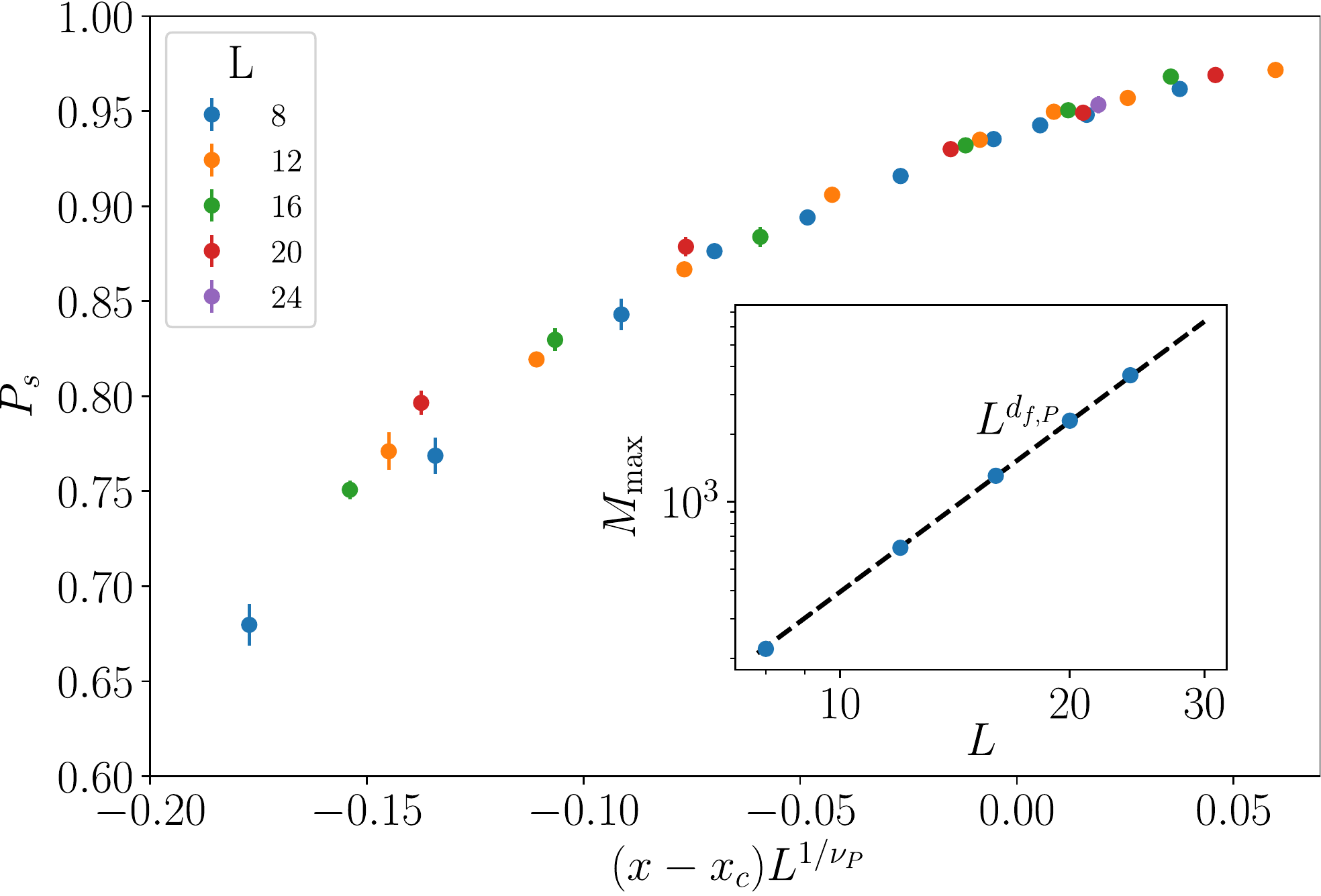}
\caption{Main Panel: scaling collapse of $P_s$ as a function of $(x-x_{cP})L^{1/\nu_P}$, for $y=1$, with  $x_{cP} = 0.0865$  and $\nu_P = 0.8762$ Inset: $M_\mathrm{max}$ as function of the system size $L$ at at $x = 0.087$. Straight line shows power-law with fractal dimension of percolation universality class, $d_{f,P} = 2.53$.}
\label{fig:PsPercolation}
\end{figure}

When the percolation transition takes place in within the thermodynamically trivial phase  we expect conventional percolation universality.\footnote{To see conventional percolation exponents here it is important that the geometrical objects we are considering are really clusters rather than strict loops: we have nodes where where $>2$  occupied links connect at a site. If we adopted a definition where the geometrical objects were strictly loop like, we would obtain a different universality class for unoriented loops \cite{nahum2013phase, nahum2012universal}.}
As an example we consider the case $y=1.$
An attempt to obtain scaling collapse of $P_s$ suggests that finite size effects are important for this range of system sizes. 
Fig.~\ref{fig:PsPercolation} shows an attempt at scaling collapse using $\nu_P=0.8762$ \cite{xu2014simultaneous}.
An estimate of the fractal dimension of the loops  from $M_\text{max}$ (inset to Fig.~\ref{fig:PsPercolation}) gives $d_f=2.56$ (to be compared with 2.53 for the percolation universality class).

\subsection{Percolation on the self-dual line}

The phase diagram Fig.~\ref{fig:phase_diag_perco} in the main text shows that we encounter several percolation transitions as we move along the self-dual line.
We have shown data close to the self-dual critical point $x_c$ in the main text. 
For smaller $x$ we encounter the first-order line where two phases coexist, one with $A>0$ and one with $A<0$. 
To separate the properties of the two coexisting phases, we may average $P_s$ separately for configurations with $A>0$ and $A<0$.
The phase with $A>0$ appears to  percolate throughout the entire range of the first-order line.
Therefore the phase with $A<0$ must also percolate for some region of the first-order line close to the critical endpoint (since the two phases become identical there).  
One possibility (at first sight the more natural) is that the phase with $A<0$ undergoes a percolation transition 
at some intermediate $x$ lying on the interior of the first-order line. 
Another possibility is that  this transition is pushed all the way to $x_c$, with the $A<0$ phase having an extremely weak but nonzero percolation order parameter for $x\lesssim x_c$. Data for small sizes do not allow us to determine which of these occurs.

\bibliography{IGTrefs}

\begin{thebibliography}{143}%
\makeatletter
\providecommand \@ifxundefined [1]{%
 \@ifx{#1\undefined}
}%
\providecommand \@ifnum [1]{%
 \ifnum #1\expandafter \@firstoftwo
 \else \expandafter \@secondoftwo
 \fi
}%
\providecommand \@ifx [1]{%
 \ifx #1\expandafter \@firstoftwo
 \else \expandafter \@secondoftwo
 \fi
}%
\providecommand \natexlab [1]{#1}%
\providecommand \enquote  [1]{``#1''}%
\providecommand \bibnamefont  [1]{#1}%
\providecommand \bibfnamefont [1]{#1}%
\providecommand \citenamefont [1]{#1}%
\providecommand \href@noop [0]{\@secondoftwo}%
\providecommand \href [0]{\begingroup \@sanitize@url \@href}%
\providecommand \@href[1]{\@@startlink{#1}\@@href}%
\providecommand \@@href[1]{\endgroup#1\@@endlink}%
\providecommand \@sanitize@url [0]{\catcode `\\12\catcode `\$12\catcode
  `\&12\catcode `\#12\catcode `\^12\catcode `\_12\catcode `\%12\relax}%
\providecommand \@@startlink[1]{}%
\providecommand \@@endlink[0]{}%
\providecommand \url  [0]{\begingroup\@sanitize@url \@url }%
\providecommand \@url [1]{\endgroup\@href {#1}{\urlprefix }}%
\providecommand \urlprefix  [0]{URL }%
\providecommand \Eprint [0]{\href }%
\providecommand \doibase [0]{http://dx.doi.org/}%
\providecommand \selectlanguage [0]{\@gobble}%
\providecommand \bibinfo  [0]{\@secondoftwo}%
\providecommand \bibfield  [0]{\@secondoftwo}%
\providecommand \translation [1]{[#1]}%
\providecommand \BibitemOpen [0]{}%
\providecommand \bibitemStop [0]{}%
\providecommand \bibitemNoStop [0]{.\EOS\space}%
\providecommand \EOS [0]{\spacefactor3000\relax}%
\providecommand \BibitemShut  [1]{\csname bibitem#1\endcsname}%
\let\auto@bib@innerbib\@empty
\bibitem [{\citenamefont {Fisher}(1974)}]{fisher1974renormalization}%
  \BibitemOpen
  \bibfield  {author} {\bibinfo {author} {\bibfnamefont {Michael~E.}\
  \bibnamefont {Fisher}},\ }\bibfield  {title} {\enquote {\bibinfo {title} {The
  renormalization group in the theory of critical behavior},}\ }\href {\doibase
  10.1103/RevModPhys.46.597} {\bibfield  {journal} {\bibinfo  {journal} {Rev.
  Mod. Phys.}\ }\textbf {\bibinfo {volume} {46}},\ \bibinfo {pages} {597--616}
  (\bibinfo {year} {1974})}\BibitemShut {NoStop}%
\bibitem [{\citenamefont {Sachdev}(2007)}]{sachdev2007quantum}%
  \BibitemOpen
  \bibfield  {author} {\bibinfo {author} {\bibfnamefont {Subir}\ \bibnamefont
  {Sachdev}},\ }\bibfield  {title} {\enquote {\bibinfo {title} {Quantum phase
  transitions},}\ }\href@noop {} {\bibfield  {journal} {\bibinfo  {journal}
  {Handbook of Magnetism and Advanced Magnetic Materials}\ } (\bibinfo {year}
  {2007})}\BibitemShut {NoStop}%
\bibitem [{\citenamefont {Wegner}(1971)}]{wegner1971duality}%
  \BibitemOpen
  \bibfield  {author} {\bibinfo {author} {\bibfnamefont {Franz~J}\ \bibnamefont
  {Wegner}},\ }\bibfield  {title} {\enquote {\bibinfo {title} {Duality in
  generalized ising models and phase transitions without local order
  parameters},}\ }\href@noop {} {\bibfield  {journal} {\bibinfo  {journal}
  {Journal of Mathematical Physics}\ }\textbf {\bibinfo {volume} {12}},\
  \bibinfo {pages} {2259--2272} (\bibinfo {year} {1971})}\BibitemShut {NoStop}%
\bibitem [{\citenamefont {Kosterlitz}\ and\ \citenamefont
  {Thouless}(1973)}]{kosterlitz1973ordering}%
  \BibitemOpen
  \bibfield  {author} {\bibinfo {author} {\bibfnamefont {John~Michael}\
  \bibnamefont {Kosterlitz}}\ and\ \bibinfo {author} {\bibfnamefont
  {David~James}\ \bibnamefont {Thouless}},\ }\bibfield  {title} {\enquote
  {\bibinfo {title} {Ordering, metastability and phase transitions in
  two-dimensional systems},}\ }\href@noop {} {\bibfield  {journal} {\bibinfo
  {journal} {Journal of Physics C: Solid State Physics}\ }\textbf {\bibinfo
  {volume} {6}},\ \bibinfo {pages} {1181} (\bibinfo {year} {1973})}\BibitemShut
  {NoStop}%
\bibitem [{\citenamefont {de~Gennes}(1972)}]{de1972exponents}%
  \BibitemOpen
  \bibfield  {author} {\bibinfo {author} {\bibfnamefont {Pierre-Gilles}\
  \bibnamefont {de~Gennes}},\ }\bibfield  {title} {\enquote {\bibinfo {title}
  {Exponents for the excluded volume problem as derived by the wilson
  method},}\ }\href@noop {} {\bibfield  {journal} {\bibinfo  {journal} {Physics
  Letters A}\ }\textbf {\bibinfo {volume} {38}},\ \bibinfo {pages} {339--340}
  (\bibinfo {year} {1972})}\BibitemShut {NoStop}%
\bibitem [{\citenamefont {Huse}\ and\ \citenamefont
  {Leibler}(1991)}]{huse1991sponge}%
  \BibitemOpen
  \bibfield  {author} {\bibinfo {author} {\bibfnamefont {David~A}\ \bibnamefont
  {Huse}}\ and\ \bibinfo {author} {\bibfnamefont {Stanislas}\ \bibnamefont
  {Leibler}},\ }\bibfield  {title} {\enquote {\bibinfo {title} {Are sponge
  phases of membranes experimental gauge-higgs systems?}}\ }\href@noop {}
  {\bibfield  {journal} {\bibinfo  {journal} {Physical review letters}\
  }\textbf {\bibinfo {volume} {66}},\ \bibinfo {pages} {437} (\bibinfo {year}
  {1991})}\BibitemShut {NoStop}%
\bibitem [{\citenamefont {Lammert}\ \emph {et~al.}(1993)\citenamefont
  {Lammert}, \citenamefont {Rokhsar},\ and\ \citenamefont
  {Toner}}]{lammert1993topology}%
  \BibitemOpen
  \bibfield  {author} {\bibinfo {author} {\bibfnamefont {Paul~E}\ \bibnamefont
  {Lammert}}, \bibinfo {author} {\bibfnamefont {Daniel~S}\ \bibnamefont
  {Rokhsar}}, \ and\ \bibinfo {author} {\bibfnamefont {John}\ \bibnamefont
  {Toner}},\ }\bibfield  {title} {\enquote {\bibinfo {title} {Topology and
  nematic ordering},}\ }\href@noop {} {\bibfield  {journal} {\bibinfo
  {journal} {Physical review letters}\ }\textbf {\bibinfo {volume} {70}},\
  \bibinfo {pages} {1650} (\bibinfo {year} {1993})}\BibitemShut {NoStop}%
\bibitem [{\citenamefont {Francesco}\ \emph {et~al.}(2012)\citenamefont
  {Francesco}, \citenamefont {Mathieu},\ and\ \citenamefont
  {S{\'e}n{\'e}chal}}]{francesco2012conformal}%
  \BibitemOpen
  \bibfield  {author} {\bibinfo {author} {\bibfnamefont {Philippe}\
  \bibnamefont {Francesco}}, \bibinfo {author} {\bibfnamefont {Pierre}\
  \bibnamefont {Mathieu}}, \ and\ \bibinfo {author} {\bibfnamefont {David}\
  \bibnamefont {S{\'e}n{\'e}chal}},\ }\href@noop {} {\emph {\bibinfo {title}
  {Conformal field theory}}}\ (\bibinfo  {publisher} {Springer Science \&
  Business Media},\ \bibinfo {year} {2012})\BibitemShut {NoStop}%
\bibitem [{\citenamefont {Fradkin}(2013)}]{fradkin2013field}%
  \BibitemOpen
  \bibfield  {author} {\bibinfo {author} {\bibfnamefont {Eduardo}\ \bibnamefont
  {Fradkin}},\ }\href@noop {} {\emph {\bibinfo {title} {Field theories of
  condensed matter physics}}}\ (\bibinfo  {publisher} {Cambridge University
  Press},\ \bibinfo {year} {2013})\BibitemShut {NoStop}%
\bibitem [{\citenamefont {Senthil}\ \emph {et~al.}(2004)\citenamefont
  {Senthil}, \citenamefont {Vishwanath}, \citenamefont {Balents}, \citenamefont
  {Sachdev},\ and\ \citenamefont {Fisher}}]{senthil2004deconfined}%
  \BibitemOpen
  \bibfield  {author} {\bibinfo {author} {\bibfnamefont {Todadri}\ \bibnamefont
  {Senthil}}, \bibinfo {author} {\bibfnamefont {Ashvin}\ \bibnamefont
  {Vishwanath}}, \bibinfo {author} {\bibfnamefont {Leon}\ \bibnamefont
  {Balents}}, \bibinfo {author} {\bibfnamefont {Subir}\ \bibnamefont
  {Sachdev}}, \ and\ \bibinfo {author} {\bibfnamefont {Matthew~PA}\
  \bibnamefont {Fisher}},\ }\bibfield  {title} {\enquote {\bibinfo {title}
  {Deconfined quantum critical points},}\ }\href@noop {} {\bibfield  {journal}
  {\bibinfo  {journal} {Science}\ }\textbf {\bibinfo {volume} {303}},\ \bibinfo
  {pages} {1490--1494} (\bibinfo {year} {2004})}\BibitemShut {NoStop}%
\bibitem [{\citenamefont {Fradkin}\ and\ \citenamefont
  {Shenker}(1979)}]{fradkin1979phase}%
  \BibitemOpen
  \bibfield  {author} {\bibinfo {author} {\bibfnamefont {Eduardo}\ \bibnamefont
  {Fradkin}}\ and\ \bibinfo {author} {\bibfnamefont {Stephen~H}\ \bibnamefont
  {Shenker}},\ }\bibfield  {title} {\enquote {\bibinfo {title} {Phase diagrams
  of lattice gauge theories with higgs fields},}\ }\href@noop {} {\bibfield
  {journal} {\bibinfo  {journal} {Physical Review D}\ }\textbf {\bibinfo
  {volume} {19}},\ \bibinfo {pages} {3682} (\bibinfo {year}
  {1979})}\BibitemShut {NoStop}%
\bibitem [{\citenamefont {Jongeward}\ \emph {et~al.}(1980)\citenamefont
  {Jongeward}, \citenamefont {Stack},\ and\ \citenamefont
  {Jayaprakash}}]{jongeward1980monte}%
  \BibitemOpen
  \bibfield  {author} {\bibinfo {author} {\bibfnamefont {Gary~A}\ \bibnamefont
  {Jongeward}}, \bibinfo {author} {\bibfnamefont {John~D}\ \bibnamefont
  {Stack}}, \ and\ \bibinfo {author} {\bibfnamefont {C}~\bibnamefont
  {Jayaprakash}},\ }\bibfield  {title} {\enquote {\bibinfo {title} {Monte carlo
  calculations on z 2 gauge-higgs theories},}\ }\href@noop {} {\bibfield
  {journal} {\bibinfo  {journal} {Physical Review D}\ }\textbf {\bibinfo
  {volume} {21}},\ \bibinfo {pages} {3360} (\bibinfo {year}
  {1980})}\BibitemShut {NoStop}%
\bibitem [{\citenamefont {Tupitsyn}\ \emph {et~al.}(2010)\citenamefont
  {Tupitsyn}, \citenamefont {Kitaev}, \citenamefont {Prokof’Ev},\ and\
  \citenamefont {Stamp}}]{TupitsynTopological}%
  \BibitemOpen
  \bibfield  {author} {\bibinfo {author} {\bibfnamefont {IS}~\bibnamefont
  {Tupitsyn}}, \bibinfo {author} {\bibfnamefont {Alexei}\ \bibnamefont
  {Kitaev}}, \bibinfo {author} {\bibfnamefont {NV}~\bibnamefont {Prokof’Ev}},
  \ and\ \bibinfo {author} {\bibfnamefont {PCE}\ \bibnamefont {Stamp}},\
  }\bibfield  {title} {\enquote {\bibinfo {title} {Topological multicritical
  point in the phase diagram of the toric code model and three-dimensional
  lattice gauge higgs model},}\ }\href@noop {} {\bibfield  {journal} {\bibinfo
  {journal} {Physical Review B}\ }\textbf {\bibinfo {volume} {82}},\ \bibinfo
  {pages} {085114} (\bibinfo {year} {2010})}\BibitemShut {NoStop}%
\bibitem [{\citenamefont {Vidal}\ \emph {et~al.}(2009)\citenamefont {Vidal},
  \citenamefont {Dusuel},\ and\ \citenamefont {Schmidt}}]{vidal2009low}%
  \BibitemOpen
  \bibfield  {author} {\bibinfo {author} {\bibfnamefont {Julien}\ \bibnamefont
  {Vidal}}, \bibinfo {author} {\bibfnamefont {S{\'e}bastien}\ \bibnamefont
  {Dusuel}}, \ and\ \bibinfo {author} {\bibfnamefont {Kai~Phillip}\
  \bibnamefont {Schmidt}},\ }\bibfield  {title} {\enquote {\bibinfo {title}
  {Low-energy effective theory of the toric code model in a parallel magnetic
  field},}\ }\href@noop {} {\bibfield  {journal} {\bibinfo  {journal} {Physical
  Review B}\ }\textbf {\bibinfo {volume} {79}},\ \bibinfo {pages} {033109}
  (\bibinfo {year} {2009})}\BibitemShut {NoStop}%
\bibitem [{\citenamefont {Kogut}(1979)}]{kogut1979introduction}%
  \BibitemOpen
  \bibfield  {author} {\bibinfo {author} {\bibfnamefont {John~B}\ \bibnamefont
  {Kogut}},\ }\bibfield  {title} {\enquote {\bibinfo {title} {An introduction
  to lattice gauge theory and spin systems},}\ }\href@noop {} {\bibfield
  {journal} {\bibinfo  {journal} {Reviews of Modern Physics}\ }\textbf
  {\bibinfo {volume} {51}},\ \bibinfo {pages} {659} (\bibinfo {year}
  {1979})}\BibitemShut {NoStop}%
\bibitem [{\citenamefont {Read}\ and\ \citenamefont
  {Chakraborty}(1989)}]{read1989statistics}%
  \BibitemOpen
  \bibfield  {author} {\bibinfo {author} {\bibfnamefont {N}~\bibnamefont
  {Read}}\ and\ \bibinfo {author} {\bibfnamefont {B}~\bibnamefont
  {Chakraborty}},\ }\bibfield  {title} {\enquote {\bibinfo {title} {Statistics
  of the excitations of the resonating-valence-bond state},}\ }\href@noop {}
  {\bibfield  {journal} {\bibinfo  {journal} {Physical Review B}\ }\textbf
  {\bibinfo {volume} {40}},\ \bibinfo {pages} {7133} (\bibinfo {year}
  {1989})}\BibitemShut {NoStop}%
\bibitem [{\citenamefont {Kivelson}(1989)}]{kivelson1989statistics}%
  \BibitemOpen
  \bibfield  {author} {\bibinfo {author} {\bibfnamefont {Steven}\ \bibnamefont
  {Kivelson}},\ }\bibfield  {title} {\enquote {\bibinfo {title} {Statistics of
  holons in the quantum hard-core dimer gas},}\ }\href@noop {} {\bibfield
  {journal} {\bibinfo  {journal} {Physical Review B}\ }\textbf {\bibinfo
  {volume} {39}},\ \bibinfo {pages} {259} (\bibinfo {year} {1989})}\BibitemShut
  {NoStop}%
\bibitem [{\citenamefont {Read}\ and\ \citenamefont
  {Sachdev}(1991)}]{read1991large}%
  \BibitemOpen
  \bibfield  {author} {\bibinfo {author} {\bibfnamefont {N}~\bibnamefont
  {Read}}\ and\ \bibinfo {author} {\bibfnamefont {Subir}\ \bibnamefont
  {Sachdev}},\ }\bibfield  {title} {\enquote {\bibinfo {title} {Large-n
  expansion for frustrated quantum antiferromagnets},}\ }\href@noop {}
  {\bibfield  {journal} {\bibinfo  {journal} {Physical review letters}\
  }\textbf {\bibinfo {volume} {66}},\ \bibinfo {pages} {1773} (\bibinfo {year}
  {1991})}\BibitemShut {NoStop}%
\bibitem [{\citenamefont {Wen}(1991)}]{wen1991mean}%
  \BibitemOpen
  \bibfield  {author} {\bibinfo {author} {\bibfnamefont {Xiao-Gang}\
  \bibnamefont {Wen}},\ }\bibfield  {title} {\enquote {\bibinfo {title}
  {Mean-field theory of spin-liquid states with finite energy gap and
  topological orders},}\ }\href@noop {} {\bibfield  {journal} {\bibinfo
  {journal} {Physical Review B}\ }\textbf {\bibinfo {volume} {44}},\ \bibinfo
  {pages} {2664} (\bibinfo {year} {1991})}\BibitemShut {NoStop}%
\bibitem [{\citenamefont {Senthil}\ and\ \citenamefont
  {Fisher}(2000)}]{senthil2000z}%
  \BibitemOpen
  \bibfield  {author} {\bibinfo {author} {\bibfnamefont {T}~\bibnamefont
  {Senthil}}\ and\ \bibinfo {author} {\bibfnamefont {Matthew~PA}\ \bibnamefont
  {Fisher}},\ }\bibfield  {title} {\enquote {\bibinfo {title} {Z 2 gauge theory
  of electron fractionalization in strongly correlated systems},}\ }\href@noop
  {} {\bibfield  {journal} {\bibinfo  {journal} {Physical Review B}\ }\textbf
  {\bibinfo {volume} {62}},\ \bibinfo {pages} {7850} (\bibinfo {year}
  {2000})}\BibitemShut {NoStop}%
\bibitem [{\citenamefont {Moessner}\ \emph {et~al.}(2001)\citenamefont
  {Moessner}, \citenamefont {Sondhi},\ and\ \citenamefont
  {Fradkin}}]{moessner2001short}%
  \BibitemOpen
  \bibfield  {author} {\bibinfo {author} {\bibfnamefont {Roderich}\
  \bibnamefont {Moessner}}, \bibinfo {author} {\bibfnamefont {Shivaji~L}\
  \bibnamefont {Sondhi}}, \ and\ \bibinfo {author} {\bibfnamefont {Eduardo}\
  \bibnamefont {Fradkin}},\ }\bibfield  {title} {\enquote {\bibinfo {title}
  {Short-ranged resonating valence bond physics, quantum dimer models, and
  ising gauge theories},}\ }\href@noop {} {\bibfield  {journal} {\bibinfo
  {journal} {Physical Review B}\ }\textbf {\bibinfo {volume} {65}},\ \bibinfo
  {pages} {024504} (\bibinfo {year} {2001})}\BibitemShut {NoStop}%
\bibitem [{\citenamefont {Kitaev}(2003)}]{kitaev2003fault}%
  \BibitemOpen
  \bibfield  {author} {\bibinfo {author} {\bibfnamefont {A~Yu}\ \bibnamefont
  {Kitaev}},\ }\bibfield  {title} {\enquote {\bibinfo {title} {Fault-tolerant
  quantum computation by anyons},}\ }\href@noop {} {\bibfield  {journal}
  {\bibinfo  {journal} {Annals of Physics}\ }\textbf {\bibinfo {volume}
  {303}},\ \bibinfo {pages} {2--30} (\bibinfo {year} {2003})}\BibitemShut
  {NoStop}%
\bibitem [{\citenamefont {Balian}\ \emph {et~al.}(1974)\citenamefont {Balian},
  \citenamefont {Drouffe},\ and\ \citenamefont {Itzykson}}]{balian1974gauge}%
  \BibitemOpen
  \bibfield  {author} {\bibinfo {author} {\bibfnamefont {Roger}\ \bibnamefont
  {Balian}}, \bibinfo {author} {\bibfnamefont {JM}~\bibnamefont {Drouffe}}, \
  and\ \bibinfo {author} {\bibfnamefont {Claude}\ \bibnamefont {Itzykson}},\
  }\bibfield  {title} {\enquote {\bibinfo {title} {Gauge fields on a lattice.
  i. general outlook},}\ }\href@noop {} {\bibfield  {journal} {\bibinfo
  {journal} {Physical Review D}\ }\textbf {\bibinfo {volume} {10}},\ \bibinfo
  {pages} {3376} (\bibinfo {year} {1974})}\BibitemShut {NoStop}%
\bibitem [{\citenamefont {Wu}\ \emph {et~al.}(2012)\citenamefont {Wu},
  \citenamefont {Deng},\ and\ \citenamefont {Prokof'ev}}]{wu2012phase}%
  \BibitemOpen
  \bibfield  {author} {\bibinfo {author} {\bibfnamefont {Fengcheng}\
  \bibnamefont {Wu}}, \bibinfo {author} {\bibfnamefont {Youjin}\ \bibnamefont
  {Deng}}, \ and\ \bibinfo {author} {\bibfnamefont {Nikolay}\ \bibnamefont
  {Prokof'ev}},\ }\bibfield  {title} {\enquote {\bibinfo {title} {Phase diagram
  of the toric code model in a parallel magnetic field},}\ }\href@noop {}
  {\bibfield  {journal} {\bibinfo  {journal} {Physical Review B}\ }\textbf
  {\bibinfo {volume} {85}},\ \bibinfo {pages} {195104} (\bibinfo {year}
  {2012})}\BibitemShut {NoStop}%
\bibitem [{\citenamefont {Huse}\ and\ \citenamefont
  {Leibler}(1988)}]{huse1988phase}%
  \BibitemOpen
  \bibfield  {author} {\bibinfo {author} {\bibfnamefont {D~A}\ \bibnamefont
  {Huse}}\ and\ \bibinfo {author} {\bibfnamefont {S}~\bibnamefont {Leibler}},\
  }\bibfield  {title} {\enquote {\bibinfo {title} {Phase behaviour of an
  ensemble of nonintersecting random fluid films},}\ }\href@noop {} {\bibfield
  {journal} {\bibinfo  {journal} {J. Phys. (Paris)}\ }\textbf {\bibinfo
  {volume} {49}},\ \bibinfo {pages} {605} (\bibinfo {year} {1988})}\BibitemShut
  {NoStop}%
\bibitem [{\citenamefont {Cates}\ \emph {et~al.}(1988)\citenamefont {Cates},
  \citenamefont {Roux}, \citenamefont {Andelman}, \citenamefont {Milner},\ and\
  \citenamefont {Safran}}]{cates1988random}%
  \BibitemOpen
  \bibfield  {author} {\bibinfo {author} {\bibfnamefont {ME}~\bibnamefont
  {Cates}}, \bibinfo {author} {\bibfnamefont {D}~\bibnamefont {Roux}}, \bibinfo
  {author} {\bibfnamefont {D}~\bibnamefont {Andelman}}, \bibinfo {author}
  {\bibfnamefont {ST}~\bibnamefont {Milner}}, \ and\ \bibinfo {author}
  {\bibfnamefont {SA}~\bibnamefont {Safran}},\ }\bibfield  {title} {\enquote
  {\bibinfo {title} {Random surface model for the l3-phase of dilute surfactant
  solutions},}\ }\href@noop {} {\bibfield  {journal} {\bibinfo  {journal} {EPL
  (Europhysics Letters)}\ }\textbf {\bibinfo {volume} {5}},\ \bibinfo {pages}
  {733} (\bibinfo {year} {1988})}\BibitemShut {NoStop}%
\bibitem [{\citenamefont {David}(1989)}]{david1989n}%
  \BibitemOpen
  \bibfield  {author} {\bibinfo {author} {\bibfnamefont {Francois}\
  \bibnamefont {David}},\ }\bibfield  {title} {\enquote {\bibinfo {title} {O
  (n) gauge models and self-avoiding random surfaces in three dimensions},}\
  }\href@noop {} {\bibfield  {journal} {\bibinfo  {journal} {EPL (Europhysics
  Letters)}\ }\textbf {\bibinfo {volume} {9}},\ \bibinfo {pages} {575}
  (\bibinfo {year} {1989})}\BibitemShut {NoStop}%
\bibitem [{\citenamefont {Roux}\ \emph {et~al.}(1990)\citenamefont {Roux},
  \citenamefont {Cates}, \citenamefont {Olsson}, \citenamefont {Ball},
  \citenamefont {Nallet},\ and\ \citenamefont {Bellocq}}]{roux1990light}%
  \BibitemOpen
  \bibfield  {author} {\bibinfo {author} {\bibfnamefont {D}~\bibnamefont
  {Roux}}, \bibinfo {author} {\bibfnamefont {ME}~\bibnamefont {Cates}},
  \bibinfo {author} {\bibfnamefont {U}~\bibnamefont {Olsson}}, \bibinfo
  {author} {\bibfnamefont {RC}~\bibnamefont {Ball}}, \bibinfo {author}
  {\bibfnamefont {F}~\bibnamefont {Nallet}}, \ and\ \bibinfo {author}
  {\bibfnamefont {AM}~\bibnamefont {Bellocq}},\ }\bibfield  {title} {\enquote
  {\bibinfo {title} {Light scattering from a surfactant “sponge” phase:
  Evidence for a hidden symmetry},}\ }\href@noop {} {\bibfield  {journal}
  {\bibinfo  {journal} {EPL (Europhysics Letters)}\ }\textbf {\bibinfo {volume}
  {11}},\ \bibinfo {pages} {229} (\bibinfo {year} {1990})}\BibitemShut
  {NoStop}%
\bibitem [{\citenamefont {Roux}\ \emph {et~al.}(1992)\citenamefont {Roux},
  \citenamefont {Coulon},\ and\ \citenamefont {Cates}}]{roux1992sponge}%
  \BibitemOpen
  \bibfield  {author} {\bibinfo {author} {\bibfnamefont {D}~\bibnamefont
  {Roux}}, \bibinfo {author} {\bibfnamefont {C}~\bibnamefont {Coulon}}, \ and\
  \bibinfo {author} {\bibfnamefont {ME}~\bibnamefont {Cates}},\ }\bibfield
  {title} {\enquote {\bibinfo {title} {Sponge phases in surfactant
  solutions},}\ }\href@noop {} {\bibfield  {journal} {\bibinfo  {journal} {The
  Journal of Physical Chemistry}\ }\textbf {\bibinfo {volume} {96}},\ \bibinfo
  {pages} {4174--4187} (\bibinfo {year} {1992})}\BibitemShut {NoStop}%
\bibitem [{\citenamefont {Roux}(1995)}]{roux1995sponge}%
  \BibitemOpen
  \bibfield  {author} {\bibinfo {author} {\bibfnamefont {Didier}\ \bibnamefont
  {Roux}},\ }\bibfield  {title} {\enquote {\bibinfo {title} {Sponge phases: An
  example of random surfaces},}\ }\href@noop {} {\bibfield  {journal} {\bibinfo
   {journal} {Physica A: Statistical Mechanics and its Applications}\ }\textbf
  {\bibinfo {volume} {213}},\ \bibinfo {pages} {168--172} (\bibinfo {year}
  {1995})}\BibitemShut {NoStop}%
\bibitem [{\citenamefont {Peliti}()}]{pelitiamphiphilic}%
  \BibitemOpen
  \bibfield  {author} {\bibinfo {author} {\bibfnamefont {L}~\bibnamefont
  {Peliti}},\ }\href@noop {} {\enquote {\bibinfo {title} {``amphiphilic
  membranes'', in ``fluctuating geometries in statistical mechanics and field
  theory'', f. david, p. ginsparg, j. zinn-justin, (eds.) les houches, session
  lxii, 1994 (amsterdam: Elsevier, 1996) 195-285},}\ }\BibitemShut {NoStop}%
\bibitem [{\citenamefont {Nandkishore}\ \emph {et~al.}(2012)\citenamefont
  {Nandkishore}, \citenamefont {Metlitski},\ and\ \citenamefont
  {Senthil}}]{nandkishore2012orthogonal}%
  \BibitemOpen
  \bibfield  {author} {\bibinfo {author} {\bibfnamefont {Rahul}\ \bibnamefont
  {Nandkishore}}, \bibinfo {author} {\bibfnamefont {Max~A}\ \bibnamefont
  {Metlitski}}, \ and\ \bibinfo {author} {\bibfnamefont {T}~\bibnamefont
  {Senthil}},\ }\bibfield  {title} {\enquote {\bibinfo {title} {Orthogonal
  metals: The simplest non-fermi liquids},}\ }\href@noop {} {\bibfield
  {journal} {\bibinfo  {journal} {Physical Review B}\ }\textbf {\bibinfo
  {volume} {86}},\ \bibinfo {pages} {045128} (\bibinfo {year}
  {2012})}\BibitemShut {NoStop}%
\bibitem [{\citenamefont {Kapustin}\ and\ \citenamefont
  {Thorngren}(2017)}]{kapustin2017higher}%
  \BibitemOpen
  \bibfield  {author} {\bibinfo {author} {\bibfnamefont {Anton}\ \bibnamefont
  {Kapustin}}\ and\ \bibinfo {author} {\bibfnamefont {Ryan}\ \bibnamefont
  {Thorngren}},\ }\bibfield  {title} {\enquote {\bibinfo {title} {Higher
  symmetry and gapped phases of gauge theories},}\ }in\ \href@noop {} {\emph
  {\bibinfo {booktitle} {Algebra, Geometry, and Physics in the 21st Century}}}\
  (\bibinfo  {publisher} {Springer},\ \bibinfo {year} {2017})\ pp.\ \bibinfo
  {pages} {177--202}\BibitemShut {NoStop}%
\bibitem [{\citenamefont {Gaiotto}\ \emph {et~al.}(2015)\citenamefont
  {Gaiotto}, \citenamefont {Kapustin}, \citenamefont {Seiberg},\ and\
  \citenamefont {Willett}}]{gaiotto2015generalized}%
  \BibitemOpen
  \bibfield  {author} {\bibinfo {author} {\bibfnamefont {Davide}\ \bibnamefont
  {Gaiotto}}, \bibinfo {author} {\bibfnamefont {Anton}\ \bibnamefont
  {Kapustin}}, \bibinfo {author} {\bibfnamefont {Nathan}\ \bibnamefont
  {Seiberg}}, \ and\ \bibinfo {author} {\bibfnamefont {Brian}\ \bibnamefont
  {Willett}},\ }\bibfield  {title} {\enquote {\bibinfo {title} {Generalized
  global symmetries},}\ }\href@noop {} {\bibfield  {journal} {\bibinfo
  {journal} {Journal of High Energy Physics}\ }\textbf {\bibinfo {volume}
  {2015}},\ \bibinfo {pages} {172} (\bibinfo {year} {2015})}\BibitemShut
  {NoStop}%
\bibitem [{\citenamefont {Wen}(2019)}]{wen2019emergent}%
  \BibitemOpen
  \bibfield  {author} {\bibinfo {author} {\bibfnamefont {Xiao-Gang}\
  \bibnamefont {Wen}},\ }\bibfield  {title} {\enquote {\bibinfo {title}
  {Emergent anomalous higher symmetries from topological order and from
  dynamical electromagnetic field in condensed matter systems},}\ }\href@noop
  {} {\bibfield  {journal} {\bibinfo  {journal} {Physical Review B}\ }\textbf
  {\bibinfo {volume} {99}},\ \bibinfo {pages} {205139} (\bibinfo {year}
  {2019})}\BibitemShut {NoStop}%
\bibitem [{\citenamefont {Newman}\ \emph {et~al.}()\citenamefont {Newman},
  \citenamefont {Barkema},\ and\ \citenamefont {Barkema}}]{newman1999monte}%
  \BibitemOpen
  \bibfield  {author} {\bibinfo {author} {\bibfnamefont {M.E.J.}\ \bibnamefont
  {Newman}}, \bibinfo {author} {\bibfnamefont {G.T.}\ \bibnamefont {Barkema}},
  \ and\ \bibinfo {author} {\bibfnamefont {I.T.P.G.T.}\ \bibnamefont
  {Barkema}},\ }\href@noop {} {\emph {\bibinfo {title} {Monte Carlo Methods in
  Statistical Physics}}}\BibitemShut {NoStop}%
\bibitem [{\citenamefont {Geraedts}\ and\ \citenamefont
  {Motrunich}(2012{\natexlab{a}})}]{geraedts2012monte}%
  \BibitemOpen
  \bibfield  {author} {\bibinfo {author} {\bibfnamefont {Scott~D}\ \bibnamefont
  {Geraedts}}\ and\ \bibinfo {author} {\bibfnamefont {Olexei~I}\ \bibnamefont
  {Motrunich}},\ }\bibfield  {title} {\enquote {\bibinfo {title} {Monte carlo
  study of a u (1)$\times$ u (1) system with $\pi$-statistical interaction},}\
  }\href@noop {} {\bibfield  {journal} {\bibinfo  {journal} {Physical Review
  B}\ }\textbf {\bibinfo {volume} {85}},\ \bibinfo {pages} {045114} (\bibinfo
  {year} {2012}{\natexlab{a}})}\BibitemShut {NoStop}%
\bibitem [{\citenamefont {Burnell}(2018)}]{burnell2018anyon}%
  \BibitemOpen
  \bibfield  {author} {\bibinfo {author} {\bibfnamefont {Fiona~J}\ \bibnamefont
  {Burnell}},\ }\bibfield  {title} {\enquote {\bibinfo {title} {Anyon
  condensation and its applications},}\ }\href@noop {} {\bibfield  {journal}
  {\bibinfo  {journal} {Annual Review of Condensed Matter Physics}\ }\textbf
  {\bibinfo {volume} {9}},\ \bibinfo {pages} {307--327} (\bibinfo {year}
  {2018})}\BibitemShut {NoStop}%
\bibitem [{\citenamefont {Geraedts}\ and\ \citenamefont
  {Motrunich}(2012{\natexlab{b}})}]{geraedts2012phases}%
  \BibitemOpen
  \bibfield  {author} {\bibinfo {author} {\bibfnamefont {Scott~D}\ \bibnamefont
  {Geraedts}}\ and\ \bibinfo {author} {\bibfnamefont {Olexei~I}\ \bibnamefont
  {Motrunich}},\ }\bibfield  {title} {\enquote {\bibinfo {title} {Phases and
  phase transitions in a u (1)$\times$ u (1) system with $\theta$= 2 $\pi$/3
  mutual statistics},}\ }\href@noop {} {\bibfield  {journal} {\bibinfo
  {journal} {Physical Review B}\ }\textbf {\bibinfo {volume} {86}},\ \bibinfo
  {pages} {045106} (\bibinfo {year} {2012}{\natexlab{b}})}\BibitemShut
  {NoStop}%
\bibitem [{\citenamefont {Lee}\ \emph {et~al.}(2016)\citenamefont {Lee},
  \citenamefont {Geraedts},\ and\ \citenamefont {Motrunich}}]{lee2016monte}%
  \BibitemOpen
  \bibfield  {author} {\bibinfo {author} {\bibfnamefont {Jong~Yeon}\
  \bibnamefont {Lee}}, \bibinfo {author} {\bibfnamefont {Scott}\ \bibnamefont
  {Geraedts}}, \ and\ \bibinfo {author} {\bibfnamefont {Olexei~I}\ \bibnamefont
  {Motrunich}},\ }\bibfield  {title} {\enquote {\bibinfo {title} {Monte carlo
  study of phase transitions out of symmetry-enriched topological phases of
  bosons in two dimensions},}\ }\href@noop {} {\bibfield  {journal} {\bibinfo
  {journal} {Physical Review B}\ }\textbf {\bibinfo {volume} {93}},\ \bibinfo
  {pages} {035103} (\bibinfo {year} {2016})}\BibitemShut {NoStop}%
\bibitem [{\citenamefont {Geraedts}\ and\ \citenamefont
  {Motrunich}(2012{\natexlab{c}})}]{geraedts2012line}%
  \BibitemOpen
  \bibfield  {author} {\bibinfo {author} {\bibfnamefont {Scott~D}\ \bibnamefont
  {Geraedts}}\ and\ \bibinfo {author} {\bibfnamefont {Olexei~I}\ \bibnamefont
  {Motrunich}},\ }\bibfield  {title} {\enquote {\bibinfo {title} {Line of
  continuous phase transitions in a three-dimensional u (1) loop model with 1/r
  2 current-current interactions},}\ }\href@noop {} {\bibfield  {journal}
  {\bibinfo  {journal} {Physical Review B}\ }\textbf {\bibinfo {volume} {85}},\
  \bibinfo {pages} {144303} (\bibinfo {year} {2012}{\natexlab{c}})}\BibitemShut
  {NoStop}%
\bibitem [{\citenamefont {Levin}\ and\ \citenamefont
  {Senthil}(2004)}]{levin2004deconfined}%
  \BibitemOpen
  \bibfield  {author} {\bibinfo {author} {\bibfnamefont {Michael}\ \bibnamefont
  {Levin}}\ and\ \bibinfo {author} {\bibfnamefont {Todadri}\ \bibnamefont
  {Senthil}},\ }\bibfield  {title} {\enquote {\bibinfo {title} {Deconfined
  quantum criticality and n{\'e}el order via dimer disorder},}\ }\href@noop {}
  {\bibfield  {journal} {\bibinfo  {journal} {Physical Review B}\ }\textbf
  {\bibinfo {volume} {70}},\ \bibinfo {pages} {220403} (\bibinfo {year}
  {2004})}\BibitemShut {NoStop}%
\bibitem [{\citenamefont {Motrunich}\ and\ \citenamefont
  {Vishwanath}(2004)}]{motrunich2004emergent}%
  \BibitemOpen
  \bibfield  {author} {\bibinfo {author} {\bibfnamefont {Olexei~I}\
  \bibnamefont {Motrunich}}\ and\ \bibinfo {author} {\bibfnamefont {Ashvin}\
  \bibnamefont {Vishwanath}},\ }\bibfield  {title} {\enquote {\bibinfo {title}
  {Emergent photons and transitions in the o (3) sigma model with hedgehog
  suppression},}\ }\href@noop {} {\bibfield  {journal} {\bibinfo  {journal}
  {Physical Review B}\ }\textbf {\bibinfo {volume} {70}},\ \bibinfo {pages}
  {075104} (\bibinfo {year} {2004})}\BibitemShut {NoStop}%
\bibitem [{\citenamefont {Senthil}\ \emph {et~al.}(2005)\citenamefont
  {Senthil}, \citenamefont {Balents}, \citenamefont {Sachdev}, \citenamefont
  {Vishwanath},\ and\ \citenamefont {Fisher}}]{senthil2005deconfined}%
  \BibitemOpen
  \bibfield  {author} {\bibinfo {author} {\bibfnamefont {Todadri}\ \bibnamefont
  {Senthil}}, \bibinfo {author} {\bibfnamefont {Leon}\ \bibnamefont {Balents}},
  \bibinfo {author} {\bibfnamefont {Subir}\ \bibnamefont {Sachdev}}, \bibinfo
  {author} {\bibfnamefont {Ashvin}\ \bibnamefont {Vishwanath}}, \ and\ \bibinfo
  {author} {\bibfnamefont {Matthew P~A}\ \bibnamefont {Fisher}},\ }\bibfield
  {title} {\enquote {\bibinfo {title} {Deconfined criticality critically
  defined},}\ }\href@noop {} {\bibfield  {journal} {\bibinfo  {journal}
  {Journal of the Physical Society of Japan}\ }\textbf {\bibinfo {volume}
  {74}},\ \bibinfo {pages} {1--9} (\bibinfo {year} {2005})}\BibitemShut
  {NoStop}%
\bibitem [{\citenamefont {Tanaka}\ and\ \citenamefont
  {Hu}(2005)}]{tanaka2005many}%
  \BibitemOpen
  \bibfield  {author} {\bibinfo {author} {\bibfnamefont {Akihiro}\ \bibnamefont
  {Tanaka}}\ and\ \bibinfo {author} {\bibfnamefont {Xiao}\ \bibnamefont {Hu}},\
  }\bibfield  {title} {\enquote {\bibinfo {title} {Many-body spin berry phases
  emerging from the $\pi$-flux state: Competition between antiferromagnetism
  and the valence-bond-solid state},}\ }\href@noop {} {\bibfield  {journal}
  {\bibinfo  {journal} {Physical review letters}\ }\textbf {\bibinfo {volume}
  {95}},\ \bibinfo {pages} {036402} (\bibinfo {year} {2005})}\BibitemShut
  {NoStop}%
\bibitem [{\citenamefont {Senthil}\ and\ \citenamefont
  {Fisher}(2006)}]{senthil2006competing}%
  \BibitemOpen
  \bibfield  {author} {\bibinfo {author} {\bibfnamefont {T}~\bibnamefont
  {Senthil}}\ and\ \bibinfo {author} {\bibfnamefont {Matthew P~A}\ \bibnamefont
  {Fisher}},\ }\bibfield  {title} {\enquote {\bibinfo {title} {Competing
  orders, nonlinear sigma models, and topological terms in quantum magnets},}\
  }\href@noop {} {\bibfield  {journal} {\bibinfo  {journal} {Physical Review
  B}\ }\textbf {\bibinfo {volume} {74}},\ \bibinfo {pages} {064405} (\bibinfo
  {year} {2006})}\BibitemShut {NoStop}%
\bibitem [{\citenamefont {Sandvik}(2007)}]{sandvik2007evidence}%
  \BibitemOpen
  \bibfield  {author} {\bibinfo {author} {\bibfnamefont {Anders~W}\
  \bibnamefont {Sandvik}},\ }\bibfield  {title} {\enquote {\bibinfo {title}
  {Evidence for deconfined quantum criticality in a two-dimensional heisenberg
  model with four-spin interactions},}\ }\href@noop {} {\bibfield  {journal}
  {\bibinfo  {journal} {Physical review letters}\ }\textbf {\bibinfo {volume}
  {98}},\ \bibinfo {pages} {227202} (\bibinfo {year} {2007})}\BibitemShut
  {NoStop}%
\bibitem [{\citenamefont {Bhattacharjee}(2011)}]{bhattacharjee2011quantum}%
  \BibitemOpen
  \bibfield  {author} {\bibinfo {author} {\bibfnamefont {Subhro}\ \bibnamefont
  {Bhattacharjee}},\ }\bibfield  {title} {\enquote {\bibinfo {title} {Quantum
  destruction of spiral order in two-dimensional frustrated magnets},}\
  }\href@noop {} {\bibfield  {journal} {\bibinfo  {journal} {Physical Review
  B}\ }\textbf {\bibinfo {volume} {84}},\ \bibinfo {pages} {104430} (\bibinfo
  {year} {2011})}\BibitemShut {NoStop}%
\bibitem [{\citenamefont {Wang}\ \emph {et~al.}(2017)\citenamefont {Wang},
  \citenamefont {Nahum}, \citenamefont {Metlitski}, \citenamefont {Xu},\ and\
  \citenamefont {Senthil}}]{wang2017deconfined}%
  \BibitemOpen
  \bibfield  {author} {\bibinfo {author} {\bibfnamefont {Chong}\ \bibnamefont
  {Wang}}, \bibinfo {author} {\bibfnamefont {Adam}\ \bibnamefont {Nahum}},
  \bibinfo {author} {\bibfnamefont {Max~A}\ \bibnamefont {Metlitski}}, \bibinfo
  {author} {\bibfnamefont {Cenke}\ \bibnamefont {Xu}}, \ and\ \bibinfo {author}
  {\bibfnamefont {T}~\bibnamefont {Senthil}},\ }\bibfield  {title} {\enquote
  {\bibinfo {title} {Deconfined quantum critical points: symmetries and
  dualities},}\ }\href@noop {} {\bibfield  {journal} {\bibinfo  {journal}
  {Physical Review X}\ }\textbf {\bibinfo {volume} {7}},\ \bibinfo {pages}
  {031051} (\bibinfo {year} {2017})}\BibitemShut {NoStop}%
\bibitem [{\citenamefont {Gazit}\ \emph {et~al.}(2018)\citenamefont {Gazit},
  \citenamefont {Assaad}, \citenamefont {Sachdev}, \citenamefont {Vishwanath},\
  and\ \citenamefont {Wang}}]{gazit2018confinement}%
  \BibitemOpen
  \bibfield  {author} {\bibinfo {author} {\bibfnamefont {Snir}\ \bibnamefont
  {Gazit}}, \bibinfo {author} {\bibfnamefont {Fakher~F}\ \bibnamefont
  {Assaad}}, \bibinfo {author} {\bibfnamefont {Subir}\ \bibnamefont {Sachdev}},
  \bibinfo {author} {\bibfnamefont {Ashvin}\ \bibnamefont {Vishwanath}}, \ and\
  \bibinfo {author} {\bibfnamefont {Chong}\ \bibnamefont {Wang}},\ }\bibfield
  {title} {\enquote {\bibinfo {title} {Confinement transition of z2 gauge
  theories coupled to massless fermions: Emergent quantum chromodynamics and
  so(5) symmetry},}\ }\href@noop {} {\bibfield  {journal} {\bibinfo  {journal}
  {Proceedings of the National Academy of Sciences}\ }\textbf {\bibinfo
  {volume} {115}},\ \bibinfo {pages} {E6987--E6995} (\bibinfo {year}
  {2018})}\BibitemShut {NoStop}%
\bibitem [{\citenamefont {Jiang}\ and\ \citenamefont
  {Motrunich}(2019)}]{jiang2019ising}%
  \BibitemOpen
  \bibfield  {author} {\bibinfo {author} {\bibfnamefont {Shenghan}\
  \bibnamefont {Jiang}}\ and\ \bibinfo {author} {\bibfnamefont {Olexei}\
  \bibnamefont {Motrunich}},\ }\bibfield  {title} {\enquote {\bibinfo {title}
  {Ising ferromagnet to valence bond solid transition in a one-dimensional spin
  chain: Analogies to deconfined quantum critical points},}\ }\href@noop {}
  {\bibfield  {journal} {\bibinfo  {journal} {Physical Review B}\ }\textbf
  {\bibinfo {volume} {99}},\ \bibinfo {pages} {075103} (\bibinfo {year}
  {2019})}\BibitemShut {NoStop}%
\bibitem [{\citenamefont {Freedman}\ \emph {et~al.}(2004)\citenamefont
  {Freedman}, \citenamefont {Nayak}, \citenamefont {Shtengel}, \citenamefont
  {Walker},\ and\ \citenamefont {Wang}}]{freedman2004class}%
  \BibitemOpen
  \bibfield  {author} {\bibinfo {author} {\bibfnamefont {Michael}\ \bibnamefont
  {Freedman}}, \bibinfo {author} {\bibfnamefont {Chetan}\ \bibnamefont
  {Nayak}}, \bibinfo {author} {\bibfnamefont {Kirill}\ \bibnamefont
  {Shtengel}}, \bibinfo {author} {\bibfnamefont {Kevin}\ \bibnamefont
  {Walker}}, \ and\ \bibinfo {author} {\bibfnamefont {Zhenghan}\ \bibnamefont
  {Wang}},\ }\bibfield  {title} {\enquote {\bibinfo {title} {A class of p,
  t-invariant topological phases of interacting electrons},}\ }\href@noop {}
  {\bibfield  {journal} {\bibinfo  {journal} {Annals of Physics}\ }\textbf
  {\bibinfo {volume} {310}},\ \bibinfo {pages} {428--492} (\bibinfo {year}
  {2004})}\BibitemShut {NoStop}%
\bibitem [{\citenamefont {Freedman}\ \emph {et~al.}(2005)\citenamefont
  {Freedman}, \citenamefont {Nayak},\ and\ \citenamefont
  {Shtengel}}]{freedman2005line}%
  \BibitemOpen
  \bibfield  {author} {\bibinfo {author} {\bibfnamefont {Michael}\ \bibnamefont
  {Freedman}}, \bibinfo {author} {\bibfnamefont {Chetan}\ \bibnamefont
  {Nayak}}, \ and\ \bibinfo {author} {\bibfnamefont {Kirill}\ \bibnamefont
  {Shtengel}},\ }\bibfield  {title} {\enquote {\bibinfo {title} {Line of
  critical points in 2+ 1 dimensions: Quantum critical loop gases and
  non-abelian gauge theory},}\ }\href@noop {} {\bibfield  {journal} {\bibinfo
  {journal} {Physical review letters}\ }\textbf {\bibinfo {volume} {94}},\
  \bibinfo {pages} {147205} (\bibinfo {year} {2005})}\BibitemShut {NoStop}%
\bibitem [{\citenamefont {Freedman}\ \emph {et~al.}(2008)\citenamefont
  {Freedman}, \citenamefont {Nayak},\ and\ \citenamefont
  {Shtengel}}]{freedman2008lieb}%
  \BibitemOpen
  \bibfield  {author} {\bibinfo {author} {\bibfnamefont {Michael}\ \bibnamefont
  {Freedman}}, \bibinfo {author} {\bibfnamefont {Chetan}\ \bibnamefont
  {Nayak}}, \ and\ \bibinfo {author} {\bibfnamefont {Kirill}\ \bibnamefont
  {Shtengel}},\ }\bibfield  {title} {\enquote {\bibinfo {title}
  {Lieb-schultz-mattis theorem for quasitopological systems},}\ }\href@noop {}
  {\bibfield  {journal} {\bibinfo  {journal} {Physical Review B}\ }\textbf
  {\bibinfo {volume} {78}},\ \bibinfo {pages} {174411} (\bibinfo {year}
  {2008})}\BibitemShut {NoStop}%
\bibitem [{\citenamefont {Troyer}\ \emph {et~al.}(2008)\citenamefont {Troyer},
  \citenamefont {Trebst}, \citenamefont {Shtengel},\ and\ \citenamefont
  {Nayak}}]{troyer2008local}%
  \BibitemOpen
  \bibfield  {author} {\bibinfo {author} {\bibfnamefont {Matthias}\
  \bibnamefont {Troyer}}, \bibinfo {author} {\bibfnamefont {Simon}\
  \bibnamefont {Trebst}}, \bibinfo {author} {\bibfnamefont {Kirill}\
  \bibnamefont {Shtengel}}, \ and\ \bibinfo {author} {\bibfnamefont {Chetan}\
  \bibnamefont {Nayak}},\ }\bibfield  {title} {\enquote {\bibinfo {title}
  {Local interactions and non-abelian quantum loop gases},}\ }\href@noop {}
  {\bibfield  {journal} {\bibinfo  {journal} {Physical review letters}\
  }\textbf {\bibinfo {volume} {101}},\ \bibinfo {pages} {230401} (\bibinfo
  {year} {2008})}\BibitemShut {NoStop}%
\bibitem [{\citenamefont {Dai}\ and\ \citenamefont
  {Nahum}(2020)}]{dai2020quantum}%
  \BibitemOpen
  \bibfield  {author} {\bibinfo {author} {\bibfnamefont {Zhehao}\ \bibnamefont
  {Dai}}\ and\ \bibinfo {author} {\bibfnamefont {Adam}\ \bibnamefont {Nahum}},\
  }\bibfield  {title} {\enquote {\bibinfo {title} {Quantum criticality of loops
  with topologically constrained dynamics},}\ }\href@noop {} {\bibfield
  {journal} {\bibinfo  {journal} {Physical Review Research}\ }\textbf {\bibinfo
  {volume} {2}},\ \bibinfo {pages} {033051} (\bibinfo {year}
  {2020})}\BibitemShut {NoStop}%
\bibitem [{\citenamefont {Nelson}\ \emph {et~al.}(2004)\citenamefont {Nelson},
  \citenamefont {Piran},\ and\ \citenamefont
  {Weinberg}}]{nelson2004statistical}%
  \BibitemOpen
  \bibfield  {author} {\bibinfo {author} {\bibfnamefont {David~R}\ \bibnamefont
  {Nelson}}, \bibinfo {author} {\bibfnamefont {Tsvi}\ \bibnamefont {Piran}}, \
  and\ \bibinfo {author} {\bibfnamefont {Steven}\ \bibnamefont {Weinberg}},\
  }\href@noop {} {\emph {\bibinfo {title} {Statistical mechanics of membranes
  and surfaces}}}\ (\bibinfo  {publisher} {World Scientific},\ \bibinfo {year}
  {2004})\BibitemShut {NoStop}%
\bibitem [{\citenamefont {Polchinski}(1998)}]{polchinski1998string}%
  \BibitemOpen
  \bibfield  {author} {\bibinfo {author} {\bibfnamefont {Joseph}\ \bibnamefont
  {Polchinski}},\ }\href@noop {} {\emph {\bibinfo {title} {String theory:
  Volume 1, an introduction to the bosonic string}}}\ (\bibinfo  {publisher}
  {Cambridge university press},\ \bibinfo {year} {1998})\BibitemShut {NoStop}%
\bibitem [{\citenamefont {Gregor}\ \emph {et~al.}(2011)\citenamefont {Gregor},
  \citenamefont {Huse}, \citenamefont {Moessner},\ and\ \citenamefont
  {Sondhi}}]{gregor2011diagnosing}%
  \BibitemOpen
  \bibfield  {author} {\bibinfo {author} {\bibfnamefont {K}~\bibnamefont
  {Gregor}}, \bibinfo {author} {\bibfnamefont {David~A}\ \bibnamefont {Huse}},
  \bibinfo {author} {\bibfnamefont {R}~\bibnamefont {Moessner}}, \ and\
  \bibinfo {author} {\bibfnamefont {Shivaji~Lal}\ \bibnamefont {Sondhi}},\
  }\bibfield  {title} {\enquote {\bibinfo {title} {Diagnosing deconfinement and
  topological order},}\ }\href@noop {} {\bibfield  {journal} {\bibinfo
  {journal} {New Journal of Physics}\ }\textbf {\bibinfo {volume} {13}},\
  \bibinfo {pages} {025009} (\bibinfo {year} {2011})}\BibitemShut {NoStop}%
\bibitem [{\citenamefont {Wang}\ \emph {et~al.}(2015)\citenamefont {Wang},
  \citenamefont {Nahum},\ and\ \citenamefont {Senthil}}]{wang2015topological}%
  \BibitemOpen
  \bibfield  {author} {\bibinfo {author} {\bibfnamefont {Chong}\ \bibnamefont
  {Wang}}, \bibinfo {author} {\bibfnamefont {Adam}\ \bibnamefont {Nahum}}, \
  and\ \bibinfo {author} {\bibfnamefont {T}~\bibnamefont {Senthil}},\
  }\bibfield  {title} {\enquote {\bibinfo {title} {Topological paramagnetism in
  frustrated spin-1 mott insulators},}\ }\href@noop {} {\bibfield  {journal}
  {\bibinfo  {journal} {Physical Review B}\ }\textbf {\bibinfo {volume} {91}},\
  \bibinfo {pages} {195131} (\bibinfo {year} {2015})}\BibitemShut {NoStop}%
\bibitem [{\citenamefont {Geraedts}\ and\ \citenamefont
  {Motrunich}()}]{geraedtsunpublished}%
  \BibitemOpen
  \bibfield  {author} {\bibinfo {author} {\bibfnamefont {S}~\bibnamefont
  {Geraedts}}\ and\ \bibinfo {author} {\bibfnamefont {O}~\bibnamefont
  {Motrunich}},\ }\bibfield  {title} {\enquote {\bibinfo {title}
  {Unpublished},}\ }\href@noop {} {\ }\BibitemShut {NoStop}%
\bibitem [{\citenamefont {Von~Keyserlingk}\ \emph {et~al.}(2013)\citenamefont
  {Von~Keyserlingk}, \citenamefont {Burnell},\ and\ \citenamefont
  {Simon}}]{von2013three}%
  \BibitemOpen
  \bibfield  {author} {\bibinfo {author} {\bibfnamefont {CW}~\bibnamefont
  {Von~Keyserlingk}}, \bibinfo {author} {\bibfnamefont {FJ}~\bibnamefont
  {Burnell}}, \ and\ \bibinfo {author} {\bibfnamefont {Steven~H}\ \bibnamefont
  {Simon}},\ }\bibfield  {title} {\enquote {\bibinfo {title} {Three-dimensional
  topological lattice models with surface anyons},}\ }\href@noop {} {\bibfield
  {journal} {\bibinfo  {journal} {Physical Review B}\ }\textbf {\bibinfo
  {volume} {87}},\ \bibinfo {pages} {045107} (\bibinfo {year}
  {2013})}\BibitemShut {NoStop}%
\bibitem [{\citenamefont {Wen}(2003)}]{wen2003quantum}%
  \BibitemOpen
  \bibfield  {author} {\bibinfo {author} {\bibfnamefont {Xiao-Gang}\
  \bibnamefont {Wen}},\ }\bibfield  {title} {\enquote {\bibinfo {title}
  {Quantum orders in an exact soluble model},}\ }\href@noop {} {\bibfield
  {journal} {\bibinfo  {journal} {Physical review letters}\ }\textbf {\bibinfo
  {volume} {90}},\ \bibinfo {pages} {016803} (\bibinfo {year}
  {2003})}\BibitemShut {NoStop}%
\bibitem [{\citenamefont {Nayak}\ \emph {et~al.}(2008)\citenamefont {Nayak},
  \citenamefont {Simon}, \citenamefont {Stern}, \citenamefont {Freedman},\ and\
  \citenamefont {Sarma}}]{nayak2008non}%
  \BibitemOpen
  \bibfield  {author} {\bibinfo {author} {\bibfnamefont {Chetan}\ \bibnamefont
  {Nayak}}, \bibinfo {author} {\bibfnamefont {Steven~H}\ \bibnamefont {Simon}},
  \bibinfo {author} {\bibfnamefont {Ady}\ \bibnamefont {Stern}}, \bibinfo
  {author} {\bibfnamefont {Michael}\ \bibnamefont {Freedman}}, \ and\ \bibinfo
  {author} {\bibfnamefont {Sankar~Das}\ \bibnamefont {Sarma}},\ }\bibfield
  {title} {\enquote {\bibinfo {title} {Non-abelian anyons and topological
  quantum computation},}\ }\href@noop {} {\bibfield  {journal} {\bibinfo
  {journal} {Reviews of Modern Physics}\ }\textbf {\bibinfo {volume} {80}},\
  \bibinfo {pages} {1083} (\bibinfo {year} {2008})}\BibitemShut {NoStop}%
\bibitem [{\citenamefont {Dusuel}\ \emph {et~al.}(2011)\citenamefont {Dusuel},
  \citenamefont {Kamfor}, \citenamefont {Or{\'u}s}, \citenamefont {Schmidt},\
  and\ \citenamefont {Vidal}}]{dusuel2011robustness}%
  \BibitemOpen
  \bibfield  {author} {\bibinfo {author} {\bibfnamefont {S{\'e}bastien}\
  \bibnamefont {Dusuel}}, \bibinfo {author} {\bibfnamefont {Michael}\
  \bibnamefont {Kamfor}}, \bibinfo {author} {\bibfnamefont {Rom{\'a}n}\
  \bibnamefont {Or{\'u}s}}, \bibinfo {author} {\bibfnamefont {Kai~Phillip}\
  \bibnamefont {Schmidt}}, \ and\ \bibinfo {author} {\bibfnamefont {Julien}\
  \bibnamefont {Vidal}},\ }\bibfield  {title} {\enquote {\bibinfo {title}
  {Robustness of a perturbed topological phase},}\ }\href@noop {} {\bibfield
  {journal} {\bibinfo  {journal} {Physical review letters}\ }\textbf {\bibinfo
  {volume} {106}},\ \bibinfo {pages} {107203} (\bibinfo {year}
  {2011})}\BibitemShut {NoStop}%
\bibitem [{\citenamefont {Nahum}\ \emph {et~al.}()\citenamefont {Nahum},
  \citenamefont {Serna},\ and\ \citenamefont {Somoza}}]{membranesforthcoming}%
  \BibitemOpen
  \bibfield  {author} {\bibinfo {author} {\bibfnamefont {A}~\bibnamefont
  {Nahum}}, \bibinfo {author} {\bibfnamefont {P}~\bibnamefont {Serna}}, \ and\
  \bibinfo {author} {\bibfnamefont {A}~\bibnamefont {Somoza}},\ }\href@noop {}
  {}\bibinfo {note} {\textit{in preparation}}\BibitemShut {NoStop}%
\bibitem [{\citenamefont {Metlitski}\ and\ \citenamefont
  {Thorngren}(2018)}]{metlitski2018intrinsic}%
  \BibitemOpen
  \bibfield  {author} {\bibinfo {author} {\bibfnamefont {Max~A}\ \bibnamefont
  {Metlitski}}\ and\ \bibinfo {author} {\bibfnamefont {Ryan}\ \bibnamefont
  {Thorngren}},\ }\bibfield  {title} {\enquote {\bibinfo {title} {Intrinsic and
  emergent anomalies at deconfined critical points},}\ }\href@noop {}
  {\bibfield  {journal} {\bibinfo  {journal} {Physical Review B}\ }\textbf
  {\bibinfo {volume} {98}},\ \bibinfo {pages} {085140} (\bibinfo {year}
  {2018})}\BibitemShut {NoStop}%
\bibitem [{\citenamefont {Heinrich}\ \emph {et~al.}(2016)\citenamefont
  {Heinrich}, \citenamefont {Burnell}, \citenamefont {Fidkowski},\ and\
  \citenamefont {Levin}}]{heinrich2016symmetry}%
  \BibitemOpen
  \bibfield  {author} {\bibinfo {author} {\bibfnamefont {Chris}\ \bibnamefont
  {Heinrich}}, \bibinfo {author} {\bibfnamefont {Fiona}\ \bibnamefont
  {Burnell}}, \bibinfo {author} {\bibfnamefont {Lukasz}\ \bibnamefont
  {Fidkowski}}, \ and\ \bibinfo {author} {\bibfnamefont {Michael}\ \bibnamefont
  {Levin}},\ }\bibfield  {title} {\enquote {\bibinfo {title} {Symmetry-enriched
  string nets: Exactly solvable models for set phases},}\ }\href@noop {}
  {\bibfield  {journal} {\bibinfo  {journal} {Physical Review B}\ }\textbf
  {\bibinfo {volume} {94}},\ \bibinfo {pages} {235136} (\bibinfo {year}
  {2016})}\BibitemShut {NoStop}%
\bibitem [{\citenamefont {Cheng}\ \emph {et~al.}(2017)\citenamefont {Cheng},
  \citenamefont {Gu}, \citenamefont {Jiang},\ and\ \citenamefont
  {Qi}}]{cheng2017exactly}%
  \BibitemOpen
  \bibfield  {author} {\bibinfo {author} {\bibfnamefont {Meng}\ \bibnamefont
  {Cheng}}, \bibinfo {author} {\bibfnamefont {Zheng-Cheng}\ \bibnamefont {Gu}},
  \bibinfo {author} {\bibfnamefont {Shenghan}\ \bibnamefont {Jiang}}, \ and\
  \bibinfo {author} {\bibfnamefont {Yang}\ \bibnamefont {Qi}},\ }\bibfield
  {title} {\enquote {\bibinfo {title} {Exactly solvable models for
  symmetry-enriched topological phases},}\ }\href@noop {} {\bibfield  {journal}
  {\bibinfo  {journal} {Physical Review B}\ }\textbf {\bibinfo {volume} {96}},\
  \bibinfo {pages} {115107} (\bibinfo {year} {2017})}\BibitemShut {NoStop}%
\bibitem [{\citenamefont {Aasen}\ \emph {et~al.}(2016)\citenamefont {Aasen},
  \citenamefont {Mong},\ and\ \citenamefont {Fendley}}]{aasen2016topological}%
  \BibitemOpen
  \bibfield  {author} {\bibinfo {author} {\bibfnamefont {David}\ \bibnamefont
  {Aasen}}, \bibinfo {author} {\bibfnamefont {Roger~SK}\ \bibnamefont {Mong}},
  \ and\ \bibinfo {author} {\bibfnamefont {Paul}\ \bibnamefont {Fendley}},\
  }\bibfield  {title} {\enquote {\bibinfo {title} {Topological defects on the
  lattice: I. the ising model},}\ }\href@noop {} {\bibfield  {journal}
  {\bibinfo  {journal} {Journal of Physics A: Mathematical and Theoretical}\
  }\textbf {\bibinfo {volume} {49}},\ \bibinfo {pages} {354001} (\bibinfo
  {year} {2016})}\BibitemShut {NoStop}%
\bibitem [{\citenamefont {Karch}\ \emph {et~al.}(2019)\citenamefont {Karch},
  \citenamefont {Tong},\ and\ \citenamefont {Turner}}]{karch2019web}%
  \BibitemOpen
  \bibfield  {author} {\bibinfo {author} {\bibfnamefont {Andreas}\ \bibnamefont
  {Karch}}, \bibinfo {author} {\bibfnamefont {David}\ \bibnamefont {Tong}}, \
  and\ \bibinfo {author} {\bibfnamefont {Carl}\ \bibnamefont {Turner}},\
  }\bibfield  {title} {\enquote {\bibinfo {title} {{A Web of 2d Dualities:
  ${\bf Z}_2$ Gauge Fields and Arf Invariants}},}\ }\href {\doibase
  10.21468/SciPostPhys.7.1.007} {\bibfield  {journal} {\bibinfo  {journal}
  {SciPost Phys.}\ }\textbf {\bibinfo {volume} {7}},\ \bibinfo {pages} {7}
  (\bibinfo {year} {2019})}\BibitemShut {NoStop}%
\bibitem [{\citenamefont {Jones}\ and\ \citenamefont
  {Metlitski}(2019)}]{jones20191d}%
  \BibitemOpen
  \bibfield  {author} {\bibinfo {author} {\bibfnamefont {Robert~A}\
  \bibnamefont {Jones}}\ and\ \bibinfo {author} {\bibfnamefont {Max~A}\
  \bibnamefont {Metlitski}},\ }\bibfield  {title} {\enquote {\bibinfo {title}
  {1d lattice models for the boundary of 2d" majorana" fermion spts:
  Kramers-wannier duality as an exact $ z\_2 $ symmetry},}\ }\href@noop {}
  {\bibfield  {journal} {\bibinfo  {journal} {arXiv preprint arXiv:1902.05957}\
  } (\bibinfo {year} {2019})}\BibitemShut {NoStop}%
\bibitem [{\citenamefont {Binder}(1981)}]{binder1981critical}%
  \BibitemOpen
  \bibfield  {author} {\bibinfo {author} {\bibfnamefont {Kurt}\ \bibnamefont
  {Binder}},\ }\bibfield  {title} {\enquote {\bibinfo {title} {Critical
  properties from monte carlo coarse graining and renormalization},}\
  }\href@noop {} {\bibfield  {journal} {\bibinfo  {journal} {Physical Review
  Letters}\ }\textbf {\bibinfo {volume} {47}},\ \bibinfo {pages} {693}
  (\bibinfo {year} {1981})}\BibitemShut {NoStop}%
\bibitem [{\citenamefont {Cardy}(1996)}]{cardy1996scaling}%
  \BibitemOpen
  \bibfield  {author} {\bibinfo {author} {\bibfnamefont {John}\ \bibnamefont
  {Cardy}},\ }\href@noop {} {\emph {\bibinfo {title} {Scaling and
  renormalization in statistical physics}}},\ Vol.~\bibinfo {volume} {5}\
  (\bibinfo  {publisher} {Cambridge university press},\ \bibinfo {year}
  {1996})\BibitemShut {NoStop}%
\bibitem [{\citenamefont {Caselle}\ \emph {et~al.}(2015)\citenamefont
  {Caselle}, \citenamefont {Costagliola},\ and\ \citenamefont
  {Magnoli}}]{caselle2015numerical}%
  \BibitemOpen
  \bibfield  {author} {\bibinfo {author} {\bibfnamefont {Michele}\ \bibnamefont
  {Caselle}}, \bibinfo {author} {\bibfnamefont {Gianluca}\ \bibnamefont
  {Costagliola}}, \ and\ \bibinfo {author} {\bibfnamefont {N}~\bibnamefont
  {Magnoli}},\ }\bibfield  {title} {\enquote {\bibinfo {title} {Numerical
  determination of the operator-product-expansion coefficients in the 3d ising
  model from off-critical correlators},}\ }\href@noop {} {\bibfield  {journal}
  {\bibinfo  {journal} {Physical Review D}\ }\textbf {\bibinfo {volume} {91}},\
  \bibinfo {pages} {061901} (\bibinfo {year} {2015})}\BibitemShut {NoStop}%
\bibitem [{\citenamefont {Hasenbusch}(2020{\natexlab{a}})}]{hasenbusch2020two}%
  \BibitemOpen
  \bibfield  {author} {\bibinfo {author} {\bibfnamefont {Martin}\ \bibnamefont
  {Hasenbusch}},\ }\bibfield  {title} {\enquote {\bibinfo {title} {Two-and
  three-point functions at criticality: Monte carlo simulations of the
  three-dimensional $(q+ 1) $-state clock model},}\ }\href@noop {} {\bibfield
  {journal} {\bibinfo  {journal} {arXiv preprint arXiv:2010.05699}\ } (\bibinfo
  {year} {2020}{\natexlab{a}})}\BibitemShut {NoStop}%
\bibitem [{\citenamefont {Glauber}(1963)}]{glauber1963time}%
  \BibitemOpen
  \bibfield  {author} {\bibinfo {author} {\bibfnamefont {Roy~J}\ \bibnamefont
  {Glauber}},\ }\bibfield  {title} {\enquote {\bibinfo {title} {Time-dependent
  statistics of the ising model},}\ }\href@noop {} {\bibfield  {journal}
  {\bibinfo  {journal} {Journal of mathematical physics}\ }\textbf {\bibinfo
  {volume} {4}},\ \bibinfo {pages} {294--307} (\bibinfo {year}
  {1963})}\BibitemShut {NoStop}%
\bibitem [{\citenamefont {Hohenberg}\ and\ \citenamefont
  {Halperin}(1977)}]{hohenberg1977theory}%
  \BibitemOpen
  \bibfield  {author} {\bibinfo {author} {\bibfnamefont {Pierre~C}\
  \bibnamefont {Hohenberg}}\ and\ \bibinfo {author} {\bibfnamefont
  {Bertrand~I}\ \bibnamefont {Halperin}},\ }\bibfield  {title} {\enquote
  {\bibinfo {title} {Theory of dynamic critical phenomena},}\ }\href@noop {}
  {\bibfield  {journal} {\bibinfo  {journal} {Reviews of Modern Physics}\
  }\textbf {\bibinfo {volume} {49}},\ \bibinfo {pages} {435} (\bibinfo {year}
  {1977})}\BibitemShut {NoStop}%
\bibitem [{\citenamefont {Wansleben}\ and\ \citenamefont
  {Landau}(1987)}]{wansleben1987dynamical}%
  \BibitemOpen
  \bibfield  {author} {\bibinfo {author} {\bibfnamefont {S}~\bibnamefont
  {Wansleben}}\ and\ \bibinfo {author} {\bibfnamefont {DP}~\bibnamefont
  {Landau}},\ }\bibfield  {title} {\enquote {\bibinfo {title} {Dynamical
  critical exponent of the 3d ising model},}\ }\href@noop {} {\bibfield
  {journal} {\bibinfo  {journal} {Journal of Applied Physics}\ }\textbf
  {\bibinfo {volume} {61}},\ \bibinfo {pages} {3968--3970} (\bibinfo {year}
  {1987})}\BibitemShut {NoStop}%
\bibitem [{\citenamefont {Wansleben}\ and\ \citenamefont
  {Landau}(1991)}]{wansleben1991monte}%
  \BibitemOpen
  \bibfield  {author} {\bibinfo {author} {\bibfnamefont {S}~\bibnamefont
  {Wansleben}}\ and\ \bibinfo {author} {\bibfnamefont {DP}~\bibnamefont
  {Landau}},\ }\bibfield  {title} {\enquote {\bibinfo {title} {Monte carlo
  investigation of critical dynamics in the three-dimensional ising model},}\
  }\href@noop {} {\bibfield  {journal} {\bibinfo  {journal} {Physical Review
  B}\ }\textbf {\bibinfo {volume} {43}},\ \bibinfo {pages} {6006} (\bibinfo
  {year} {1991})}\BibitemShut {NoStop}%
\bibitem [{\citenamefont {M{\"u}nkel}\ \emph {et~al.}(1993)\citenamefont
  {M{\"u}nkel}, \citenamefont {Heermann}, \citenamefont {Adler}, \citenamefont
  {Gofman},\ and\ \citenamefont {Stauffer}}]{munkel1993dynamical}%
  \BibitemOpen
  \bibfield  {author} {\bibinfo {author} {\bibfnamefont {Christian}\
  \bibnamefont {M{\"u}nkel}}, \bibinfo {author} {\bibfnamefont {Dieter~W}\
  \bibnamefont {Heermann}}, \bibinfo {author} {\bibfnamefont {Joan}\
  \bibnamefont {Adler}}, \bibinfo {author} {\bibfnamefont {Misha}\ \bibnamefont
  {Gofman}}, \ and\ \bibinfo {author} {\bibfnamefont {Dietrich}\ \bibnamefont
  {Stauffer}},\ }\bibfield  {title} {\enquote {\bibinfo {title} {The dynamical
  critical exponent of the two-, three-and five-dimensional kinetic ising
  model},}\ }\href@noop {} {\bibfield  {journal} {\bibinfo  {journal} {Physica
  A: Statistical Mechanics and its Applications}\ }\textbf {\bibinfo {volume}
  {193}},\ \bibinfo {pages} {540--552} (\bibinfo {year} {1993})}\BibitemShut
  {NoStop}%
\bibitem [{\citenamefont {Ito}(1993)}]{ito1993non}%
  \BibitemOpen
  \bibfield  {author} {\bibinfo {author} {\bibfnamefont {Nobuyasu}\
  \bibnamefont {Ito}},\ }\bibfield  {title} {\enquote {\bibinfo {title}
  {Non-equilibrium critical relaxation of the three-dimensional ising model},}\
  }\href@noop {} {\bibfield  {journal} {\bibinfo  {journal} {Physica A:
  Statistical Mechanics and its Applications}\ }\textbf {\bibinfo {volume}
  {192}},\ \bibinfo {pages} {604--616} (\bibinfo {year} {1993})}\BibitemShut
  {NoStop}%
\bibitem [{\citenamefont {Grassberger}(1995)}]{grassberger1995damage}%
  \BibitemOpen
  \bibfield  {author} {\bibinfo {author} {\bibfnamefont {Peter}\ \bibnamefont
  {Grassberger}},\ }\bibfield  {title} {\enquote {\bibinfo {title} {Damage
  spreading and critical exponents for “model a” ising dynamics},}\
  }\href@noop {} {\bibfield  {journal} {\bibinfo  {journal} {Physica A:
  Statistical Mechanics and its Applications}\ }\textbf {\bibinfo {volume}
  {214}},\ \bibinfo {pages} {547--559} (\bibinfo {year} {1995})}\BibitemShut
  {NoStop}%
\bibitem [{\citenamefont {Jaster}\ \emph {et~al.}(1999)\citenamefont {Jaster},
  \citenamefont {Mainville}, \citenamefont {Sch{\"u}lke},\ and\ \citenamefont
  {Zheng}}]{jaster1999short}%
  \BibitemOpen
  \bibfield  {author} {\bibinfo {author} {\bibfnamefont {A}~\bibnamefont
  {Jaster}}, \bibinfo {author} {\bibfnamefont {J}~\bibnamefont {Mainville}},
  \bibinfo {author} {\bibfnamefont {L}~\bibnamefont {Sch{\"u}lke}}, \ and\
  \bibinfo {author} {\bibfnamefont {B}~\bibnamefont {Zheng}},\ }\bibfield
  {title} {\enquote {\bibinfo {title} {Short-time critical dynamics of the
  three-dimensional ising model},}\ }\href@noop {} {\bibfield  {journal}
  {\bibinfo  {journal} {Journal of Physics A: Mathematical and General}\
  }\textbf {\bibinfo {volume} {32}},\ \bibinfo {pages} {1395} (\bibinfo {year}
  {1999})}\BibitemShut {NoStop}%
\bibitem [{\citenamefont {Ito}\ \emph {et~al.}(2000)\citenamefont {Ito},
  \citenamefont {Hukushima}, \citenamefont {Ogawa},\ and\ \citenamefont
  {Ozeki}}]{ito2000nonequilibrium}%
  \BibitemOpen
  \bibfield  {author} {\bibinfo {author} {\bibfnamefont {Nobuyasu}\
  \bibnamefont {Ito}}, \bibinfo {author} {\bibfnamefont {Koji}\ \bibnamefont
  {Hukushima}}, \bibinfo {author} {\bibfnamefont {Keita}\ \bibnamefont
  {Ogawa}}, \ and\ \bibinfo {author} {\bibfnamefont {Yukiyasu}\ \bibnamefont
  {Ozeki}},\ }\bibfield  {title} {\enquote {\bibinfo {title} {Nonequilibrium
  relaxation of fluctuations of physical quantities},}\ }\href@noop {}
  {\bibfield  {journal} {\bibinfo  {journal} {Journal of the Physical Society
  of Japan}\ }\textbf {\bibinfo {volume} {69}},\ \bibinfo {pages} {1931--1934}
  (\bibinfo {year} {2000})}\BibitemShut {NoStop}%
\bibitem [{\citenamefont {Murase}\ and\ \citenamefont
  {Ito}(2007)}]{murase2007dynamic}%
  \BibitemOpen
  \bibfield  {author} {\bibinfo {author} {\bibfnamefont {Yohsuke}\ \bibnamefont
  {Murase}}\ and\ \bibinfo {author} {\bibfnamefont {Nobuyasu}\ \bibnamefont
  {Ito}},\ }\bibfield  {title} {\enquote {\bibinfo {title} {Dynamic critical
  exponents of three-dimensional ising models and two-dimensional three-states
  potts models},}\ }\href@noop {} {\bibfield  {journal} {\bibinfo  {journal}
  {Journal of the Physical Society of Japan}\ }\textbf {\bibinfo {volume}
  {77}},\ \bibinfo {pages} {014002} (\bibinfo {year} {2007})}\BibitemShut
  {NoStop}%
\bibitem [{\citenamefont {Collura}(2010)}]{collura2010off}%
  \BibitemOpen
  \bibfield  {author} {\bibinfo {author} {\bibfnamefont {Mario}\ \bibnamefont
  {Collura}},\ }\bibfield  {title} {\enquote {\bibinfo {title} {Off-equilibrium
  relaxational dynamics with an improved ising hamiltonian},}\ }\href@noop {}
  {\bibfield  {journal} {\bibinfo  {journal} {Journal of Statistical Mechanics:
  Theory and Experiment}\ }\textbf {\bibinfo {volume} {2010}},\ \bibinfo
  {pages} {P12036} (\bibinfo {year} {2010})}\BibitemShut {NoStop}%
\bibitem [{\citenamefont {Niermann}\ \emph {et~al.}(2015)\citenamefont
  {Niermann}, \citenamefont {Grams}, \citenamefont {Becker}, \citenamefont
  {Bohat{\`y}}, \citenamefont {Schenck},\ and\ \citenamefont
  {Hemberger}}]{niermann2015critical}%
  \BibitemOpen
  \bibfield  {author} {\bibinfo {author} {\bibfnamefont {D}~\bibnamefont
  {Niermann}}, \bibinfo {author} {\bibfnamefont {CP}~\bibnamefont {Grams}},
  \bibinfo {author} {\bibfnamefont {P}~\bibnamefont {Becker}}, \bibinfo
  {author} {\bibfnamefont {L}~\bibnamefont {Bohat{\`y}}}, \bibinfo {author}
  {\bibfnamefont {H}~\bibnamefont {Schenck}}, \ and\ \bibinfo {author}
  {\bibfnamefont {J}~\bibnamefont {Hemberger}},\ }\bibfield  {title} {\enquote
  {\bibinfo {title} {Critical slowing down near the multiferroic phase
  transition in mnwo 4},}\ }\href@noop {} {\bibfield  {journal} {\bibinfo
  {journal} {Physical Review Letters}\ }\textbf {\bibinfo {volume} {114}},\
  \bibinfo {pages} {037204} (\bibinfo {year} {2015})}\BibitemShut {NoStop}%
\bibitem [{\citenamefont {Mesterhazy}\ \emph {et~al.}(2015)\citenamefont
  {Mesterhazy}, \citenamefont {Stockemer},\ and\ \citenamefont
  {Tanizaki}}]{mesterhazy2015quantum}%
  \BibitemOpen
  \bibfield  {author} {\bibinfo {author} {\bibfnamefont {David}\ \bibnamefont
  {Mesterhazy}}, \bibinfo {author} {\bibfnamefont {Jan~H}\ \bibnamefont
  {Stockemer}}, \ and\ \bibinfo {author} {\bibfnamefont {Yuya}\ \bibnamefont
  {Tanizaki}},\ }\bibfield  {title} {\enquote {\bibinfo {title} {From quantum
  to classical dynamics: The relativistic o (n) model in the framework of the
  real-time functional renormalization group},}\ }\href@noop {} {\bibfield
  {journal} {\bibinfo  {journal} {Physical Review D}\ }\textbf {\bibinfo
  {volume} {92}},\ \bibinfo {pages} {076001} (\bibinfo {year}
  {2015})}\BibitemShut {NoStop}%
\bibitem [{\citenamefont {Duclut}\ and\ \citenamefont
  {Delamotte}(2017)}]{duclut2017frequency}%
  \BibitemOpen
  \bibfield  {author} {\bibinfo {author} {\bibfnamefont {Charlie}\ \bibnamefont
  {Duclut}}\ and\ \bibinfo {author} {\bibfnamefont {Bertrand}\ \bibnamefont
  {Delamotte}},\ }\bibfield  {title} {\enquote {\bibinfo {title} {Frequency
  regulators for the nonperturbative renormalization group: A general study and
  the model a as a benchmark},}\ }\href@noop {} {\bibfield  {journal} {\bibinfo
   {journal} {Physical Review E}\ }\textbf {\bibinfo {volume} {95}},\ \bibinfo
  {pages} {012107} (\bibinfo {year} {2017})}\BibitemShut {NoStop}%
\bibitem [{\citenamefont {Adzhemyan}\ \emph {et~al.}(2018)\citenamefont
  {Adzhemyan}, \citenamefont {Ivanova}, \citenamefont {Kompaniets},\ and\
  \citenamefont {Vorobyeva}}]{adzhemyan2018diagram}%
  \BibitemOpen
  \bibfield  {author} {\bibinfo {author} {\bibfnamefont {L~Ts}\ \bibnamefont
  {Adzhemyan}}, \bibinfo {author} {\bibfnamefont {EV}~\bibnamefont {Ivanova}},
  \bibinfo {author} {\bibfnamefont {MV}~\bibnamefont {Kompaniets}}, \ and\
  \bibinfo {author} {\bibfnamefont {S~Ye}\ \bibnamefont {Vorobyeva}},\
  }\bibfield  {title} {\enquote {\bibinfo {title} {Diagram reduction in problem
  of critical dynamics of ferromagnets: 4-loop approximation},}\ }\href@noop {}
  {\bibfield  {journal} {\bibinfo  {journal} {Journal of Physics A:
  Mathematical and Theoretical}\ }\textbf {\bibinfo {volume} {51}},\ \bibinfo
  {pages} {155003} (\bibinfo {year} {2018})}\BibitemShut {NoStop}%
\bibitem [{\citenamefont
  {Hasenbusch}(2020{\natexlab{b}})}]{hasenbusch2020dynamic}%
  \BibitemOpen
  \bibfield  {author} {\bibinfo {author} {\bibfnamefont {Martin}\ \bibnamefont
  {Hasenbusch}},\ }\bibfield  {title} {\enquote {\bibinfo {title} {Dynamic
  critical exponent z of the three-dimensional ising universality class: Monte
  carlo simulations of the improved blume-capel model},}\ }\href@noop {}
  {\bibfield  {journal} {\bibinfo  {journal} {Physical Review E}\ }\textbf
  {\bibinfo {volume} {101}},\ \bibinfo {pages} {022126} (\bibinfo {year}
  {2020}{\natexlab{b}})}\BibitemShut {NoStop}%
\bibitem [{\citenamefont {Shi}\ \emph {et~al.}(2011)\citenamefont {Shi},
  \citenamefont {Lamacraft},\ and\ \citenamefont {Fendley}}]{shi2011boson}%
  \BibitemOpen
  \bibfield  {author} {\bibinfo {author} {\bibfnamefont {Yifei}\ \bibnamefont
  {Shi}}, \bibinfo {author} {\bibfnamefont {Austen}\ \bibnamefont {Lamacraft}},
  \ and\ \bibinfo {author} {\bibfnamefont {Paul}\ \bibnamefont {Fendley}},\
  }\bibfield  {title} {\enquote {\bibinfo {title} {Boson pairing and unusual
  criticality in a generalized x y model},}\ }\href@noop {} {\bibfield
  {journal} {\bibinfo  {journal} {Physical review letters}\ }\textbf {\bibinfo
  {volume} {107}},\ \bibinfo {pages} {240601} (\bibinfo {year}
  {2011})}\BibitemShut {NoStop}%
\bibitem [{\citenamefont {Hastings}\ and\ \citenamefont
  {Wen}(2005)}]{hastings2005quasiadiabatic}%
  \BibitemOpen
  \bibfield  {author} {\bibinfo {author} {\bibfnamefont {Matthew~B}\
  \bibnamefont {Hastings}}\ and\ \bibinfo {author} {\bibfnamefont {Xiao-Gang}\
  \bibnamefont {Wen}},\ }\bibfield  {title} {\enquote {\bibinfo {title}
  {Quasiadiabatic continuation of quantum states: The stability of topological
  ground-state degeneracy and emergent gauge invariance},}\ }\href@noop {}
  {\bibfield  {journal} {\bibinfo  {journal} {Physical review b}\ }\textbf
  {\bibinfo {volume} {72}},\ \bibinfo {pages} {045141} (\bibinfo {year}
  {2005})}\BibitemShut {NoStop}%
\bibitem [{\citenamefont {Fredenhagen}\ and\ \citenamefont
  {Marcu}(1986)}]{fredenhagen1986confinement}%
  \BibitemOpen
  \bibfield  {author} {\bibinfo {author} {\bibfnamefont {Klaus}\ \bibnamefont
  {Fredenhagen}}\ and\ \bibinfo {author} {\bibfnamefont {Mihail}\ \bibnamefont
  {Marcu}},\ }\bibfield  {title} {\enquote {\bibinfo {title} {Confinement
  criterion for qcd with dynamical quarks},}\ }\href@noop {} {\bibfield
  {journal} {\bibinfo  {journal} {Physical review letters}\ }\textbf {\bibinfo
  {volume} {56}},\ \bibinfo {pages} {223} (\bibinfo {year} {1986})}\BibitemShut
  {NoStop}%
\bibitem [{\citenamefont {Nahum}\ and\ \citenamefont
  {Chalker}(2012)}]{nahum2012universal}%
  \BibitemOpen
  \bibfield  {author} {\bibinfo {author} {\bibfnamefont {Adam}\ \bibnamefont
  {Nahum}}\ and\ \bibinfo {author} {\bibfnamefont {JT}~\bibnamefont
  {Chalker}},\ }\bibfield  {title} {\enquote {\bibinfo {title} {Universal
  statistics of vortex lines},}\ }\href@noop {} {\bibfield  {journal} {\bibinfo
   {journal} {Physical Review E}\ }\textbf {\bibinfo {volume} {85}},\ \bibinfo
  {pages} {031141} (\bibinfo {year} {2012})}\BibitemShut {NoStop}%
\bibitem [{\citenamefont {Winter}\ \emph {et~al.}(2008)\citenamefont {Winter},
  \citenamefont {Janke},\ and\ \citenamefont {Schakel}}]{winter2008geometric}%
  \BibitemOpen
  \bibfield  {author} {\bibinfo {author} {\bibfnamefont {Frank}\ \bibnamefont
  {Winter}}, \bibinfo {author} {\bibfnamefont {Wolfhard}\ \bibnamefont
  {Janke}}, \ and\ \bibinfo {author} {\bibfnamefont {Adriaan~MJ}\ \bibnamefont
  {Schakel}},\ }\bibfield  {title} {\enquote {\bibinfo {title} {Geometric
  properties of the three-dimensional ising and x y models},}\ }\href@noop {}
  {\bibfield  {journal} {\bibinfo  {journal} {Physical Review E}\ }\textbf
  {\bibinfo {volume} {77}},\ \bibinfo {pages} {061108} (\bibinfo {year}
  {2008})}\BibitemShut {NoStop}%
\bibitem [{\citenamefont {Kompaniets}\ and\ \citenamefont
  {Wiese}(2020)}]{kompaniets2020fractal}%
  \BibitemOpen
  \bibfield  {author} {\bibinfo {author} {\bibfnamefont {Mikhail}\ \bibnamefont
  {Kompaniets}}\ and\ \bibinfo {author} {\bibfnamefont {Kay~J{\"o}rg}\
  \bibnamefont {Wiese}},\ }\bibfield  {title} {\enquote {\bibinfo {title}
  {Fractal dimension of critical curves in the o (n)-symmetric $\phi$ 4 model
  and crossover exponent at 6-loop order: Loop-erased random walks,
  self-avoiding walks, ising, x y, and heisenberg models},}\ }\href@noop {}
  {\bibfield  {journal} {\bibinfo  {journal} {Physical Review E}\ }\textbf
  {\bibinfo {volume} {101}},\ \bibinfo {pages} {012104} (\bibinfo {year}
  {2020})}\BibitemShut {NoStop}%
\bibitem [{\citenamefont {Nienhuis}(1982)}]{nienhuis1982exact}%
  \BibitemOpen
  \bibfield  {author} {\bibinfo {author} {\bibfnamefont {Bernard}\ \bibnamefont
  {Nienhuis}},\ }\bibfield  {title} {\enquote {\bibinfo {title} {Exact critical
  point and critical exponents of o (n) models in two dimensions},}\
  }\href@noop {} {\bibfield  {journal} {\bibinfo  {journal} {Physical Review
  Letters}\ }\textbf {\bibinfo {volume} {49}},\ \bibinfo {pages} {1062}
  (\bibinfo {year} {1982})}\BibitemShut {NoStop}%
\bibitem [{\citenamefont {Kang}\ \emph {et~al.}(2020)\citenamefont {Kang},
  \citenamefont {Parameswaran}, \citenamefont {Potter}, \citenamefont
  {Vasseur},\ and\ \citenamefont {Gazit}}]{kang2020superuniversality}%
  \BibitemOpen
  \bibfield  {author} {\bibinfo {author} {\bibfnamefont {Byungmin}\
  \bibnamefont {Kang}}, \bibinfo {author} {\bibfnamefont {SA}~\bibnamefont
  {Parameswaran}}, \bibinfo {author} {\bibfnamefont {Andrew~C}\ \bibnamefont
  {Potter}}, \bibinfo {author} {\bibfnamefont {Romain}\ \bibnamefont
  {Vasseur}}, \ and\ \bibinfo {author} {\bibfnamefont {Snir}\ \bibnamefont
  {Gazit}},\ }\bibfield  {title} {\enquote {\bibinfo {title} {Superuniversality
  from disorder at two-dimensional topological phase transitions},}\
  }\href@noop {} {\bibfield  {journal} {\bibinfo  {journal} {arXiv preprint
  arXiv:2008.09617}\ } (\bibinfo {year} {2020})}\BibitemShut {NoStop}%
\bibitem [{\citenamefont {Svetitsky}(1986)}]{svetitsky1986symmetry}%
  \BibitemOpen
  \bibfield  {author} {\bibinfo {author} {\bibfnamefont {Benjamin}\
  \bibnamefont {Svetitsky}},\ }\bibfield  {title} {\enquote {\bibinfo {title}
  {Symmetry aspects of finite-temperature confinement transitions},}\
  }\href@noop {} {\bibfield  {journal} {\bibinfo  {journal} {Physics Reports}\
  }\textbf {\bibinfo {volume} {132}},\ \bibinfo {pages} {1--53} (\bibinfo
  {year} {1986})}\BibitemShut {NoStop}%
\bibitem [{\citenamefont {Genovese}\ \emph {et~al.}(2003)\citenamefont
  {Genovese}, \citenamefont {Gliozzi}, \citenamefont {Rago},\ and\
  \citenamefont {Torrero}}]{genovese2003phase}%
  \BibitemOpen
  \bibfield  {author} {\bibinfo {author} {\bibfnamefont {Luigi}\ \bibnamefont
  {Genovese}}, \bibinfo {author} {\bibfnamefont {Ferdinando}\ \bibnamefont
  {Gliozzi}}, \bibinfo {author} {\bibfnamefont {Antonio}\ \bibnamefont {Rago}},
  \ and\ \bibinfo {author} {\bibfnamefont {Christian}\ \bibnamefont
  {Torrero}},\ }\bibfield  {title} {\enquote {\bibinfo {title} {The phase
  diagram of the three-dimensionalz2 gauge higgs system at zero and finite
  temperature},}\ }\href@noop {} {\bibfield  {journal} {\bibinfo  {journal}
  {Nuclear Physics B-Proceedings Supplements}\ }\textbf {\bibinfo {volume}
  {119}},\ \bibinfo {pages} {894--899} (\bibinfo {year} {2003})}\BibitemShut
  {NoStop}%
\bibitem [{\citenamefont {Zhu}\ and\ \citenamefont
  {Zhang}(2019)}]{zhu2019gapless}%
  \BibitemOpen
  \bibfield  {author} {\bibinfo {author} {\bibfnamefont {Guo-Yi}\ \bibnamefont
  {Zhu}}\ and\ \bibinfo {author} {\bibfnamefont {Guang-Ming}\ \bibnamefont
  {Zhang}},\ }\bibfield  {title} {\enquote {\bibinfo {title} {Gapless coulomb
  state emerging from a self-dual topological tensor-network state},}\
  }\href@noop {} {\bibfield  {journal} {\bibinfo  {journal} {Physical review
  letters}\ }\textbf {\bibinfo {volume} {122}},\ \bibinfo {pages} {176401}
  (\bibinfo {year} {2019})}\BibitemShut {NoStop}%
\bibitem [{\citenamefont {Pujari}\ \emph {et~al.}(2015)\citenamefont {Pujari},
  \citenamefont {Alet},\ and\ \citenamefont {Damle}}]{pujari2015transitions}%
  \BibitemOpen
  \bibfield  {author} {\bibinfo {author} {\bibfnamefont {Sumiran}\ \bibnamefont
  {Pujari}}, \bibinfo {author} {\bibfnamefont {Fabien}\ \bibnamefont {Alet}}, \
  and\ \bibinfo {author} {\bibfnamefont {Kedar}\ \bibnamefont {Damle}},\
  }\bibfield  {title} {\enquote {\bibinfo {title} {Transitions to valence-bond
  solid order in a honeycomb lattice antiferromagnet},}\ }\href@noop {}
  {\bibfield  {journal} {\bibinfo  {journal} {Physical Review B}\ }\textbf
  {\bibinfo {volume} {91}},\ \bibinfo {pages} {104411} (\bibinfo {year}
  {2015})}\BibitemShut {NoStop}%
\bibitem [{\citenamefont {Shao}\ \emph {et~al.}(2020)\citenamefont {Shao},
  \citenamefont {Guo},\ and\ \citenamefont {Sandvik}}]{shao2020monte}%
  \BibitemOpen
  \bibfield  {author} {\bibinfo {author} {\bibfnamefont {Hui}\ \bibnamefont
  {Shao}}, \bibinfo {author} {\bibfnamefont {Wenan}\ \bibnamefont {Guo}}, \
  and\ \bibinfo {author} {\bibfnamefont {Anders~W}\ \bibnamefont {Sandvik}},\
  }\bibfield  {title} {\enquote {\bibinfo {title} {Monte carlo renormalization
  flows in the space of relevant and irrelevant operators: Application to
  three-dimensional clock models},}\ }\href@noop {} {\bibfield  {journal}
  {\bibinfo  {journal} {Physical Review Letters}\ }\textbf {\bibinfo {volume}
  {124}},\ \bibinfo {pages} {080602} (\bibinfo {year} {2020})}\BibitemShut
  {NoStop}%
\bibitem [{\citenamefont {Chester}\ \emph {et~al.}(2020)\citenamefont
  {Chester}, \citenamefont {Landry}, \citenamefont {Liu}, \citenamefont
  {Poland}, \citenamefont {Simmons-Duffin}, \citenamefont {Su},\ and\
  \citenamefont {Vichi}}]{chester2020carving}%
  \BibitemOpen
  \bibfield  {author} {\bibinfo {author} {\bibfnamefont {Shai~M}\ \bibnamefont
  {Chester}}, \bibinfo {author} {\bibfnamefont {Walter}\ \bibnamefont
  {Landry}}, \bibinfo {author} {\bibfnamefont {Junyu}\ \bibnamefont {Liu}},
  \bibinfo {author} {\bibfnamefont {David}\ \bibnamefont {Poland}}, \bibinfo
  {author} {\bibfnamefont {David}\ \bibnamefont {Simmons-Duffin}}, \bibinfo
  {author} {\bibfnamefont {Ning}\ \bibnamefont {Su}}, \ and\ \bibinfo {author}
  {\bibfnamefont {Alessandro}\ \bibnamefont {Vichi}},\ }\bibfield  {title}
  {\enquote {\bibinfo {title} {Carving out ope space and precise o(2) model
  critical exponents},}\ }\href@noop {} {\bibfield  {journal} {\bibinfo
  {journal} {Journal of High Energy Physics}\ }\textbf {\bibinfo {volume}
  {2020}},\ \bibinfo {pages} {1--52} (\bibinfo {year} {2020})}\BibitemShut
  {NoStop}%
\bibitem [{\citenamefont {Hasenbusch}\ and\ \citenamefont
  {Vicari}(2011)}]{hasenbusch2011anisotropic}%
  \BibitemOpen
  \bibfield  {author} {\bibinfo {author} {\bibfnamefont {Martin}\ \bibnamefont
  {Hasenbusch}}\ and\ \bibinfo {author} {\bibfnamefont {Ettore}\ \bibnamefont
  {Vicari}},\ }\bibfield  {title} {\enquote {\bibinfo {title} {Anisotropic
  perturbations in three-dimensional o (n)-symmetric vector models},}\
  }\href@noop {} {\bibfield  {journal} {\bibinfo  {journal} {Physical Review
  B}\ }\textbf {\bibinfo {volume} {84}},\ \bibinfo {pages} {125136} (\bibinfo
  {year} {2011})}\BibitemShut {NoStop}%
\bibitem [{\citenamefont {Hasenbusch}(2019)}]{hasenbusch2019monte}%
  \BibitemOpen
  \bibfield  {author} {\bibinfo {author} {\bibfnamefont {Martin}\ \bibnamefont
  {Hasenbusch}},\ }\bibfield  {title} {\enquote {\bibinfo {title} {Monte carlo
  study of an improved clock model in three dimensions},}\ }\href@noop {}
  {\bibfield  {journal} {\bibinfo  {journal} {Physical Review B}\ }\textbf
  {\bibinfo {volume} {100}},\ \bibinfo {pages} {224517} (\bibinfo {year}
  {2019})}\BibitemShut {NoStop}%
\bibitem [{\citenamefont {Wilson}(1974)}]{wilson1974confinement}%
  \BibitemOpen
  \bibfield  {author} {\bibinfo {author} {\bibfnamefont {Kenneth~G}\
  \bibnamefont {Wilson}},\ }\bibfield  {title} {\enquote {\bibinfo {title}
  {Confinement of quarks},}\ }\href@noop {} {\bibfield  {journal} {\bibinfo
  {journal} {Physical review D}\ }\textbf {\bibinfo {volume} {10}},\ \bibinfo
  {pages} {2445} (\bibinfo {year} {1974})}\BibitemShut {NoStop}%
\bibitem [{\citenamefont {Edwards}(1977)}]{edwards1977theory}%
  \BibitemOpen
  \bibfield  {author} {\bibinfo {author} {\bibfnamefont {SF}~\bibnamefont
  {Edwards}},\ }\bibfield  {title} {\enquote {\bibinfo {title} {The theory of
  rubber elasticity},}\ }\href@noop {} {\bibfield  {journal} {\bibinfo
  {journal} {British Polymer Journal}\ }\textbf {\bibinfo {volume} {9}},\
  \bibinfo {pages} {140--143} (\bibinfo {year} {1977})}\BibitemShut {NoStop}%
\bibitem [{\citenamefont {De~Gennes}\ and\ \citenamefont
  {Gennes}(1979)}]{de1979scaling}%
  \BibitemOpen
  \bibfield  {author} {\bibinfo {author} {\bibfnamefont {Pierre-Gilles}\
  \bibnamefont {De~Gennes}}\ and\ \bibinfo {author} {\bibfnamefont
  {Pierre-Gilles}\ \bibnamefont {Gennes}},\ }\href@noop {} {\emph {\bibinfo
  {title} {Scaling concepts in polymer physics}}}\ (\bibinfo  {publisher}
  {Cornell university press},\ \bibinfo {year} {1979})\BibitemShut {NoStop}%
\bibitem [{\citenamefont {des Cloizeaux}(1981)}]{des1981ring}%
  \BibitemOpen
  \bibfield  {author} {\bibinfo {author} {\bibfnamefont {Jacques}\ \bibnamefont
  {des Cloizeaux}},\ }\bibfield  {title} {\enquote {\bibinfo {title} {Ring
  polymers in solution: topological effects},}\ }\href@noop {} {\bibfield
  {journal} {\bibinfo  {journal} {Journal de Physique Lettres}\ }\textbf
  {\bibinfo {volume} {42}},\ \bibinfo {pages} {433--436} (\bibinfo {year}
  {1981})}\BibitemShut {NoStop}%
\bibitem [{\citenamefont {Khokhlov}\ and\ \citenamefont
  {Nechaev}(1985)}]{khokhlov1985polymer}%
  \BibitemOpen
  \bibfield  {author} {\bibinfo {author} {\bibfnamefont {AR}~\bibnamefont
  {Khokhlov}}\ and\ \bibinfo {author} {\bibfnamefont {SK}~\bibnamefont
  {Nechaev}},\ }\bibfield  {title} {\enquote {\bibinfo {title} {Polymer chain
  in an array of obstacles},}\ }\href@noop {} {\bibfield  {journal} {\bibinfo
  {journal} {Physics Letters A}\ }\textbf {\bibinfo {volume} {112}},\ \bibinfo
  {pages} {156--160} (\bibinfo {year} {1985})}\BibitemShut {NoStop}%
\bibitem [{\citenamefont {Rubinstein}(1986)}]{rubinstein1986dynamics}%
  \BibitemOpen
  \bibfield  {author} {\bibinfo {author} {\bibfnamefont {Michael}\ \bibnamefont
  {Rubinstein}},\ }\bibfield  {title} {\enquote {\bibinfo {title} {Dynamics of
  ring polymers in the presence of fixed obstacles},}\ }\href@noop {}
  {\bibfield  {journal} {\bibinfo  {journal} {Physical review letters}\
  }\textbf {\bibinfo {volume} {57}},\ \bibinfo {pages} {3023} (\bibinfo {year}
  {1986})}\BibitemShut {NoStop}%
\bibitem [{\citenamefont {Cates}\ and\ \citenamefont
  {Deutsch}(1986)}]{cates1986j}%
  \BibitemOpen
  \bibfield  {author} {\bibinfo {author} {\bibfnamefont {ME}~\bibnamefont
  {Cates}}\ and\ \bibinfo {author} {\bibfnamefont {JM}~\bibnamefont
  {Deutsch}},\ }\bibfield  {title} {\enquote {\bibinfo {title} {J phys
  paris},}\ }\href@noop {} {\bibfield  {journal} {\bibinfo  {journal} {J Phys
  (Paris)}\ }\textbf {\bibinfo {volume} {47}},\ \bibinfo {pages} {2121}
  (\bibinfo {year} {1986})}\BibitemShut {NoStop}%
\bibitem [{\citenamefont {Halverson}\ \emph {et~al.}(2014)\citenamefont
  {Halverson}, \citenamefont {Smrek}, \citenamefont {Kremer},\ and\
  \citenamefont {Grosberg}}]{halverson2014melt}%
  \BibitemOpen
  \bibfield  {author} {\bibinfo {author} {\bibfnamefont {Jonathan~D}\
  \bibnamefont {Halverson}}, \bibinfo {author} {\bibfnamefont {Jan}\
  \bibnamefont {Smrek}}, \bibinfo {author} {\bibfnamefont {Kurt}\ \bibnamefont
  {Kremer}}, \ and\ \bibinfo {author} {\bibfnamefont {Alexander~Y}\
  \bibnamefont {Grosberg}},\ }\bibfield  {title} {\enquote {\bibinfo {title}
  {From a melt of rings to chromosome territories: the role of topological
  constraints in genome folding},}\ }\href@noop {} {\bibfield  {journal}
  {\bibinfo  {journal} {Reports on Progress in Physics}\ }\textbf {\bibinfo
  {volume} {77}},\ \bibinfo {pages} {022601} (\bibinfo {year}
  {2014})}\BibitemShut {NoStop}%
\bibitem [{\citenamefont {Imakaev}\ \emph {et~al.}(2015)\citenamefont
  {Imakaev}, \citenamefont {Tchourine}, \citenamefont {Nechaev},\ and\
  \citenamefont {Mirny}}]{imakaev2015effects}%
  \BibitemOpen
  \bibfield  {author} {\bibinfo {author} {\bibfnamefont {Maxim~V}\ \bibnamefont
  {Imakaev}}, \bibinfo {author} {\bibfnamefont {Konstantin~M}\ \bibnamefont
  {Tchourine}}, \bibinfo {author} {\bibfnamefont {Sergei~K}\ \bibnamefont
  {Nechaev}}, \ and\ \bibinfo {author} {\bibfnamefont {Leonid~A}\ \bibnamefont
  {Mirny}},\ }\bibfield  {title} {\enquote {\bibinfo {title} {Effects of
  topological constraints on globular polymers},}\ }\href@noop {} {\bibfield
  {journal} {\bibinfo  {journal} {Soft matter}\ }\textbf {\bibinfo {volume}
  {11}},\ \bibinfo {pages} {665--671} (\bibinfo {year} {2015})}\BibitemShut
  {NoStop}%
\bibitem [{\citenamefont {Serna}\ \emph {et~al.}(2015)\citenamefont {Serna},
  \citenamefont {Bunin},\ and\ \citenamefont {Nahum}}]{serna2015topological}%
  \BibitemOpen
  \bibfield  {author} {\bibinfo {author} {\bibfnamefont {Pablo}\ \bibnamefont
  {Serna}}, \bibinfo {author} {\bibfnamefont {Guy}\ \bibnamefont {Bunin}}, \
  and\ \bibinfo {author} {\bibfnamefont {Adam}\ \bibnamefont {Nahum}},\
  }\bibfield  {title} {\enquote {\bibinfo {title} {Topological constraints in
  directed polymer melts},}\ }\href@noop {} {\bibfield  {journal} {\bibinfo
  {journal} {Physical review letters}\ }\textbf {\bibinfo {volume} {115}},\
  \bibinfo {pages} {228303} (\bibinfo {year} {2015})}\BibitemShut {NoStop}%
\bibitem [{\citenamefont {De~Forcrand}(1995)}]{de1995progress}%
  \BibitemOpen
  \bibfield  {author} {\bibinfo {author} {\bibfnamefont {Philippe}\
  \bibnamefont {De~Forcrand}},\ }\bibfield  {title} {\enquote {\bibinfo {title}
  {Progress on lattice qcd algorithms},}\ }\href@noop {} {\bibfield  {journal}
  {\bibinfo  {journal} {arXiv preprint hep-lat/9509082}\ } (\bibinfo {year}
  {1995})}\BibitemShut {NoStop}%
\bibitem [{\citenamefont {Peardon}(2002)}]{peardon2002progress}%
  \BibitemOpen
  \bibfield  {author} {\bibinfo {author} {\bibfnamefont {Mike}\ \bibnamefont
  {Peardon}},\ }\bibfield  {title} {\enquote {\bibinfo {title} {Progress in
  lattice algorithms},}\ }\href@noop {} {\bibfield  {journal} {\bibinfo
  {journal} {arXiv preprint hep-lat/0201003}\ } (\bibinfo {year}
  {2002})}\BibitemShut {NoStop}%
\bibitem [{\citenamefont {Schaefer}(2011)}]{schaefer2011algorithms}%
  \BibitemOpen
  \bibfield  {author} {\bibinfo {author} {\bibfnamefont {Stefan}\ \bibnamefont
  {Schaefer}},\ }\bibfield  {title} {\enquote {\bibinfo {title} {Algorithms for
  lattice qcd: progress and challenges},}\ }in\ \href@noop {} {\emph {\bibinfo
  {booktitle} {AIP Conference Proceedings}}},\ Vol.\ \bibinfo {volume} {1343}\
  (\bibinfo {organization} {American Institute of Physics},\ \bibinfo {year}
  {2011})\ pp.\ \bibinfo {pages} {93--98}\BibitemShut {NoStop}%
\bibitem [{\citenamefont {Ishikawa}(2009)}]{ishikawa2009recent}%
  \BibitemOpen
  \bibfield  {author} {\bibinfo {author} {\bibfnamefont {Ken-Ichi}\
  \bibnamefont {Ishikawa}},\ }\bibfield  {title} {\enquote {\bibinfo {title}
  {Recent algorithm and machine developments for lattice qcd},}\ }in\
  \href@noop {} {\emph {\bibinfo {booktitle} {The XXVI International Symposium
  on Lattice Field Theory}}},\ Vol.~\bibinfo {volume} {66}\ (\bibinfo
  {organization} {SISSA Medialab},\ \bibinfo {year} {2009})\ p.\ \bibinfo
  {pages} {013}\BibitemShut {NoStop}%
\bibitem [{\citenamefont {Liu}\ \emph {et~al.}(2017)\citenamefont {Liu},
  \citenamefont {Qi}, \citenamefont {Meng},\ and\ \citenamefont
  {Fu}}]{liu2017selflearningmontecarlomethod}%
  \BibitemOpen
  \bibfield  {author} {\bibinfo {author} {\bibfnamefont {Junwei}\ \bibnamefont
  {Liu}}, \bibinfo {author} {\bibfnamefont {Yang}\ \bibnamefont {Qi}}, \bibinfo
  {author} {\bibfnamefont {Zi~Yang}\ \bibnamefont {Meng}}, \ and\ \bibinfo
  {author} {\bibfnamefont {Liang}\ \bibnamefont {Fu}},\ }\bibfield  {title}
  {\enquote {\bibinfo {title} {Self-learning monte carlo method},}\ }\href
  {\doibase 10.1103/PhysRevB.95.041101} {\bibfield  {journal} {\bibinfo
  {journal} {Phys. Rev. B}\ }\textbf {\bibinfo {volume} {95}},\ \bibinfo
  {pages} {041101} (\bibinfo {year} {2017})}\BibitemShut {NoStop}%
\bibitem [{\citenamefont {Huang}\ and\ \citenamefont
  {Wang}(2017)}]{huang2017accelerated}%
  \BibitemOpen
  \bibfield  {author} {\bibinfo {author} {\bibfnamefont {Li}~\bibnamefont
  {Huang}}\ and\ \bibinfo {author} {\bibfnamefont {Lei}\ \bibnamefont {Wang}},\
  }\bibfield  {title} {\enquote {\bibinfo {title} {Accelerated monte carlo
  simulations with restricted boltzmann machines},}\ }\href@noop {} {\bibfield
  {journal} {\bibinfo  {journal} {Physical Review B}\ }\textbf {\bibinfo
  {volume} {95}},\ \bibinfo {pages} {035105} (\bibinfo {year}
  {2017})}\BibitemShut {NoStop}%
\bibitem [{\citenamefont {Suwa}\ \emph {et~al.}(2016)\citenamefont {Suwa},
  \citenamefont {Sen},\ and\ \citenamefont {Sandvik}}]{suwa2016level}%
  \BibitemOpen
  \bibfield  {author} {\bibinfo {author} {\bibfnamefont {Hidemaro}\
  \bibnamefont {Suwa}}, \bibinfo {author} {\bibfnamefont {Arnab}\ \bibnamefont
  {Sen}}, \ and\ \bibinfo {author} {\bibfnamefont {Anders~W}\ \bibnamefont
  {Sandvik}},\ }\bibfield  {title} {\enquote {\bibinfo {title} {Level
  spectroscopy in a two-dimensional quantum magnet: Linearly dispersing spinons
  at the deconfined quantum critical point},}\ }\href@noop {} {\bibfield
  {journal} {\bibinfo  {journal} {Physical Review B}\ }\textbf {\bibinfo
  {volume} {94}},\ \bibinfo {pages} {144416} (\bibinfo {year}
  {2016})}\BibitemShut {NoStop}%
\bibitem [{\citenamefont {Vanderstraeten}\ \emph {et~al.}(2015)\citenamefont
  {Vanderstraeten}, \citenamefont {Mari{\"e}n}, \citenamefont {Verstraete},\
  and\ \citenamefont {Haegeman}}]{vanderstraeten2015excitations}%
  \BibitemOpen
  \bibfield  {author} {\bibinfo {author} {\bibfnamefont {Laurens}\ \bibnamefont
  {Vanderstraeten}}, \bibinfo {author} {\bibfnamefont {Micha{\"e}l}\
  \bibnamefont {Mari{\"e}n}}, \bibinfo {author} {\bibfnamefont {Frank}\
  \bibnamefont {Verstraete}}, \ and\ \bibinfo {author} {\bibfnamefont {Jutho}\
  \bibnamefont {Haegeman}},\ }\bibfield  {title} {\enquote {\bibinfo {title}
  {Excitations and the tangent space of projected entangled-pair states},}\
  }\href@noop {} {\bibfield  {journal} {\bibinfo  {journal} {Physical Review
  B}\ }\textbf {\bibinfo {volume} {92}},\ \bibinfo {pages} {201111} (\bibinfo
  {year} {2015})}\BibitemShut {NoStop}%
\bibitem [{\citenamefont {Vanderstraeten}\ \emph {et~al.}(2019)\citenamefont
  {Vanderstraeten}, \citenamefont {Haegeman},\ and\ \citenamefont
  {Verstraete}}]{vanderstraeten2019simulating}%
  \BibitemOpen
  \bibfield  {author} {\bibinfo {author} {\bibfnamefont {Laurens}\ \bibnamefont
  {Vanderstraeten}}, \bibinfo {author} {\bibfnamefont {Jutho}\ \bibnamefont
  {Haegeman}}, \ and\ \bibinfo {author} {\bibfnamefont {Frank}\ \bibnamefont
  {Verstraete}},\ }\bibfield  {title} {\enquote {\bibinfo {title} {Simulating
  excitation spectra with projected entangled-pair states},}\ }\href@noop {}
  {\bibfield  {journal} {\bibinfo  {journal} {Physical Review B}\ }\textbf
  {\bibinfo {volume} {99}},\ \bibinfo {pages} {165121} (\bibinfo {year}
  {2019})}\BibitemShut {NoStop}%
\bibitem [{\citenamefont {Dalmonte}\ and\ \citenamefont
  {Montangero}(2016)}]{dalmonte2016lattice}%
  \BibitemOpen
  \bibfield  {author} {\bibinfo {author} {\bibfnamefont {Marcello}\
  \bibnamefont {Dalmonte}}\ and\ \bibinfo {author} {\bibfnamefont {Simone}\
  \bibnamefont {Montangero}},\ }\bibfield  {title} {\enquote {\bibinfo {title}
  {Lattice gauge theory simulations in the quantum information era},}\
  }\href@noop {} {\bibfield  {journal} {\bibinfo  {journal} {Contemporary
  Physics}\ }\textbf {\bibinfo {volume} {57}},\ \bibinfo {pages} {388--412}
  (\bibinfo {year} {2016})}\BibitemShut {NoStop}%
\bibitem [{\citenamefont {Zohar}\ \emph
  {et~al.}(2017{\natexlab{a}})\citenamefont {Zohar}, \citenamefont {Farace},
  \citenamefont {Reznik},\ and\ \citenamefont {Cirac}}]{zohar2017digital}%
  \BibitemOpen
  \bibfield  {author} {\bibinfo {author} {\bibfnamefont {Erez}\ \bibnamefont
  {Zohar}}, \bibinfo {author} {\bibfnamefont {Alessandro}\ \bibnamefont
  {Farace}}, \bibinfo {author} {\bibfnamefont {Benni}\ \bibnamefont {Reznik}},
  \ and\ \bibinfo {author} {\bibfnamefont {J~Ignacio}\ \bibnamefont {Cirac}},\
  }\bibfield  {title} {\enquote {\bibinfo {title} {Digital lattice gauge
  theories},}\ }\href@noop {} {\bibfield  {journal} {\bibinfo  {journal}
  {Physical Review A}\ }\textbf {\bibinfo {volume} {95}},\ \bibinfo {pages}
  {023604} (\bibinfo {year} {2017}{\natexlab{a}})}\BibitemShut {NoStop}%
\bibitem [{\citenamefont {Zohar}\ \emph
  {et~al.}(2017{\natexlab{b}})\citenamefont {Zohar}, \citenamefont {Farace},
  \citenamefont {Reznik},\ and\ \citenamefont {Cirac}}]{zohar2017digital2}%
  \BibitemOpen
  \bibfield  {author} {\bibinfo {author} {\bibfnamefont {Erez}\ \bibnamefont
  {Zohar}}, \bibinfo {author} {\bibfnamefont {Alessandro}\ \bibnamefont
  {Farace}}, \bibinfo {author} {\bibfnamefont {Benni}\ \bibnamefont {Reznik}},
  \ and\ \bibinfo {author} {\bibfnamefont {J~Ignacio}\ \bibnamefont {Cirac}},\
  }\bibfield  {title} {\enquote {\bibinfo {title} {Digital quantum simulation
  of z2 lattice gauge theories with dynamical fermionic matter},}\ }\href@noop
  {} {\bibfield  {journal} {\bibinfo  {journal} {Physical Review Letters}\
  }\textbf {\bibinfo {volume} {118}},\ \bibinfo {pages} {070501} (\bibinfo
  {year} {2017}{\natexlab{b}})}\BibitemShut {NoStop}%
\bibitem [{\citenamefont {Schweizer}\ \emph {et~al.}(2019)\citenamefont
  {Schweizer}, \citenamefont {Grusdt}, \citenamefont {Berngruber},
  \citenamefont {Barbiero}, \citenamefont {Demler}, \citenamefont {Goldman},
  \citenamefont {Bloch},\ and\ \citenamefont
  {Aidelsburger}}]{schweizer2019floquet}%
  \BibitemOpen
  \bibfield  {author} {\bibinfo {author} {\bibfnamefont {Christian}\
  \bibnamefont {Schweizer}}, \bibinfo {author} {\bibfnamefont {Fabian}\
  \bibnamefont {Grusdt}}, \bibinfo {author} {\bibfnamefont {Moritz}\
  \bibnamefont {Berngruber}}, \bibinfo {author} {\bibfnamefont {Luca}\
  \bibnamefont {Barbiero}}, \bibinfo {author} {\bibfnamefont {Eugene}\
  \bibnamefont {Demler}}, \bibinfo {author} {\bibfnamefont {Nathan}\
  \bibnamefont {Goldman}}, \bibinfo {author} {\bibfnamefont {Immanuel}\
  \bibnamefont {Bloch}}, \ and\ \bibinfo {author} {\bibfnamefont {Monika}\
  \bibnamefont {Aidelsburger}},\ }\bibfield  {title} {\enquote {\bibinfo
  {title} {Floquet approach to ℤ 2 lattice gauge theories with ultracold
  atoms in optical lattices},}\ }\href@noop {} {\bibfield  {journal} {\bibinfo
  {journal} {Nature Physics}\ }\textbf {\bibinfo {volume} {15}},\ \bibinfo
  {pages} {1168--1173} (\bibinfo {year} {2019})}\BibitemShut {NoStop}%
\bibitem [{\citenamefont {Barbiero}\ \emph {et~al.}(2019)\citenamefont
  {Barbiero}, \citenamefont {Schweizer}, \citenamefont {Aidelsburger},
  \citenamefont {Demler}, \citenamefont {Goldman},\ and\ \citenamefont
  {Grusdt}}]{barbiero2019coupling}%
  \BibitemOpen
  \bibfield  {author} {\bibinfo {author} {\bibfnamefont {Luca}\ \bibnamefont
  {Barbiero}}, \bibinfo {author} {\bibfnamefont {Christian}\ \bibnamefont
  {Schweizer}}, \bibinfo {author} {\bibfnamefont {Monika}\ \bibnamefont
  {Aidelsburger}}, \bibinfo {author} {\bibfnamefont {Eugene}\ \bibnamefont
  {Demler}}, \bibinfo {author} {\bibfnamefont {Nathan}\ \bibnamefont
  {Goldman}}, \ and\ \bibinfo {author} {\bibfnamefont {Fabian}\ \bibnamefont
  {Grusdt}},\ }\bibfield  {title} {\enquote {\bibinfo {title} {Coupling
  ultracold matter to dynamical gauge fields in optical lattices: From flux
  attachment to ℤ2 lattice gauge theories},}\ }\href@noop {} {\bibfield
  {journal} {\bibinfo  {journal} {Science advances}\ }\textbf {\bibinfo
  {volume} {5}},\ \bibinfo {pages} {eaav7444} (\bibinfo {year}
  {2019})}\BibitemShut {NoStop}%
\bibitem [{\citenamefont {Ba{\~n}uls}\ and\ \citenamefont
  {Cichy}(2020)}]{banuls2020review}%
  \BibitemOpen
  \bibfield  {author} {\bibinfo {author} {\bibfnamefont {Mari~Carmen}\
  \bibnamefont {Ba{\~n}uls}}\ and\ \bibinfo {author} {\bibfnamefont
  {Krzysztof}\ \bibnamefont {Cichy}},\ }\bibfield  {title} {\enquote {\bibinfo
  {title} {Review on novel methods for lattice gauge theories},}\ }\href@noop
  {} {\bibfield  {journal} {\bibinfo  {journal} {Reports on Progress in
  Physics}\ }\textbf {\bibinfo {volume} {83}},\ \bibinfo {pages} {024401}
  (\bibinfo {year} {2020})}\BibitemShut {NoStop}%
\bibitem [{\citenamefont {Davoudi}\ \emph {et~al.}(2020)\citenamefont
  {Davoudi}, \citenamefont {Hafezi}, \citenamefont {Monroe}, \citenamefont
  {Pagano}, \citenamefont {Seif},\ and\ \citenamefont
  {Shaw}}]{davoudi2020towards}%
  \BibitemOpen
  \bibfield  {author} {\bibinfo {author} {\bibfnamefont {Zohreh}\ \bibnamefont
  {Davoudi}}, \bibinfo {author} {\bibfnamefont {Mohammad}\ \bibnamefont
  {Hafezi}}, \bibinfo {author} {\bibfnamefont {Christopher}\ \bibnamefont
  {Monroe}}, \bibinfo {author} {\bibfnamefont {Guido}\ \bibnamefont {Pagano}},
  \bibinfo {author} {\bibfnamefont {Alireza}\ \bibnamefont {Seif}}, \ and\
  \bibinfo {author} {\bibfnamefont {Andrew}\ \bibnamefont {Shaw}},\ }\bibfield
  {title} {\enquote {\bibinfo {title} {Towards analog quantum simulations of
  lattice gauge theories with trapped ions},}\ }\href@noop {} {\bibfield
  {journal} {\bibinfo  {journal} {Physical Review Research}\ }\textbf {\bibinfo
  {volume} {2}},\ \bibinfo {pages} {023015} (\bibinfo {year}
  {2020})}\BibitemShut {NoStop}%
\bibitem [{\citenamefont {Cui}\ \emph {et~al.}(2020)\citenamefont {Cui},
  \citenamefont {Shi},\ and\ \citenamefont {Yang}}]{cui2020circuit}%
  \BibitemOpen
  \bibfield  {author} {\bibinfo {author} {\bibfnamefont {Xiaopeng}\
  \bibnamefont {Cui}}, \bibinfo {author} {\bibfnamefont {Yu}~\bibnamefont
  {Shi}}, \ and\ \bibinfo {author} {\bibfnamefont {Ji-Chong}\ \bibnamefont
  {Yang}},\ }\bibfield  {title} {\enquote {\bibinfo {title} {Circuit-based
  digital adiabatic quantum simulation and pseudoquantum simulation as new
  approaches to lattice gauge theory},}\ }\href@noop {} {\bibfield  {journal}
  {\bibinfo  {journal} {Journal of High Energy Physics}\ }\textbf {\bibinfo
  {volume} {2020}},\ \bibinfo {pages} {1--35} (\bibinfo {year}
  {2020})}\BibitemShut {NoStop}%
\bibitem [{\citenamefont {Paulson}\ \emph {et~al.}(2020)\citenamefont
  {Paulson}, \citenamefont {Dellantonio}, \citenamefont {Haase}, \citenamefont
  {Celi}, \citenamefont {Kan}, \citenamefont {Jena}, \citenamefont {Kokail},
  \citenamefont {van Bijnen}, \citenamefont {Jansen}, \citenamefont {Zoller}
  \emph {et~al.}}]{paulson2020towards}%
  \BibitemOpen
  \bibfield  {author} {\bibinfo {author} {\bibfnamefont {Danny}\ \bibnamefont
  {Paulson}}, \bibinfo {author} {\bibfnamefont {Luca}\ \bibnamefont
  {Dellantonio}}, \bibinfo {author} {\bibfnamefont {Jan~F}\ \bibnamefont
  {Haase}}, \bibinfo {author} {\bibfnamefont {Alessio}\ \bibnamefont {Celi}},
  \bibinfo {author} {\bibfnamefont {Angus}\ \bibnamefont {Kan}}, \bibinfo
  {author} {\bibfnamefont {Andrew}\ \bibnamefont {Jena}}, \bibinfo {author}
  {\bibfnamefont {Christian}\ \bibnamefont {Kokail}}, \bibinfo {author}
  {\bibfnamefont {Rick}\ \bibnamefont {van Bijnen}}, \bibinfo {author}
  {\bibfnamefont {Karl}\ \bibnamefont {Jansen}}, \bibinfo {author}
  {\bibfnamefont {Peter}\ \bibnamefont {Zoller}},  \emph {et~al.},\ }\bibfield
  {title} {\enquote {\bibinfo {title} {Towards simulating 2d effects in lattice
  gauge theories on a quantum computer},}\ }\href@noop {} {\bibfield  {journal}
  {\bibinfo  {journal} {arXiv preprint arXiv:2008.09252}\ } (\bibinfo {year}
  {2020})}\BibitemShut {NoStop}%
\bibitem [{\citenamefont {Poland}\ \emph {et~al.}(2019)\citenamefont {Poland},
  \citenamefont {Rychkov},\ and\ \citenamefont {Vichi}}]{poland2019conformal}%
  \BibitemOpen
  \bibfield  {author} {\bibinfo {author} {\bibfnamefont {David}\ \bibnamefont
  {Poland}}, \bibinfo {author} {\bibfnamefont {Slava}\ \bibnamefont {Rychkov}},
  \ and\ \bibinfo {author} {\bibfnamefont {Alessandro}\ \bibnamefont {Vichi}},\
  }\bibfield  {title} {\enquote {\bibinfo {title} {The conformal bootstrap:
  Theory, numerical techniques, and applications},}\ }\href@noop {} {\bibfield
  {journal} {\bibinfo  {journal} {Reviews of Modern Physics}\ }\textbf
  {\bibinfo {volume} {91}},\ \bibinfo {pages} {015002} (\bibinfo {year}
  {2019})}\BibitemShut {NoStop}%
\bibitem [{\citenamefont {El-Showk}\ \emph {et~al.}(2012)\citenamefont
  {El-Showk}, \citenamefont {Paulos}, \citenamefont {Poland}, \citenamefont
  {Rychkov}, \citenamefont {Simmons-Duffin},\ and\ \citenamefont
  {Vichi}}]{el2012solving}%
  \BibitemOpen
  \bibfield  {author} {\bibinfo {author} {\bibfnamefont {Sheer}\ \bibnamefont
  {El-Showk}}, \bibinfo {author} {\bibfnamefont {Miguel~F}\ \bibnamefont
  {Paulos}}, \bibinfo {author} {\bibfnamefont {David}\ \bibnamefont {Poland}},
  \bibinfo {author} {\bibfnamefont {Slava}\ \bibnamefont {Rychkov}}, \bibinfo
  {author} {\bibfnamefont {David}\ \bibnamefont {Simmons-Duffin}}, \ and\
  \bibinfo {author} {\bibfnamefont {Alessandro}\ \bibnamefont {Vichi}},\
  }\bibfield  {title} {\enquote {\bibinfo {title} {Solving the 3d ising model
  with the conformal bootstrap},}\ }\href@noop {} {\bibfield  {journal}
  {\bibinfo  {journal} {Physical Review D}\ }\textbf {\bibinfo {volume} {86}},\
  \bibinfo {pages} {025022} (\bibinfo {year} {2012})}\BibitemShut {NoStop}%
\bibitem [{\citenamefont {Kos}\ \emph {et~al.}(2014)\citenamefont {Kos},
  \citenamefont {Poland},\ and\ \citenamefont
  {Simmons-Duffin}}]{kos2014bootstrapping}%
  \BibitemOpen
  \bibfield  {author} {\bibinfo {author} {\bibfnamefont {Filip}\ \bibnamefont
  {Kos}}, \bibinfo {author} {\bibfnamefont {David}\ \bibnamefont {Poland}}, \
  and\ \bibinfo {author} {\bibfnamefont {David}\ \bibnamefont
  {Simmons-Duffin}},\ }\bibfield  {title} {\enquote {\bibinfo {title}
  {Bootstrapping mixed correlators in the 3d ising model},}\ }\href@noop {}
  {\bibfield  {journal} {\bibinfo  {journal} {Journal of High Energy Physics}\
  }\textbf {\bibinfo {volume} {2014}},\ \bibinfo {pages} {109} (\bibinfo {year}
  {2014})}\BibitemShut {NoStop}%
\bibitem [{\citenamefont {Kos}\ \emph {et~al.}(2016)\citenamefont {Kos},
  \citenamefont {Poland}, \citenamefont {Simmons-Duffin},\ and\ \citenamefont
  {Vichi}}]{kos2016precision}%
  \BibitemOpen
  \bibfield  {author} {\bibinfo {author} {\bibfnamefont {Filip}\ \bibnamefont
  {Kos}}, \bibinfo {author} {\bibfnamefont {David}\ \bibnamefont {Poland}},
  \bibinfo {author} {\bibfnamefont {David}\ \bibnamefont {Simmons-Duffin}}, \
  and\ \bibinfo {author} {\bibfnamefont {Alessandro}\ \bibnamefont {Vichi}},\
  }\bibfield  {title} {\enquote {\bibinfo {title} {Precision islands in the
  ising and o (n) models},}\ }\href@noop {} {\bibfield  {journal} {\bibinfo
  {journal} {Journal of High Energy Physics}\ }\textbf {\bibinfo {volume}
  {2016}},\ \bibinfo {pages} {36} (\bibinfo {year} {2016})}\BibitemShut
  {NoStop}%
\bibitem [{\citenamefont {Ferrenberg}\ \emph {et~al.}(2018)\citenamefont
  {Ferrenberg}, \citenamefont {Xu},\ and\ \citenamefont
  {Landau}}]{ferrenberg2018pushing}%
  \BibitemOpen
  \bibfield  {author} {\bibinfo {author} {\bibfnamefont {Alan~M}\ \bibnamefont
  {Ferrenberg}}, \bibinfo {author} {\bibfnamefont {Jiahao}\ \bibnamefont {Xu}},
  \ and\ \bibinfo {author} {\bibfnamefont {David~P}\ \bibnamefont {Landau}},\
  }\bibfield  {title} {\enquote {\bibinfo {title} {Pushing the limits of monte
  carlo simulations for the three-dimensional ising model},}\ }\href@noop {}
  {\bibfield  {journal} {\bibinfo  {journal} {Physical Review E}\ }\textbf
  {\bibinfo {volume} {97}},\ \bibinfo {pages} {043301} (\bibinfo {year}
  {2018})}\BibitemShut {NoStop}%
\bibitem [{\citenamefont {Nahum}\ \emph {et~al.}(2013)\citenamefont {Nahum},
  \citenamefont {Chalker}, \citenamefont {Serna}, \citenamefont {Ortuno},\ and\
  \citenamefont {Somoza}}]{nahum2013phase}%
  \BibitemOpen
  \bibfield  {author} {\bibinfo {author} {\bibfnamefont {Adam}\ \bibnamefont
  {Nahum}}, \bibinfo {author} {\bibfnamefont {J.~T.}\ \bibnamefont {Chalker}},
  \bibinfo {author} {\bibfnamefont {P.}~\bibnamefont {Serna}}, \bibinfo
  {author} {\bibfnamefont {M.}~\bibnamefont {Ortuno}}, \ and\ \bibinfo {author}
  {\bibfnamefont {A.~M.}\ \bibnamefont {Somoza}},\ }\bibfield  {title}
  {\enquote {\bibinfo {title} {Phase transitions in three-dimensional loop
  models and the c p n- 1 sigma model},}\ }\href@noop {} {\bibfield  {journal}
  {\bibinfo  {journal} {Physical Review B}\ }\textbf {\bibinfo {volume} {88}},\
  \bibinfo {pages} {134411} (\bibinfo {year} {2013})}\BibitemShut {NoStop}%
\bibitem [{\citenamefont {Xu}\ \emph {et~al.}(2014)\citenamefont {Xu},
  \citenamefont {Wang}, \citenamefont {Lv},\ and\ \citenamefont
  {Deng}}]{xu2014simultaneous}%
  \BibitemOpen
  \bibfield  {author} {\bibinfo {author} {\bibfnamefont {Xiao}\ \bibnamefont
  {Xu}}, \bibinfo {author} {\bibfnamefont {Junfeng}\ \bibnamefont {Wang}},
  \bibinfo {author} {\bibfnamefont {Jian-Ping}\ \bibnamefont {Lv}}, \ and\
  \bibinfo {author} {\bibfnamefont {Youjin}\ \bibnamefont {Deng}},\ }\bibfield
  {title} {\enquote {\bibinfo {title} {Simultaneous analysis of
  three-dimensional percolation models},}\ }\href@noop {} {\bibfield  {journal}
  {\bibinfo  {journal} {Frontiers of Physics}\ }\textbf {\bibinfo {volume}
  {9}},\ \bibinfo {pages} {113--119} (\bibinfo {year} {2014})}\BibitemShut
  {NoStop}%
\end{thebibliography}%
\end{document}